\documentclass{lmcs}

\usepackage{tikz}
\usetikzlibrary{positioning,shapes.geometric}

\title[Boolean Basis and Succinctness of ML and CTL]{Boolean Basis and Succinctness of Modal Logic and Computation Tree Logic via Hella-Vilander games} 

\author[S. Pfau]{Sebastian Pfau}

\address{Technische Universität Ilmenau, Germany}

\email{sebastian.pfau@tu-ilmenau.de}


\begin{document}

\tikzset{
 formula/.style={
    baseline={([yshift={-\ht\strutbox}]current bounding box.north)},
    every node/.append style={inner sep=2pt},
    level distance=.65cm},
 graph/.style={circle, draw=black!60, minimum size=1.5cm}
}

\begin{abstract}
The Hella-Vilander game for modal logic is a model comparison game that captures the formula size necessary to separate sets of pointed Kripke structures. We introduce the M-ON game for modal logic as a modification of this game. Our game captures the necessary number of modal operators, i.e., Diamonds and Boxes instead of formula size. We use our game to show that the bi-implication, sometimes also called equivalence, enables us to write modal logic formula with significantly fewer modal operators. With this we show, that with bi-implications we can also write significantly shorter modal logic formulas. This result holds even if only special classes of Kripke structures are considered.
To be more precise we show that there is an exponential succinctness gap between modal logic and its extension with bi-implication on the class of structures with a transitive and reflexive accessibility relation, as well as on the class of structures with a symmetrical and reflexive accessibility relation. Next we show that for the class of structures with a transitive and symmetrical accessibility relation this succinctness gap disappears. Then we introduce a version of our game for computation tree logic and use it to show that with bi-implications we can also write formulas in this logic with exponentially less temporal operators. Lastly we show that for both logics without bi-implication the representation through DAGs or circuits is exponentially more succinct than the representation with formulas.
\end{abstract}

\maketitle

\section{Introduction}
Model comparison games, like the Ehrenfeucht-Fra\"{\i}ss\'{e} game \cite{Ehrenfeucht, Fraisse}, are a type of logic games. They can be used to prove that it is difficult to express certain properties in a logic. For example the game defined by Adler and Immerman in \cite{A-I} has been used to prove lower bounds for formula size in FO$^2$(TC) and the logic CTL. Extended syntax trees \cite{Grohe2005} can be seen as an equivalent method to this game, even though they drop the game setting. The game and the syntax tree method can also be modified to work on other logics, see \cite{French, VanDerHoeck, VanDitmarsch2014139} for examples. In \cite{HellaVilander} Hella and Vilander introduce their own game, which builds on the Adler-Immerman game. This Hella-Vilander game is defined for modal logic and the $\mu$-calculus, but it can also be adapted for other logics. The main idea of this game is as follows: players $S$ (\emph{Spoiler, Samson}) and $D$ (\emph{Duplicator, Delilah}) play on two sets of structures $\mathbb{A}$ and $\mathbb{B}$ with a resource parameter $k$. Player $S$ spends the parameter $k$ in order to make his moves, while $D$ answers. $S$ has to play a winning move before $k$ runs out. In such a play $S$ has a winning strategy if, and only if, there is a formula $\varphi$ of size at most $k$ of the corresponding logic, that separates $\mathbb{A}$ and $\mathbb{B}$.

There are also model comparison games that only count the number of occurrences of specific characters, instead of the more general formula size. See \cite{Carmosino, Fagin2025} for examples. As part of their conclusion Hella and Vilander suggest that their game can be modified to count a more specific parameter as well.
One such possible parameter is the number of modal operators used in a modal logic formula, i.e., $\Diamond$ and $\Box$. This can be done, because every move of the Hella-Vilander game corresponds to some character of a separating formula. So by simply adjusting the way the parameter $k$ has to be spend, we can create a game for the number of modal operators instead of formula size. The main problem with this approach is that some moves do not reduce the parameter at all. Firstly this opens up the possibility of endless games, where $S$ cannot win, but he can ensure he does not lose either. Secondly it makes the game inefficient, since moves that do not change the parameter are inherently less interesting.

In order to deal with these problems we look at another model-comparison game used by Vinall-Smeeth in \cite{Vinall-Smeeth2024}. He calls this game the $k$-QVT game. However in this paper we use the parameter $k$ with a different meaning. So in order to avoid confusion we will use the name $l$-QVT game when referring to this game instead. The $l$-QVT counts the number of quantifiers in formulas of the $l$-variable fragment of first-order logic. It can be seen as a modification of the Adler-Immerman game, where after every move all structures without a partially isomorphic counterpart in the respective other set are removed. This is possible, because these are exactly the structures that can be removed in the Adler-Immerman game without playing any quantifier-moves. With this idea we can create our own game for counting the number of modal operators necessary in modal logic formulas. We call this game the $\mathcal{M}$-\emph{operator}-\emph{number} game for modal logic. Afterwards we examine a different logic called computation tree logic. This logic can be seen as an extension of modal logic by some more complicated temporal operators. Then we modify our game into the $\mathcal{M}$-operator-number game for computation tree logic. This game counts the number of necessary temporal operators in a separating computation tree logic formula.

Afterwards we use our new games in order to look at how different sets of Boolean connectives can be used to write more succinct formulas. This topic was already researched by Berkholz, Kuske and Schwarz in \cite{Schwarz}. They show that all complete operator sets, which only contain so-called \emph{locally monotone} operators, have the same succinctness up to a polynomial. On the other hand all sets that contain some non locally monotone operator also coincide in succinctness. Their findings hold for many logics, including modal logic and computation tree logic. In this paper we use the De Morgan basis, consisting of $\land$, $\lor$ and $\lnot$, as our set of locally monotone Boolean connectives. Then we use the De Morgan basis extended by the bi-implication $\leftrightarrow$, sometimes also called equivalence, as the set with a non locally monotone connective.
Berkholz et al. also show that for modal logic there is an exponential succinctness gap between these two classes. They prove this by working directly with formulas and using intricate and ad hoc arguments. This gap still exists if only reflexive Kripke structures are taken into consideration. However if the focus is tightened even further to only Kripke structures with an equivalence relation as accessibility relation, then this succinctness gap is closed.

From these findings some interesting follow-up questions emerge: Is it possible to remove $\leftrightarrow$ without significantly increasing the number of modal operators? This question is interesting because modal logic formulas with few modal operators are easier to work with. However we answer this question negatively by showing that there is an exponential operator-succinctness gap between modal logic and modal logic with bi-implication as well. Another interesting follow-up question is the following: What about other classes of Kripke-structures? In this paper we want to answer this question for some other important classes. In particular we show that for the class of Kripke structures with a reflexive and transitive accessibility relation, as well as the class of structures with a reflexive and symmetrical relation this exponential succinctness gap exists. This directly implies that the gap also exists for the class of transitive and the class of symmetrical structures. For all of these classes there is an exponential operator-succinctness gap as well. Only for the class of structures with a transitive and symmetrical relation this gap disappears. 
In this way our work offers a more complete and systematic examination of the connection between succinctness and the basis of boolean connectives in modal logic on special classes of Kripke structures. 

Of course it is also interesting to ask similar questions for different logics. In this paper we also examine computation tree logic. Using our $\mathcal{M}$-Operator-Number game for computation tree logic we show that the bi-implication also allows us to write exponentially shorter formulas and formulas with exponentially fewer temporal operators in this logic as well. We also show that these succinctness gaps also exists when only graphs with a serial and transitive accessibility relation are considered.

In many practical applications logic formulas are represented by directed acyclic graphs, called DAGs or circuits. In such a DAG the bi-implication can be removed with only minimal size increase. Because of this every time we show that we can write exponentially shorter formulas by using the bi-implication, we also show that there are DAGs for which the shortest equivalent formula is still exponentially larger then the DAG.

A shorter version of this paper was published as part of the CSL-conference\cite{confVer}. In that version we only considered modal logic. Everything regarding computation tree logic is a new addition to this paper. Also new to this version is the examination of DAGs.

\section{Preliminaries}
\subsection{Kripke structures}
Let $P$ be an infinite set of propositional variables, also called \emph{labels}.
Also let $\mathcal{M}=(W,R,V)$ be a connected, directed, labelled graph, where $W$ is the set of possible worlds or nodes, $R \subseteq W \times W$ is a binary accessibility relation and $V: W \rightarrow \mathcal{P}(P)$ is a labelling function, assigning sets of propositional variables to each node. For a node $w \in W$ we call $V(w)\subseteq P$ the \emph{colour} of $w$. If $V(W):=\bigcup\limits_{w \in W} V(w)$, the set of all labels used in $\mathcal{M}$, is finite, we say $\mathcal{M}$ is a \emph{finitely labelled graph}.
 The tuple $(\mathcal{M},w)$ is called a \emph{pointed Kripke structure} or just \emph{Kripke structure}. It should be noted here, that our naming conventions might differ from the norm. In the literature the labelled graph $\mathcal{M}$ is sometimes referred to as a Kripke structure, while $(\mathcal{M},w)$ is called a pointed Kripke structure. However in this paper we do not view $\mathcal{M}$ as a structure. Also our conventions allow us to use the terms pointed Kripke structure, Kripke structure or sometimes just structure interchangeably when referring to $(\mathcal{M},w)$.
 
 Let $A \subseteq W$ be a set of nodes. We write $(\mathcal{M},A)$ to mean the class of structures $\{(\mathcal{M},w) \mid w \in A\}$. We define $\Box w =\{w' \in W \mid (w,w') \in R\}$ as the set of all successors of a node $w$.
 This notation can also be extended to sets of nodes with $\Box A = \bigcup\limits_{w \in A} \Box w$ and structures with $\Box (\mathcal{M},w)=(\mathcal{M},\Box w)= \{(\mathcal{M},w') \mid w' \in \Box w\}$. The extension to classes of structures is defined in the obvious way.
 
An infinite path $(w_i)_{0 \leq i}$ is an infinite sequence of nodes $w_0,w_1,...\in W$ with $(w_i,w_{i+1})\in R$ for all $i\in \mathbb{N}$. Similarly a finite path $(w_i)_{0 \leq i \leq \ell}$ is a finite sequence of nodes $w_0,w_1,...,w_{\ell}$ with $(w_i,w_{i+1})\in R$ for all $i \in \mathbb{N}$ with $i < \ell$.

\subsection{Modal logic with Bi-implication}\label{Modal Logik}
We define ML$[\leftrightarrow]$ as the set of all modal logic formulas over the set of propositional variables $P$ using $\top$ and $\bot$ as constants, $\Diamond$ and $\Box$ as modal operators, as well as $\lnot$, $\lor$, $\land$ and $\leftrightarrow$ as Boolean connectives. Formally ML$[\leftrightarrow]$ is the set of all formulas generated by the following grammar:
\begin{equation*}
\varphi::= \bot \mid \top \mid p \mid \lnot \varphi \mid (\varphi \lor \varphi) \mid (\varphi \land \varphi) \mid (\varphi \leftrightarrow \varphi) \mid \Diamond \varphi \mid \Box \varphi
\end{equation*}
where $p \in P$ is a propositional variable. Now modal logic can be defined as a restriction of modal logic with bi-implication. So ML is defined as the set of all formulas from ML$[\leftrightarrow]$ that do not contain any occurrences of $\leftrightarrow$.
Next we define the semantics of modal logic with bi-implication.

Let $(\mathcal{M},w)$ be some Kripke structure. The satisfaction relation $\models$ is defined inductively as follows:\\

\begin{itemize}
\item $(\mathcal{M},w)\models \top$
\item $(\mathcal{M},w)\not\models \bot$
\item $(\mathcal{M},w)\models p $ iff, $ p \in V(w)$
\item $(\mathcal{M},w)\models \lnot\varphi $ iff, $ (\mathcal{M},w)\not\models \varphi$
\item $(\mathcal{M},w)\models (\varphi \lor \psi) $ iff, $ (\mathcal{M},w) \models \varphi$ or $(\mathcal{M},w)\models \psi$
\item $(\mathcal{M},w)\models (\varphi \land \psi) $ iff, $ (\mathcal{M},w) \models \varphi$ and $(\mathcal{M},w)\models \psi$
\item $(\mathcal{M},w)\models \varphi \leftrightarrow \psi$ iff, $(\mathcal{M},w)\models (\varphi \land \psi) \lor (\lnot \varphi \land \lnot \psi)$
\item $(\mathcal{M},w)\models \Diamond \varphi $ iff, $ (\mathcal{M},v) \models \varphi$ holds for some $v\in \Box w$
\item $(\mathcal{M},w)\models \Box \varphi $ iff, $ (\mathcal{M},v) \models \varphi$ holds for all $v\in \Box w$
\end{itemize}

\subsection{Computation tree logic with Bi-implication}
We define CTL$[\leftrightarrow]$ as the set of all computation tree logic formulas over $P$ using $\top$ and $\bot$ as constants, $EX$, $EG$, $EF$, $EU$, $AX$, $AG$, $AF$ and $AU$ as temporal operators, as well as $\lnot$, $\lor$, $\land$ and $\leftrightarrow$ as Boolean connectives. Formally CTL$[\leftrightarrow]$ is the set of all formulas generated by the following grammar:
\begin{equation*}
\begin{split}
\varphi::=& \bot \mid \top \mid p \mid \lnot \varphi \mid (\varphi \lor \varphi) \mid (\varphi \land \varphi) \mid (\varphi \leftrightarrow \varphi) \mid \\ & EX\varphi \mid EG\varphi \mid EF\varphi \mid \varphi EU \varphi \mid AX\varphi \mid AG\varphi \mid AF\varphi \mid \varphi AU \varphi
\end{split}
\end{equation*}
where $p \in P$ is a propositional variable. Then CTL is defined as the set of all formulas from CTL$[\leftrightarrow]$ that do not contain any occurrences of $\leftrightarrow$.

Let $(\mathcal{M},w)$ be some Kripke structure. The satisfaction relation $\models$ is defined inductively as follows:\\

\begin{itemize}
\item $(\mathcal{M},w)\models \top$
\item $(\mathcal{M},w)\not\models \bot$
\item $(\mathcal{M},w)\models p $ iff, $ p \in V(w)$ 
\item $(\mathcal{M},w)\models \lnot\varphi $ iff, $ (\mathcal{M},w)\not\models \varphi$
\item $(\mathcal{M},w)\models (\varphi \lor \psi) $ iff, $ (\mathcal{M},w) \models \varphi$ or $(\mathcal{M},w)\models \psi$
\item $(\mathcal{M},w)\models (\varphi \land \psi) $ iff, $ (\mathcal{M},w) \models \varphi$ and $(\mathcal{M},w)\models \psi$
\item $(\mathcal{M},w)\models \varphi \leftrightarrow \psi$ iff, $(\mathcal{M},w)\models (\varphi \land \psi) \lor (\lnot \varphi \land \lnot \psi)$
\item $(\mathcal{M},w)\models EX \varphi $ iff, $ (\mathcal{M},v) \models \varphi$ holds for some $v\in \Box w$
\item $(\mathcal{M},w)\models EG \varphi $ iff, there is a path $(w_i)_{0 \leq i}$ with $w_0=w$ and $(\mathcal{M},w_i)\models \varphi$ for all $i \geq 0$
\item $(\mathcal{M},w)\models EF \varphi$ iff, there is a path $(w_i)_{0 \leq i}$ and a $j\geq 0$ with $w_0=w$ and $(\mathcal{M},w_j)\models \varphi$
\item $(\mathcal{M},w)\models \varphi EU \psi$ iff, there is a path $(w_i)_{0 \leq i}$ and a $j \geq 0$ with $w_0=w$, $(\mathcal{M},w_i)\models \varphi$ for all $0 \leq i < j$ and $(\mathcal{M},w_j)\models \psi$
\item $(\mathcal{M},w)\models AX \varphi $ iff, $ (\mathcal{M},v) \models \varphi$ holds for all $v\in \Box w$
\item $(\mathcal{M},w)\models AG \varphi$ iff, every path $(w_i)_{0 \leq i}$ with $w_0=w$ fulfils $(\mathcal{M},w_i)\models \varphi$ for all $i \geq 0$
\item $(\mathcal{M},w)\models AF \varphi$ iff, for every path $(w_i)_{0 \leq i}$ with $w_0=w$ there is a $j \geq 0$ with $(\mathcal{M},w_j)\models \varphi$
\item $(\mathcal{M},w)\models \varphi AU \psi$ iff, for every path $(w_i)_{0 \leq i}$ with $w_0=w$ there is a $j \geq 0$ so that $(\mathcal{M},w_i)\models \varphi$ for all $0 \leq i < j$ and $(\mathcal{M},w_j)\models \psi$
\end{itemize}

Since $EX$ is identical to $\Diamond$ and $AX$ is identical to $\Box$, it is easy to see that ML$[\leftrightarrow]$ is a syntactic fragment of CTL$[\leftrightarrow]$.

\subsection{Succinctness}
Let $\varphi$ be some formula. Then the \emph{size} $\vert\varphi\vert$ of $\varphi$ is the number of symbols needed to write it, excluding brackets. For $\varphi \in$ML$[\leftrightarrow]$ or $\varphi\in$CTL$[\leftrightarrow]$ this can be expressed quite easily using the following inductive definition:\\

\begin{itemize}
\item $\vert \varphi \vert = 1$, if $\varphi \in \{\top,\bot\} \cup P$
\item $\vert \varphi \vert = \vert \psi \vert + 1$, if $\varphi=\lnot \psi$, $\varphi = \Diamond \psi$, $\varphi = \Box \psi$, $\varphi=EX \psi$, $\varphi=EG \psi$, $\varphi=EF \psi$, $\varphi = AX \psi$, $\varphi=AG \psi$ or $\varphi=AF \psi$
\item $\vert \varphi \vert = \vert\psi\vert + \vert\chi\vert +1$, if $\varphi=\psi \land \chi$, $\varphi=\psi \lor \chi$, $\varphi=\psi \leftrightarrow \chi$, $\varphi= \psi EU \chi$ or $\varphi= \psi AU \chi$
\end{itemize}

Similarly the \emph{operator-number} $\vert\varphi\vert_O$ of $\varphi$ is the number of modal or temporal operators used in the formula. For modal logic we consider $\Diamond$ and $\Box$ as modal operators. For CTL $EX$, $EG$, $EF$, $EU$, $AX$, $AG$, $AF$ and $AU$ are considered as temporal operators. The formal definition of operator-number for CTL$[\leftrightarrow]$ and ML$[\leftrightarrow]$ is as follows:\\

\begin{itemize}
\item $\vert \varphi \vert_O = 0$, if $\varphi \in \{\top,\bot\} \cup P$
\item $\vert \varphi \vert_O = \vert \psi \vert_O$, if $\varphi=\lnot \psi$
\item $\vert \varphi \vert_O = \vert \psi \vert_O + \vert \chi \vert_O$, if $\varphi=\psi \lor \chi$, $\varphi = \psi \land \chi$, $\varphi=\psi \leftrightarrow \chi$
\item $\vert \varphi \vert_O = \vert \psi \vert_O + 1$, if $\varphi= \Diamond \psi$, $\varphi = \Box \psi$, $\varphi=EX \psi$, $\varphi=EG \psi$, $\varphi=EF \psi$, $\varphi = AX \psi$, $\varphi=AG \psi$ or $\varphi=AF \psi$
\item $\vert \varphi \vert_O = \vert \psi \vert_O + \vert \chi \vert_O+1$, if $\varphi=\psi EU \chi$ or $\varphi = \psi AU \chi$
\end{itemize}

We say that two formulas $\varphi$ and $\psi$ are \emph{equivalent}, in writing $\varphi \equiv \psi$, if every pointed Kripke structure $(\mathcal{M},w)$ satisfies $\varphi$ if, and only if, it satisfies $\psi$. Let $\mathcal{C}$ be some class of Kripke structures. Then two formulas $\varphi$ and $\psi$ are \emph{equivalent on $\mathcal{C}$}, in writing $\varphi \equiv_{\mathcal{C}} \psi$, if every pointed Kripke structure $(\mathcal{M},w)\in \mathcal{C}$ from $\mathcal{C}$ satisfies $\varphi$ if, and only if, it satisfies $\psi$.

The term succinctness is used to compare logics based on the length of formulas needed to express the same property.
Given two logics $L_1$, $L_2$, where $L_2$ is at least as expressive as $L_1$ and a function $f:\mathbb{N}\rightarrow\mathbb{N}$, we say $L_1$ is \emph{$f$-times more succinct} than $L_2$ if, and only if, for every $n\in \mathbb{N}$ there is a formula $\varphi \in L_1$ with $\vert\varphi\vert \geq n$, so that $\vert \psi \vert \geq f(\vert \varphi \vert)$ for every $L_2$-formula $\psi$ with $\psi \equiv \varphi$. If $f$ is an exponential function, we also say $L_1$ is exponentially more succinct than $L_2$. Let $\mathcal{C}$ be a class of structures, then succinctness on $\mathcal{C}$ is defined by taking the definition of succinctness and replacing $\equiv$ with $\equiv_{\mathcal{C}}$.

Succinctness is based on formula size, but we can define a corresponding measurement for operator-number. We say that $L_1$ is \emph{$f$-times more operator-succinct} than $L_2$ if, and only if, for every $n \in \mathbb{N}$ there is a formula $\varphi \in L_1$ with $\vert \varphi \vert_O\geq n$, so that $\vert \psi \vert_O \geq f(\vert \varphi \vert_O)$ for every $L_2$-formula $\psi$ with $\psi \equiv \varphi$. Exponential operator-succinctness and operator-succinctness on special classes of structures are defined in the obvious way.

Counter intuitively it is theoretically possible that there are two logics $L_1$ and $L_2$ where $L_1$ is exponentially more succinct than $L_2$ and $L_2$ is exponentially more succinct than $L_1$. This is because there might be some properties that can be expressed more succinctly in $L_1$, while some other properties can be expressed more succinctly in $L_2$. However this cannot happen for almost all succinctness gaps examined in this paper. Since every ML formula is also an ML$[\leftrightarrow]$ formula, ML cannot be more succinct than ML$[\leftrightarrow]$. The same goes for CTL and CTL$[\leftrightarrow]$.

Next we want to extend the satisfaction relation $\models$ from structures to classes of structures. Let $\mathfrak{A}$ be a class of pointed Kripke structures and let $\varphi$ be some formula from ML$[\leftrightarrow]$ or CTL$[\leftrightarrow]$. Then $\mathfrak{A}\models \varphi$ if, and only if, $(\mathcal{M},w)\models \varphi$ for all $(\mathcal{M},w) \in \mathfrak{A}$.
We say a formula $\varphi$ separates two classes of pointed Kripke structures $\mathfrak{A}$ and $\mathfrak{B}$ if either $\mathfrak{A} \models \varphi$ and $\mathfrak{B}\models\lnot\varphi$, or $\mathfrak{B}\models\varphi$ and $\mathfrak{A}\models\lnot\varphi$. So $\varphi$ must be true for all structures of one class and for no structure of the other class. In this case we will also call $\varphi$ a \emph{separating formula}. It should be noted here that our definition differs slightly from the definition used by Hella and Vilander. In their definition $\varphi$ is a separating formula only if $\mathfrak{A}\models \varphi$ and $\mathfrak{B}\models\lnot\varphi$. However it is easy to see that if $\varphi$ is a separating formula by our definition then $\varphi$ or $\lnot \varphi$ is a separating formula by their definition. For $\mathfrak{A}=\{\mathcal{A}\}$ and $\mathfrak{B}=\{\mathcal{B}\}$ we might also say $\varphi$ separates $\mathcal{A}$ and $\mathcal{B}$ instead of $\mathfrak{A}$ and $\mathfrak{B}$.

\subsection{Directed acyclic graphs}
Directed acyclic graphs or \emph{DAGs} can be used as an alternative way of representing logic formulas. The main difference to formulas is that in a DAG every copy of a sub-formula is represented by the same sub-graph. Any ML- or CTL-formula $\varphi$ can be easily turned into a DAG. For each sub-formula $\psi$ of $\varphi$ we simply create a new node that is labelled with the outermost symbol of $\psi$ and connect it to the nodes corresponding to the main sub-formulas of $\psi$. We call the resulting graph $DAG(\varphi)$. Of course from this DAG we can simply reconstruct the original formula by making copies of sub-formulas that appear multiple times. Instead of giving proper semantics for DAGs we simply state $(\mathcal{M},w)\models DAG(\varphi) \Leftrightarrow (\mathcal{M},w)\models \varphi$ for every Kripke structure $(\mathcal{M},\varphi)$. The size of a DAG is defined as the number of its nodes. For $DAG(\varphi)$ this is equivalent to the number of sub-formulas of $\varphi$. The set of all of these sub-formulas $sub(\varphi)$ is inductively defined as follows:

\begin{itemize}
\item $sub(\varphi)=\{\varphi\}$, if $\varphi\in \{\top,\bot\} \cup P$
\item $sub(\varphi)=\{\varphi\}\cup sub(\psi)$, if $\varphi=\lnot \psi$, $\varphi = \Diamond \psi$, $\varphi = \Box \psi$, $\varphi=EX \psi$, $\varphi=EG \psi$, $\varphi=EF \psi$, $\varphi = AX \psi$, $\varphi=AG \psi$ or $\varphi=AF \psi$
\item $sub(\varphi)=\{\varphi\}\cup sub(\psi) \cup sub(\chi)$, if $\varphi=\psi EU \chi$ or $\varphi = \psi AU \chi$
\end{itemize}

Then the size of $DAG(\varphi)$ can be defined by $\vert DAG(\varphi) \vert = \vert sub(\varphi) \vert$. We define the two new logics DAG(ML)$=\{DAG(\varphi) \mid \varphi\in \text{ML}\}$ and DAG(CTL)$=\{DAG(\varphi) \mid \varphi\in \text{CTL}\}$ as the set of all DAGs that correspond to ML- and CTL-formulas respectively.

\section{The operator-number game for modal logic}\label{section game for ML}
Let $\mathcal{M}=(W,R,V)$ be a finitely labelled graph and let $A,B \subseteq W$ be two sets of nodes of  the graph. Then $\{A,B\}$ is an \emph{unordered pair}. From this point on, we will use just the term pair to refer to unordered pairs.
We say such a pair $\{A,B\}$ is \emph{tidy}, if for every $a \in A$ there is a $b \in B$ with $V(a)=V(b)$ and vice versa. For any sets $A$ and $B$, $\{A,B\}$ can be tidied up by removing all nodes $a\in A$ without a fitting node in $B$ and all $b \in B$ without a fitting node in $A$. Let $t(\{A,B\})=\{A',B'\}$, where $A' \subseteq A$ and $B' \subseteq B$ are the largest subsets of $A$ and $B$ respectively, so that $\{A',B'\}$ is tidy.

\begin{defi}
Let $\mathcal{M}=(W,R,V)$ be a finitely labelled graph, $A_0,B_0 \subseteq W$ two sets of nodes and $k_0\in \mathbb{N}$. The $\mathcal{M}$-\emph{Operator}-\emph{Number} game for modal logic, or $\mathcal{M}$-ON game for ML for short, played from $A_0$ and $B_0$ with parameter $k_0$ has two players $S$ and $D$. An \emph{$S$-position} $(\{A,B\},k)$ of the game consists of a tidy pair of sets of nodes $A,B \subseteq W$, as well as a parameter $k \in \mathbb{N}$. The starting position is $(t(\{A_0,B_0\}),k_0)$. A \emph{turn for $S$} starts from an $S$-position and proceeds in one of the following two ways:
\begin{enumerate}
\item \textbf{Split-move:} $S$ may only play this type of move if $k \geq 2$. He picks a set $X \in \{A,B\}$ and splits it into two disjoint and non-empty sets $X_1$ and $X_2$. Then he chooses two parameters $k_1,k_2\geq 1$ with $k_1+k_2=k$. Let $Y$ be the set that fulfils $\{X,Y\}=\{A,B\}$. This play by $S$ creates the $D$-position $(X_1,X_2,Y,k_1,k_2)$.
Afterwards $D$ plays by taking the set $Y$ and picking a subset $Y' \subseteq Y$ of it. Then she chooses $i \in \{1,2\}$ and the game continues from the $S$-position $(t(\{X_i,Y'\}),k_i)$.

\item \textbf{Operator-move:} $S$ may only play this type of move if $k\geq 1$. He picks a set $X \in \{A,B\}$ and then picks a function $f:X\rightarrow W$, where $f(x)\in \Box x$ for every $x\in X$. Let $X'=\{f(x)\mid x \in X\}$ be the image of $X$ under $f$ and let $Y$ be the set with $\{X,Y\}=\{A,B\}$. Then the $D$-position after this move by $S$ is $(X',Y,k-1)$. Afterwards $D$ plays by choosing a set $Y'\subseteq \Box Y$ of successors of the nodes from $Y$. The game continues from the $S$-position $(t(\{X',Y'\}),k-1)$.
\end{enumerate}
Then our game is a reachability game, where $S$ has to reach a position $(\{A,B\},k)$ with $A=B=\emptyset$.
\end{defi}

At a later point in this paper we will define another version of the $\mathcal{M}-ON$ game. To differentiate these two versions of the game we use the addition of 'for ML' to the name of the game. However if it is clear from context what version we are talking about, we will omit this part of the name.
When looking at an operator-move we might say \emph{$S$ picks $x'$ as a successor of $x$} instead of \emph{$S$ picks a function $f$ with $f(x)=x'$}. We usually make $D$ play in such a way that no nodes are removed during tidy-up. For this reason we will only mention the tidy-up process if nodes are getting removed. The form of a $D$-position depends on the type of move played by $S$. However in contrast to $S$-positions we mainly view $D$-positions as intermediate positions. So it is not necessary to give a full definition for them.

The parameter $k$ is reduced in every move and the play cannot continue, once the parameter is depleted. So a play with starting parameter $k$ can go on for $k$ rounds at most. In particular every play has to be finite. This implies in every play either player $S$ or player $D$ wins.
So our game is a finite, chance-less two-player game with complete information, that cannot end in a draw. Then by Zermelo's Theorem for every position of the $\mathcal{M}$-ON game one of the players, $S$ or $D$, has a positional winning strategy.

Such a positional strategy for a player, $S$ or $D$, is a partial map $f_S$ or $f_D$. For $S$ a strategy $f_S$ takes as input an $S$-position and outputs a $D$-position, that can be reached from the input position via a single move by $S$. It is a winning strategy, if every play where $S$ plays according to $f_S$ is won by him. Similarly a strategy $f_D$ for $D$ takes a $D$-position as input and gives an $S$-position, reachable in one move by $D$, as output. It is a winning strategy, if $D$ wins every play in which she plays $f_D$. If player $S$ has a winning strategy for some position $(\{A,B\},k)$, we will say this position is a winning position for player $S$. The definition of winning positions for $D$ is done in the obvious way.

As previously stated, the main inspiration for our game is the Hella-Vilander game from \cite{HellaVilander}, which in turn builds on the Adler-Immerman game from \cite{A-I}. Before we show that our game can be used to prove lower bounds for operator-number, we want to discuss how our game differs from these games.

The $\mathcal{M}$-ON game is played by two players, $S$ (Samson) and $D$ (Delilah), just like the game defined by Hella and Vilander. However instead of playing on two sets of structures, in our game the players play on unordered pairs of sets of nodes. This shows two significant differences between our game and their game. Firstly we are using nodes, where they used structures. This change however is only notational, because we will still implicitly talk about pointed Kripke structures. We will use the node $w$ as a representation of the structure $(\mathcal{M},w)$. In order to do so, we have to fix a single graph $\mathcal{M}$ for the entire game. So every pointed Kripke structure that appears in a play must be based on the same labelled graph. While this appears like a real restriction, it is possible to merge multiple labelled graphs into one without changing the validity of formulas at any world.

The second significant difference is that we use unordered pairs instead of ordered pairs or tuples. This change allows us to switch around the elements of our pairs at will. In the game by Adler and Immerman, such a switch is used as a not-move. So basically in our game, we can simulate a not-move without actually making a move. This is possible, because we are only interested in the number of operators and not in the number of negations. We also add an additional step to every move, where we tidy up the pair of our position. By doing so we remove all elements, that could have been removed in the Hella-Vilander game by $S$ playing a number of turns without playing an operator-move. A similar approach is used in the $l$-QVT game from \cite{Vinall-Smeeth2024}. In this game structures without a partially isomorphic counterpart in the other set are removed after every step. A problem with this method occurs when tidying up would skip an infinite number of moves. Vinall-Smeeth solves this problem by only considering finite structures. However for our purposes a weaker condition is sufficient. We only demand that our graphs are finitely labelled.

As a last major difference: In the Hella-Vilander game, player $D$ is forced to play following the so-called \emph{oblivious} strategy. This means she always has to answer with the largest possible set. This works, because this oblivious strategy is always optimal. So if $D$ can win the game, she can do so by playing obliviously. We decided not to force $D$ to follow this strategy because of a design decision we explain in the following passage.

We intend to mainly use our game to show that there is no separating formula with a small number of operators. For this we need to show that player $D$ has a winning strategy in our game. The most direct way to do this, is giving a strategy $f_D$ for her and then showing that this strategy wins against every possible strategy of player $S$. So it is advantageous if player $D$ has a lot of options. Then we have more possible choices when deciding on $f_D$, even if some choices are worse than others. For this reason we do not force her to play obliviously. On the other hand we want to give $S$ only a small number of options, thus reducing the number of his strategies we need to check. For this reason we force $S$ to pick disjoint subsets during a split-move, even though the game would still work for any $X_1,X_2$ with $X_1 \cup X_2=X$. Similarly during an operator-move we force $S$ to pick exactly one successor per node, even though he could pick multiple successors.

Our aim is to prove that the $\mathcal{M}$-ON game can be used to characterize the number of operators needed in order to separate sets of nodes. As a stepping stone towards this goal, we state the following two lemmas.

\begin{lem}\label{tidy separating formulas}
Let $\mathcal{M}=(W,R,V)$ be a finitely labelled graph and $A,B,A',B' \subseteq W$ be sets of nodes with $t(\{A,B\})=\{A',B'\}$. If there is a separating ML-formula $\varphi'$ for $(\mathcal{M},A')$ and $(\mathcal{M},B')$, then there is also a separating ML-formula $\varphi$ for $(\mathcal{M},A)$ and $(\mathcal{M},B)$ with $\vert \varphi \vert_O=\vert \varphi'\vert_O$.
\end{lem}

\begin{proof} 
Without loss of generality we assume $A' \subseteq A$ and $B' \subseteq B$. In order to prove this lemma, we look at the sets $A''=A \setminus A'$ and $B''=B \setminus B'$. We know that no node in $A''$ has the same colour as any node in $B$ and no node in $B''$ has the same color as any node in $A$. Two pointed Kripke structures based on differently coloured initial nodes can be separated by a literal. In this case let $\psi_{a,b}$ be a literal with $(\mathcal{M},a)\models \psi_{a,b}$ and $(\mathcal{M},b)\not\models \psi_{a,b}$. Then for every $a \in A''$ we define $\Psi_a:=\{\psi_{a,b} \mid b \in B\}$. All nodes in $W$ agree on all labels that are not used in $\mathcal{M}$. Then every formula in $\Psi_a$ must be either an atomic formula from $V(W)$ or the negation of such an atomic formula.
So even though $B$ might be infinite, $\Psi_a$ will always be finite, since $V(W)$ is finite. So we can construct the formula $\psi_a:=\bigwedge\limits_{\psi_{a,b}\in \Psi_a}\psi_{a,b}$, which separates $(\mathcal{M},a)$ from $(\mathcal{M},B)$. Next we look at the set $\Psi:=\{\psi_a \mid a\in A''\}$ and the formula $\psi:=\bigvee\limits_{\psi_a \in \Psi}\psi_a$. This formula $\psi$ holds in $(\mathcal{M},A'')$, while its negation holds in $(\mathcal{M},B)$.

In a similar fashion we define $X_b:=\{ \psi_{a,b}\mid a \in A\}$ and $\chi_b:=\bigvee\limits_{\psi_{a,b}\in X_b}\psi_{a,b}$ for every $b \in B''$. Next we construct $X:=\{\chi_b \mid b\in B''\}$ and $\chi:= \bigwedge\limits_{\chi_b \in X}\chi_b$.
This formula holds in $(\mathcal{M},A)$ and its negation holds in $(\mathcal{M},B'')$. With this we can finally define the wanted formula $\varphi$. Without loss of generality we may assume $(\mathcal{M},A')\models \varphi'$, since otherwise we could use $\lnot \varphi'$ instead. Then $\varphi=(\varphi' \land \chi) \lor \psi$ is a separating formula for $(\mathcal{M},A)$ and $(\mathcal{M},B)$. Since neither $\psi$ nor $\chi$ contain any operators, $\vert \varphi \vert_O=\vert \varphi' \vert_O$ holds as well.
\end{proof}

\begin{lem}\label{inseparable tidy pairs}
Let $\mathcal{M}=(W,R,V)$ be a finitely labelled graph, $A,B \subseteq W$ be sets of nodes that form a tidy pair and let $\varphi$ be an ML-formula without any modal operators. Then $(\mathcal{M},A)\models \varphi$ if, and only if, $(\mathcal{M},B)\models \varphi$.
\end{lem}

\begin{proof} 
If $A=B=\emptyset$, then the lemma is trivially true. Otherwise we assume $(\mathcal{M},A)\models \varphi$. Because $\{A,B\}$ is tidy, for every $b \in B$ there has to be a $a\in A$ with $V(a)=V(b)$. Then $a$ and $b$ agree on all atomic formulas and thus on all operator-free formulas, so especially on $\varphi$. This means $(\mathcal{M},b)\models \varphi$ for every $b \in B$ and thus $(\mathcal{M},B)\models \varphi$. The converse implication follows from $\{A,B\}=\{B,A\}$.
\end{proof}

Now we want to prove the following Correctness- and Completeness-Theorem for our game. We do so by proving both directions separately, beginning with the proof of correctness.

\begin{lem}\label{from strategy to formula}
Let $\mathcal{M}=(W,R,V)$ be a finitely labelled graph, $A,B \subseteq W$ two sets of nodes and $k\in \mathbb{N}$.
If $S$ has a winning strategy in the $\mathcal{M}$-ON game played from $A$ and $B$ with parameter $k$, then there is an ML-formula $\varphi$ separating $(\mathcal{M},A)$ and $(\mathcal{M},B)$ with $\vert \varphi \vert_O \leq k$.
\end{lem}

\begin{proof}
For this proof we assume that $S$ has a winning strategy for the $\mathcal{M}$-ON game played from $A$ and $B$ with parameter $k$. Because of Lemma~\ref{tidy separating formulas} we may assume that $A$ and $B$ already form a tidy pair. Then the starting position of this game is $(\{A,B\},k)$. We construct the separating formula $\varphi$ by induction over $k$. First we look at the base case $k=0$. Since the parameter is already used up, $S$ cannot make a move at all. So the only possible way for him to win is, if the current position already fulfils $\{A,B\}=\{\emptyset,\emptyset\}$. Then $\varphi=\top$ is a trivial separating formula with no modal operators. 

We now assume, the implication holds for all $S$-positions $(\{A',B'\},k')$ with $k' < k$ and we look at some winning strategy $f_S$ for the $\mathcal{M}$-ON game starting from $(\{A,B\},k)$. We have to differentiate between the types of moves $f_S$ prescribes as the next move from this position. 

We begin by assuming that $f_S$ prescribes a split-move. In this case $S$ picks a set from $\{A,B\}$, without loss of generality he picks $A$. Then $S$ splits it into two sets $A_1$ and $A_2$ and he splits $k$ into $k_1$ and $k_2$, both of which have to be at most $k-1$. Since $f_S$ is a winning strategy for $S$, it has to win against every possible strategy of $D$, in particular against the following ones: $D$ answers with the set $B$ itself and then picks either $1$ or $2$. The game continues from $(t(\{A_1,B\}),k_1)$ or $(t(\{A_2,B\}),k_2)$ respectively. From both positions $f_S$ is still a winning strategy. So thanks to our induction hypothesis and Lemma~\ref{tidy separating formulas}, for $i\in\{1,2\}$ we have separating formulas $\varphi_i$ for $(\mathcal{M},A_i)$ and $(\mathcal{M},B)$ with $\vert \varphi_i \vert_O\leq k_i$. We may assume $(\mathcal{M},A_i)\models\varphi_i$, because otherwise we could use $\lnot \varphi_i$ instead. Then $\varphi=\varphi_1 \lor \varphi_2$ is a separating formula for $(\mathcal{M},A)$ and $(\mathcal{M},B)$ with $\vert\varphi\vert_O=\vert\varphi_1\vert_O+\vert\varphi_2\vert_O\leq k_1+k_2 = k$.

Next we look into $f_S$ prescribing an operator-move. Once again we may assume that $S$ plays on $A$. For every node in $A$ he picks a successor and forms $A'$. The strategy $f_S$ has to win against a strategy of $D$ that simply picks $B'=\Box B$. So $S$ can also win from the position $(t(\{A',\Box B\}),k-1)$. This means there has to be a formula $\varphi'$ with $(\mathcal{M},A')\models \varphi'$, $(\mathcal{M},\Box B)\models \lnot\varphi'$ and $\vert\varphi'\vert_O < k$. So every node in $A$ has at least one successor, namely those picked by $S$ during the construction of $A'$, that satisfies $\varphi'$. On the other hand no node in $B$ can have such a successor, since all of them are part of $\Box B$. So $\varphi = \Diamond \varphi'$ is a separating formula for $(\mathcal{M},A)$ and $(\mathcal{M},B)$ with at most $k$ operators. This finishes our proof of the first implication.\end{proof}

Next we prove the completeness of our game.

\begin{lem}\label{from formula to strategy}
Let $\mathcal{M}=(W,R,V)$ be a finitely labelled graph, $\varphi$ an ML-formula, $A,B \subseteq W$ two sets of nodes and $k\in \mathbb{N}$, so that $\varphi$ is a separating formula for $(\mathcal{M},A)$ and $(\mathcal{M},B)$ with $\vert \varphi \vert_O\leq k$. Then $S$ has a winning strategy in the $\mathcal{M}$-ON game played from $A$ and $B$ with parameter $k$.
\end{lem}

\begin{proof}
To begin with we may assume that the separating formula $\varphi$ only contains the connectives $\lor$ and $\lnot$, as well as the operator $\Diamond$, since otherwise we can use the well known equivalences $\psi \land \chi\equiv \lnot(\lnot\psi\lor\lnot\chi)$ and $\Box \psi \equiv \lnot \Diamond \lnot \psi$ in order to remove all occurrences of $\land$ and $\Box$. None of these changes increase the number of modal operators, so $\vert\varphi\vert_O\leq k$ still holds. The $\mathcal{M}$-ON game played from $A$ and $B$ with parameter $k$ has the starting position $(t(\{A,B\}),k)$. If $\varphi$ is a separating formula for $(\mathcal{M},A)$ and $(\mathcal{M},B)$, it is also a separating formula for the tidied up subsets of $A$ and $B$. So we may assume that $\{A,B\}$ is already tidy, meaning $t(\{A,B\})=\{A,B\}$. We construct $f_S$, the winning strategy of $S$, by playing according to $\varphi$. This construction is done by induction over the formula size of $\varphi$.

For the base case, we look at all atomic formulas, so $\varphi\in P \cup \{\top,\bot\}$. Because of Lemma~\ref{inseparable tidy pairs}, $(\mathcal{M},A)$ and $(\mathcal{M},B)$ agree on $\varphi$. So the only way for $\varphi$ to be a separating formula is $A=B=\emptyset$. In this case $(\{A,B\},k)$ is already a winning $S$-position and $S$ does not have to make a move in order to win the play.

We now assume the implication holds for all formulas $\psi$ with $\vert\psi\vert\leq i$ and we look at a separating formula $\varphi$ for $(\mathcal{M},A)$ and $(\mathcal{M},B)$ with $\vert \varphi \vert =i+1$ and $\vert\varphi\vert_O\leq k$. We have to differentiate based on the outermost operator of $\varphi$.

We begin with $\varphi=\lnot\psi$. Because of our definition of separating formula, $\psi$ also is a separating formula for $(\mathcal{M},A)$ and $(\mathcal{M},B)$ with $\vert \psi \vert=i$ and $\vert \psi \vert_O \leq k$. So thanks to our induction hypothesis, $S$ already has a winning strategy for the $\mathcal{M}$-ON game played on $A$ and $B$ with parameter $k$.

Next we look at $\varphi=\Diamond\psi$ with $\vert\psi\vert=i$ and $\vert\psi\vert_O\leq k-1$. This also implies $k\geq 1$. So every node in one of the sets, w.l.o.g. we assume $A$, has a successor that satisfies $\psi$, while no node in the other set, $B$, has such a successor. Then $f_S$ prescribes an operator-move, where $S$ picks for every $a\in A$ a successor $a'$ with $(\mathcal{M},a')\models\psi$ and collects them into the set $A'$. Then $D$ answers by gathering some nodes $b'\in\Box B$ into a set $B'$. However none of the nodes picked by $D$ can satisfy $\psi$. The game continues from the position $(t(\{A',B'\}),k-1)$, where $\psi$ is a separating formula for $A'$ and $B'$ and thus also for the sets after tidying up. So because of our induction hypothesis, for every choice made by $D$ during this turn, $S$ has a winning strategy from the resulting position. So $f_S$ simply copies the fitting strategy for $D$'s choice of $B'$.

For $\varphi=\psi\lor\chi$ we have to make an additional differentiation. It is possible that one of the sub-formulas, w.l.o.g. we assume $\chi$, does not contain any modal operators. Then one of the sets, we assume $A$, satisfies $\psi\lor\chi$, while the other satisfies its negation, so $(\mathcal{M},B)\models \lnot\psi \land \lnot\chi$. This implies $(\mathcal{M},B)\models\lnot \psi$ and $(\mathcal{M},B)\models\lnot\chi$. So because of Lemma~\ref{inseparable tidy pairs}, we also know $(\mathcal{M},A)\models\lnot\chi$, which in turn implies $(\mathcal{M},A)\models \psi$. This means $\psi$ is also a separating formula for $A$ and $B$ with $\vert \psi \vert \leq i$ and $\vert \psi \vert_O \leq k$. So our hypothesis offers a winning strategy for $S$ in $(\{A,B\},k)$.

Lastly we look at $\varphi=\psi\lor\chi$, where both $\psi$ and $\chi$ contain at least one modal operator. This implies $k\geq 2$. Once again we assume $(\mathcal{M},A)\models \psi\lor\chi$ and thus $(\mathcal{M},B)\models \lnot\psi\land\lnot\chi$. Let $A_1$ and $A_2$ be the disjoint subsets of $A$ with $A_1:=\{a \in A \mid (\mathcal{M},a)\models \psi\}$ and $A_2:=\{a\in A \mid (\mathcal{M},a) \models \lnot \psi \land \chi\}=A\setminus A_1$. If $A_2$ is empty, then $A=A_1$ and thus $(\mathcal{M},A)\models \psi$ holds. So $\psi$ is a shorter separating formula for $(\mathcal{M},A)$ and $(\mathcal{M},B)$ and according to our induction hypothesis $S$ has a winning strategy for the position $(\{A,B\},k)$. Similarly if $A_1$ is empty, then $\chi$ is a shorter separating formula. Otherwise we know $A_1 \neq \emptyset$ and $A_2 \neq \emptyset$. So $S$ may play a split-move where he splits $A$ into $A_1$ and $A_2$. Then he picks the new parameters as $k_1= \vert \psi \vert_O$ and $k_2=k-k_1$, which implies $k_2 \geq \vert \chi \vert_O$. Player $D$ answers by picking a set $B'\subseteq B$. Independent from her choice $(\mathcal{M},B')\models \lnot \psi \land \lnot \chi$ holds. The game continues from either $(t(\{A_1,B'\}),k_1)$ or $(t(\{A_2,B'\}),k_2)$. For the first case, $S$ can use $\psi$ as a separating formula for $A_1$ and $B'$. Similarly $\chi$ is a separating formula for $A_2$ and $B'$. So by our induction hypothesis every possible follow-up position is a winning position for $S$ and thus $(\{A,B\},k)$ is as well.
\end{proof}

Lemma~\ref{from strategy to formula} and Lemma~\ref{from formula to strategy} can be combined into the following Correctness- and Completeness-Theorem.

\begin{thm}\label{correctness theorem}
Let $\mathcal{M}=(W,R,V)$ be a finitely labelled graph, $A,B \subseteq W$ two sets of nodes and $k\in \mathbb{N}$.
Then $S$ has a winning strategy in the $\mathcal{M}$-ON game for ML played from $A$ and $B$ with parameter $k$ if, and only if, there is an ML-formula $\varphi$ with $\vert \varphi \vert_O \leq k$, that is a separating formula for $(\mathcal{M},A)$ and $(\mathcal{M},B)$.
\end{thm}

We end this segment by quickly proving the following corollary, which will be helpful when actually playing the game.

\begin{cor}\label{loosing condition for S}
Let $\mathcal{M}=(W,R,V)$ be a finitely labelled graph, $A,B\subseteq W$ be two finite sets of nodes with $A \cap B \neq \emptyset$. Then $D$ has a winning strategy for the $\mathcal{M}$-ON game for ML played on $A$ and $B$ with any parameter $k \in \mathbb{N}$.
\end{cor}

\begin{proof} We prove this corollary by contradiction. Let us assume that $D$ does not have a winning strategy for some parameter $k$. Then by Zermelos Theorem $S$ must have a winning strategy instead. Thus by Theorem~\ref{correctness theorem} there has to exist some ML-formula $\varphi$ that separates $(\mathcal{M},A)$ and $(\mathcal{M},B)$. However this is impossible, since these two sets of structures are not disjoint. So our assumption has to be wrong and $D$ must have a winning strategy.\end{proof}

We can use this knowledge in order to limit the options of $S$. Every strategy $f_S$ that picks a node, which can be picked by $D$ as well, cannot be a winning strategy for him. So when looking for a winning strategy, we may already discard all moves of this type.

\section{Succinctness between ML[$\leftrightarrow$] and ML}\label{ML}

We begin this section by proving some simple upper bounds for succinctness.

\begin{lem}\label{exponential upper bound}
The logic ML$[\leftrightarrow]$ is at most exponentially more succinct and at most exponentially more operator-succinct than ML.
\end{lem}

\begin{proof} It is quite easy to see that the succinctness gap between ML and ML$[\leftrightarrow]$ is at most exponential. We can simply replace every sub-formula $\varphi \leftrightarrow \psi$ with $(\varphi \land \psi) \lor (\lnot \varphi \land \lnot \psi)$. Every time this is done, the formula size is doubled at most. Since the number of bi-implications cannot exceed the formula size, this procedure leads to an exponential size increase at worst.

However this method cannot provide an upper bound for the operator-succinctness. While every such substitution  only doubles the number of operators as well, we cannot bound the number of operators in the resulting formula by some function of the number of operators in the original formula. E.g. for a formula $\varphi$ with only one modal operator that appears in the scope of $n$ nested bi-implications this method would result in a formula with $2^n$ modal operators.

So instead we have to use another method in order to guarantee an at most exponential increase in operator-number. This method is based on the construction used by Spira in \cite{spira}.
We begin by replacing every sub-formula $\Box \varphi$ with the equivalent $\lnot \Diamond \lnot \varphi$. Now we can assume that all operators occurring in our formula are $\Diamond$.
Next we recursively define the function $f\colon$ML$[\leftrightarrow]\rightarrow$ ML$[\leftrightarrow]$. The base case consist of all formulas $\varphi$ without any operators and in this case $f(\varphi):=\varphi$. For any other formula $\varphi$ there must be a formula $\psi$ so that $\Diamond\psi$ appears in $\varphi$ outside of the scope of any other operator. Let $\varphi_\top$ be the formula we obtain by replacing the sub-formula $\Diamond \psi$ in $\varphi$ with a $\top$. Similarly $\varphi_\bot$ is created by replacing $\Diamond \psi$ with $\bot$ instead.
Then we define $f(\varphi)$ as $f(\varphi):=(f(\varphi_\top)\land\Diamond f(\psi)) \lor (f(\varphi_\bot) \land \lnot \Diamond f(\psi))$.

Next we want to show that $f(\varphi)$ fulfils the following three conditions for every ML$[\leftrightarrow]$ formula $\varphi$: Firstly $f(\varphi) \equiv \varphi$, secondly no operator in $f(\varphi)$ appears in the scope of any bi-implication and thirdly $\vert f(\varphi)\vert_O \leq 2^{\vert \varphi \vert_O} \cdot \vert \varphi \vert_O$. We do this by induction on the number of operators in $\varphi$. As our base case we look at all ML$[\leftrightarrow]$ formulas $\varphi$ with no modal operators. Then we know $f(\varphi)=\varphi$, so the first condition is obviously fulfilled. Because $f(\varphi)$ does not contain any operators, the second condition is also trivially true and because of $\vert f(\varphi) \vert_O = \vert \varphi \vert_O=0$, the third holds as well.

For the induction step we look at all formulas $\varphi$ with $\vert \varphi \vert_O=i$ for some $i \geq 1$. In this case let $\psi$, $\varphi_\top$ and $\varphi_\bot$ be the formulas that were used when constructing $f(\varphi)$. All three of these formulas contain less modal operator than $\varphi$. So we can assume that these formulas fulfil our three conditions. Then we know $\varphi \equiv (\varphi_\top \land \Diamond \psi) \lor (\varphi_\bot \land \lnot \Diamond \psi)$. Because of $f(\psi)\equiv\psi$, $f(\varphi_\top)\equiv \varphi_\top$ and $f(\varphi_\bot) \equiv \varphi_\bot$, this is equivalent to $(f(\varphi_\top) \land \Diamond f(\psi)) \lor (f(\varphi_\bot) \land \lnot \Diamond f(\psi))=f(\varphi)$. No operator in $f(\psi)$, in $f(\varphi_\top)$ or in $f(\varphi_\bot)$ appears in the scope of any bi-implication. So the same must hold for $f(\varphi)$, since no new bi-implication is added during its construction. For the number of operators we know $\vert \varphi \vert_O=\vert \psi \vert_O + \vert \varphi_\top \vert_O + 1$ and $\vert f(\varphi) \vert_O = 2 \cdot \vert f(\psi) \vert_O + \vert f(\varphi_\top) \vert_O + \vert f(\varphi_\bot) \vert_O+ 2$. So by our induction hypothesis and with $\vert \varphi_\top \vert_O = \vert \varphi_\bot \vert_O$ we know the following:

\begin{equation*}
\begin{split}
\vert f(\varphi) \vert_O &\leq 2^{\vert \psi \vert_O+1} \cdot \vert \psi \vert_O +2^{\vert \varphi_\top \vert_O+1}\cdot\vert \varphi_\top \vert_O + 2\\
&= 2 \cdot (2^{\vert \psi \vert_O} \cdot \vert \psi \vert_O +2^{\vert \varphi_\top \vert_O}\cdot\vert \varphi_\top \vert_O + 1)\\
&\leq 2 \cdot (2^{\vert \varphi \vert_O - 1} \cdot \vert \psi \vert_O + 2^{\vert \varphi \vert_O-1} \cdot \vert \varphi_\top \vert_O + 1)\\
&\leq 2 \cdot(2^{\vert \varphi \vert_O-1} \cdot (\vert \psi \vert_O + \vert \varphi_\top \vert_O +1 ) )\\
&= 2^{\vert \varphi \vert_O} \cdot \vert \varphi \vert_O
\end{split}
\end{equation*}

So the third condition holds as well.

In total we have shown that for every ML$[\leftrightarrow]$-formula $\varphi$ there exists an equivalent ML$[\leftrightarrow]$-formula $f(\varphi)$ where no operator appears in the scope of a bi-implication and the number of operators in $\varphi'$ is at most exponential in the number of operators in $\varphi$. Lastly we can replace every sub-formula $\psi\leftrightarrow\chi$ in $f(\varphi)$ with $(\psi \land \chi)\lor (\lnot\psi \land \lnot\chi)$, in order to transform $f(\varphi)$ into an ML-formula, without increasing the number of operators any further. This construction proves, that the operator-succinctness gap between ML and ML$[\leftrightarrow]$ is at most exponential.
\end{proof}

In the following Section~\ref{section T,R} we show that there is indeed an exponential succinctness gap between ML and ML$[\leftrightarrow]$ on the class of Kripke structures with a transitive and reflexive accessibility relation. Afterwards in Section~\ref{section S,R} we show that this result also holds, if we replace transitivity with symmetry. Since these two claims are proven in similar ways, we want to use this opportunity to give a general overview of these proofs.

To begin with we construct an infinite graph $\mathcal{M}$ as well as two infinite sequences of nodes $a_0,a_1,...$ and $b_0,b_1,...$. Then for every $n \in \mathbb{N}$ we give an ML$[\leftrightarrow]$-formula $\varphi_n$. This formula separates $(\mathcal{M},a_n)$ and $(\mathcal{M},b_n)$. The number of modal operators in $\varphi_n$ is polynomial in $n$. Then we give some parameter $k_n$ that is exponential in $n$ and show that $(\{\{a_n\},\{b_n\}\},k_n)$ is a winning position for player $D$ in the $\mathcal{M}-ON$ game. It follows that $S$ cannot have a winning strategy for this position. So by Theorem~\ref{correctness theorem}, there cannot be an ML-formula $\psi$ that separates $(\mathcal{M},a_n)$ from $(\mathcal{M},b_n)$ with $\vert\psi\vert_O\leq k_n$ implying that every ML-formula equivalent to $\varphi_n$ must be of size exponential in $n$ at least. This proves an exponential succinctness gap between ML and ML$[\leftrightarrow]$ for all classes of Kripke structures $\mathcal{M}$ belongs to.

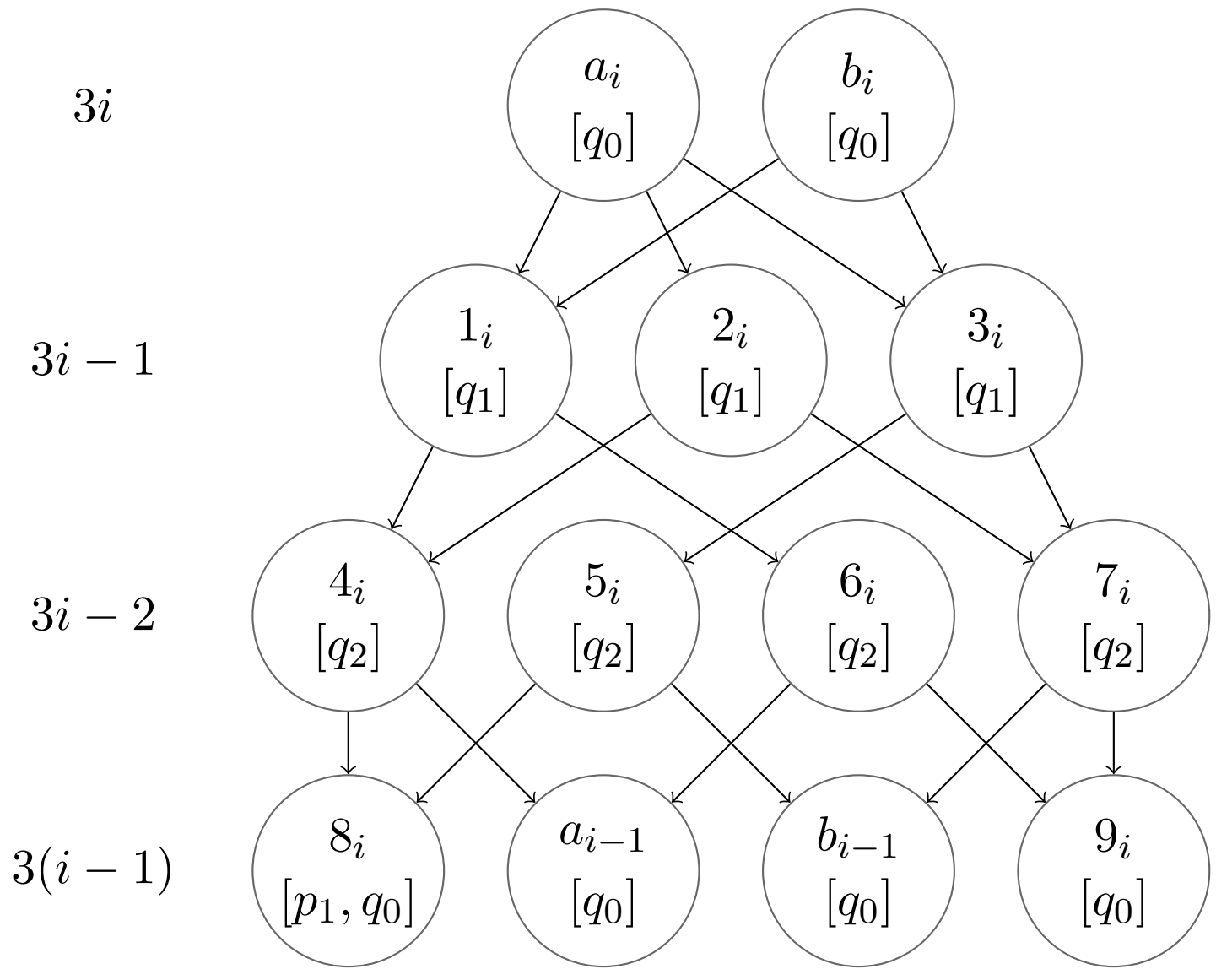
\begin{figure}[h]
\begin{tikzpicture}[on grid]
\node[graph](q1)															{};
\node[formula](q1.1)					[above=.25cm of q1]					{$4_i$};
\node[formula](q1.2)					[below=.25cm of q1]					{$[q_2]$};
\node[graph](q2)						[right=2cm of q1]					{};
\node[formula](q2.1)					[above=.25cm of q2]					{$5_i$};
\node[formula](q2.2)					[below=.25cm of q2]					{$[q_2]$};
\node[graph](q3)						[right=2cm of q2]					{};
\node[formula](q3.1)					[above=.25cm of q3]					{$6_i$};
\node[formula](q3.2)					[below=.25cm of q3]					{$[q_2]$};
\node[graph](q4)						[right=2cm of q3]					{};
\node[formula](q4.1)					[above=.25cm of q4]					{$7_i$};
\node[formula](q4.2)					[below=.25cm of q4]					{$[q_2]$};
\node[graph](q5)						[below=2cm of q1]					{};
\node[formula](q5.1)					[above=.25cm of q5]					{$8_i$};
\node[formula](q5.2)					[below=.25cm of q5]					{$[p_1,q_0]$};
\node[graph](q6)						[below=2cm of q2]					{};
\node[formula](q6.1)					[above=.25cm of q6]					{$a_{i-1}$};
\node[formula](q6.2)					[below=.25cm of q6]					{$[q_0]$};
\node[graph](q7)						[below=2cm of q3]					{};
\node[formula](q7.1)					[above=.25cm of q7]					{$b_{i-1}$};
\node[formula](q7.2)					[below=.25cm of q7]					{$[q_0]$};
\node[graph](q8)						[below=2cm of q4]					{};
\node[formula](q8.1)					[above=.25cm of q8]					{$9_i$};
\node[formula](q8.2)					[below=.25cm of q8]					{$[q_0]$};
\node[graph](q9)						[above right=2cm and 1cm of q1]		{};
\node[formula](q9.1)					[above=.25cm of q9]					{$1_i$};
\node[formula](q9.2)					[below=.25cm of q9]					{$[q_1]$};
\node[graph](q10)					[right=2cm of q9]					{};
\node[formula](q10.1)				[above=.25cm of q10]					{$2_i$};
\node[formula](q10.2)				[below=.25cm of q10]					{$[q_1]$};
\node[graph](q11)					[right=2cm of q10]					{};
\node[formula](q11.1)				[above=.25cm of q11]					{$3_i$};
\node[formula](q11.2)				[below=.25cm of q11]					{$[q_1]$};
\node[graph](q12)					[above right=2cm and 1cm of q9]		{};
\node[formula](q12.1)				[above=.25cm of q12]					{$a_i$};
\node[formula](q12.2)				[below=.25cm of q12]					{$[q_0]$};
\node[graph](q13)					[above right=2cm and 1cm of q10]		{};
\node[formula](q13.1)				[above=.25cm of q13]					{$b_i$};
\node[formula](q13.2)				[below=.25cm of q13]					{$[q_0]$};
\node[formula](q14)					[left=2cm of q1]						{$3i -2$};
\node[formula](q15)					[above=2cm of q14]					{$3i -1$};
\node[formula](q16)					[above=2cm of q15]					{$3i$};
\node[formula](q17)					[below=2cm of q14]					{$3(i-1)$};

\draw[->] 	(q12)	edge 	(q9)
			(q12)	edge 	(q10)
			(q12)	edge 	(q11)
			(q13)	edge 	(q9)
			(q13)	edge 	(q11)
			(q9)		edge 	(q1)
			(q9) 	edge 	(q3)
			(q10) 	edge 	(q1)
			(q10) 	edge 	(q4)
			(q11) 	edge 	(q2)
			(q11) 	edge 	(q4)
			(q1) 	edge		(q5)
			(q1) 	edge 	(q6)
			(q2) 	edge 	(q5)
			(q2) 	edge 	(q7)
			(q3) 	edge 	(q6)
			(q3) 	edge 	(q8)
			(q4) 	edge 	(q7)
			(q4) 	edge 	(q8);
\end{tikzpicture}
\caption{The $i$th segment of $\mathcal{M}$ including labels}
\label{ith segment}
\end{figure}

Now we construct the infinite labelled graph $\mathcal{M}=(W,R,V)$, confer Figure \ref{ith segment}. This graph will be the basis for all of our other graphs. The nodes of $\mathcal{M}$ are defined as $W:=\{a_0,b_0\}\cup \bigcup\limits_{i \geq 1}\{a_i,b_i,1_i,2_i,3_i,4_i,5_i,6_i,7_i,8_i,9_i\}$. The edges of this graph are as follows:

\begin{equation*}
R=\bigcup\limits_{i\geq 1}\left\{ 
\begin{array}{l}
(a_i,1_i),(a_i,2_i),(a_i,3_i),(b_i,1_i),(b_i,3_i),(1_i,4_i),(1_i,6_i),(2_i,4_i),(2_i,7_i),(3_i,5_i),\\
(3_i,7_i),(4_i,8_i),(4_i,a_{i-1}),(5_i,8_i),(5_i,b_{i-1}),(6_i,a_{i-1}),(6_i,9_i),(7_i,b_{i-1}),(7_i,9_i) 
\end{array}
\right\}
\end{equation*}

The graph $\mathcal{M}$ can be separated into segments. Then segment $i$ contains all nodes with index $i$, as well as $a_{i-1}$ and $b_{i-1}$. So the latter two nodes are shared nodes between segment $i$ and segment $i-1$. Figure~\ref{ith segment} only shows the nodes and edges of one such segment $i$. However all edges of $\mathcal{M}$ connect nodes within a segment and all segments have equivalent edges. So this figure fully represents the nodes and edges of $\mathcal{M}$. The numbers on the left side of the figure count the layers of the graph. Segment $i$ contains all nodes of layers $3i-1$, $3i-2$ and $3(i-1)$, as well as the nodes $a_i$ and $b_i$ from layer $3i$. Lastly we define the labelling function of $\mathcal{M}$. The labels we use are $\{p_0,p_1,q_0,q_1,q_2\}\subseteq P$. The only node labelled with $p_0$ is $a_0$. For every $i\geq 1$ the node $8_i$ is the only node of segment $i$ labelled with $p_1$. Lastly we use the remaining labels $q_0$, $q_1$ and $q_2$ in order to label the layers of the graph in an alternating pattern. More precisely for every $i\in \mathbb{N}$ we label all nodes of layer $3i$ with $q_0$, all nodes of layer $3i-1$ with $q_1$ and all nodes of layer $3i-2$ with $q_2$. The formal definition of the labelling function is given by the following list:

\begin{itemize}
\item $V(9_i)=V(b_{i-1})=\{q_0\}$ for all $i \geq 1$,
\item $V(8_i)=\{p_1, q_0\}$ for all $i \geq 1$ and 
\item $V(a_{i-1})= 	\begin{cases}
						\{p_0,q_0\} & \text{for } i =1\\
						\{q_0\} & \text{for all } i>1
					\end{cases}$
\item $V(1_i)=V(2_i)=V(3_i)=\{q_1\}$ for all $i \geq 1$,
\item $V(4_i)=V(5_i)=V(6_i)=V(7_i)=\{q_2\}$ for all $i \geq 1$,
\end{itemize}

\begin{clm}\label{phi_i}
Let $\varphi_i$ be the ML formula inductively defined by $\varphi_0:=p_0$ and $\varphi_i:=\Diamond \Box ((\Diamond p_1)\leftrightarrow (\Diamond \varphi_{i-1}))$. Then $\varphi_i$ is a separating formula for $(\mathcal{M},a_i)$ and $(\mathcal{M},b_i)$  with $(\mathcal{M},a_i)\models \varphi_i$ for every $i\in\mathbb{N}$.
\end{clm}

\begin{proof}
This proof is done by induction on $i$. For the base case $i=0$ the formula $\varphi_i=p_0$ holds only in $(\mathcal{M},a_0)$, because $a_0$ is the only node labelled with $p_0$. Next let $i\geq 1$ be some positive number. We assume that our claim holds for $i-1$. So $(\mathcal{M},a_{i-1})\models\varphi_{i-1}$ and $(\mathcal{M},b_{i-1})\models\lnot\varphi_{i-1}$. Now we look at the other nodes of segment $i$. We know $(\mathcal{M},8_i)\models p_1$, while no other node of layer $3(i-1)$ is labelled with $p_1$. This implies $(\mathcal{M},4_i)\models (\Diamond p_1)\land (\Diamond \varphi_{i-1})$, $(\mathcal{M},5_i)\models (\Diamond p_1) \land \lnot (\Diamond \varphi_{i-1})$, $(\mathcal{M},6_i)\models \lnot(\Diamond p_1) \land (\Diamond\varphi_{i-1})$, as well as $(\mathcal{M},7_i)\models \lnot(\Diamond p_1) \land \lnot (\Diamond \varphi_{i-1})$. So $(\mathcal{M},4_i)$ and $(\mathcal{M},7_i)$ model $(\Diamond p_1)\leftrightarrow(\Diamond\varphi_{i-1})$, while $(\mathcal{M},5_i)$ and $(\mathcal{M},6_i)$ do not. Because $4_i$ and $7_i$ are the only successors of $2_i$, this also means $(\mathcal{M},2_i)\models\Box((\Diamond p_1)\leftrightarrow(\Diamond \varphi_{i-1}))$. On the other hand, $1_i$ is a predecessor of $6_i$ and $3_i$ is a predecessor of $5_i$. So neither $(\mathcal{M},1_i)$ nor $(\mathcal{M},3_i)$ satisfy this formula. Lastly $2_i$ is a successor of $a_i$, so $(\mathcal{M},a_i)\models\Diamond \Box ((\Diamond p_1)\leftrightarrow (\Diamond \varphi_{i-1}))=\varphi_i$. But $1_i$ and $3_i$ are the only successors of $b_i$, so $(\mathcal{M},b_i)\models \lnot \varphi_i$, which proves the claim.
\end{proof}

It should be noted here that the formula $\varphi_i$ does not contain any occurrences of either $q_0$, $q_1$ or $q_2$. However we need those labels in order to adapt $\varphi_i$ to the modified graphs we use in the following proofs.

\subsection{Transitive and reflexive graphs}\label{section T,R}
The graph $\mathcal{M}_{t,r}=(W,R_{t,r},V)$ is defined as an extension of $\mathcal{M}=(W,R,V)$. The nodes and labels stay the same. First we add a self-loop for every node. So $(x,x)$ is a new edge for every node $x$. Then we add a new edge from a node $x$ to a node $y$, for all $x$ and $y$ where $x$ is a node on layer $i$ and $y$ is a node on layer $j$ with $i \geq j+2$. So every node is now a predecessor to all nodes two or more layers below. So formally $R_{t,r}=R \cup \{(x,x)\mid x \in W\} \cup \{(x,y)\mid x\text{ belongs to layer $i$, $y$ belongs to layer $j$, } i \geq j+2\}$.
It is easy to see that $\mathcal{M}_{t,r}$ is a reflexive graph. On top of that it is also transitive. Let $(x,y)$ and $(y,z)$ be edges of the graph. If either of them is a self loop, then $(x,z)$ is the same as the other edge. Otherwise $y$ must be on a lower layer than $x$ and $z$ must be on a lower layer than $y$. So $z$ is at least two layers below $x$ and $(x,z)$ is an edge in $\mathcal{M}_{t,r}$.

\begin{lem}\label{biimplication T,R}
There are two polynomial functions $g_1,g_2 \colon \mathbb{N}\rightarrow \mathbb{N}$, so that for every $n \in \mathbb{N}$ there is an ML$[\leftrightarrow]$-formula $\psi_n$, which separates $(\mathcal{M}_{t,r},a_n)$ from $(\mathcal{M}_{t,r},b_n)$, with $\vert \psi_n\vert_O=g_1(n) \geq n$ and $\vert \psi_n \vert=g_2(n) \geq n$.
\end{lem}

\begin{proof} 
This is proven, by simply adapting the formula $\varphi_n$ from Claim~\ref{phi_i} to separate $a_n$ and $b_n$ in $\mathcal{M}_{t,r}$ instead of in $\mathcal{M}$. In $\mathcal{M}$ every edge goes down exactly one layer. But all edges added to $R$ when constructing $R_{t,r}$ go down at least two layers or are self-loops. So we need a way to determine what layer a node belongs to. We look at the formula $\chi_i^+$ defined by $\chi_0^+:=\top$ and $\chi_{i+1}^+:=\Diamond (q_j \land \chi_i^+)$ with $i \in \mathbb{N}$ and $j=i \mod 3$. This formula states, that there is a path of length $i$ that starts at the current node and on this path the labels $q_0$, $q_1$ and $q_2$ appear in an alternating pattern. Such a path cannot use any self-loops, so every edge has to move down at least one layer. Then only nodes of layer $i$ or above satisfy this formula. Similarly only nodes of layer $i$ or below satisfy $\chi_i^-:=\lnot \chi_{i+1}^+$. Now we can use the formula $\chi_i:=\chi_i^+\land \chi_i^-$ in order to check whether or not a node belongs to layer $i$.

Then for any node $w$ of layer $i\geq 1$ $(\mathcal{M},w)\models \Diamond (\chi_{i-1} \land \varphi)$ holds if, and only if, $w$ has a successor exactly one layer below that satisfies $\varphi$. Similarly $(\mathcal{M},w)\models\Box(\lnot \chi_{i-1}\lor\varphi)$ holds if, and only if, every successor of $w$ that is exactly one layer below the current layer satisfies $\varphi$. So we can add conditions like these to $\varphi_n$ in order to only consider edges that go down exactly one layer, which are all edges from $R$, while ignoring all edges from $R_{t,r}\setminus R$. We define $\psi_n$ inductively with $\psi_0:=p_0$ and $\psi_i:=\Diamond(\chi_{3i-1}\land \Box(\lnot \chi_{3i-2}\lor((\Diamond(p_1\land \chi_{3(i-1)})\leftrightarrow (\Diamond(\psi_{i-1}\land \chi_{3(i-1)})))))$. Then for all nodes $w$ of layer $3n$, $(\mathcal{M}_{t,r},w)\models \psi_n$ holds if, and only if $(\mathcal{M},w)\models \varphi_n$. So by Claim~\ref{phi_i}, $\psi_n$ is a separating formula for $(\mathcal{M}_{t,r},a_n)$ and $(\mathcal{M}_{t,r},b_n)$. For a rough estimation of the size of $\psi_n$ we can assert that the number of times a sub-formula $\chi_i$ occurs in $\psi_n$ is linear in $n$. Also the size of such a formula is linear in $i$, which is bounded from above by $3n-1$. So the overall size of $\psi_n$ is quadratic in $n$, which is polynomial. A more in-depth examination leads to the findings $\vert\psi_n\vert_O=24\cdot n^2+10\cdot n=:g_1(n)$ and  $\vert \psi_n \vert =36\cdot n^2+17\cdot n+1=:g_2(n)$.
\end{proof}

\begin{lem}\label{no biimplication T,R}
There is an exponential function $h: \mathbb{N} \rightarrow \mathbb{Z}$, so that for every $n \in \mathbb{N}$ and for every ML-formula $\chi_n$ which separates $(\mathcal{M}_{t,r},a_n)$ from $(\mathcal{M}_{t,r},b_n)$, we have $\vert \chi_n \vert_O > h(n)$.
\end{lem}

\begin{proof} First we pick $h(n)=2^n-2$. Any ML-formula $\chi_0$ separating $a_0$ from $b_0$ satisfies $\vert \chi_0 \vert > -1 = h(0)$. So from now on we may assume $n \geq 1$.
Then we show that there is no separating ML-formula for $(\mathcal{M}_{t,r},a_n)$ and $(\mathcal{M}_{t,r},b_n)$ with at most $h(n)$ modal operators. This is done by playing the $\mathcal{M}_{t,r}$-ON game on $\{a_n\}$ and $\{b_n\}$ with parameter $h(n)$ and showing that $D$ has a winning strategy $f_D$ for these games. From this it follows that $S$ cannot have a winning strategy and with Theorem~\ref{correctness theorem} we know that no separating formula with $h(n)$ or less operators exists. The construction of the winning strategy is done by induction on $n$.

The base case is $n=1$. So we look at the game on $\{a_1\}$ and $\{b_1\}$ with parameter $h(1)=0$. Since $a_1$ and $b_1$ have the same colour, neither of them is removed when tidying up. So the starting position is $(\{\{a_1\},\{b_1\}\},0)$, which does not fulfil $A=B=\emptyset$. So $S$ does not win immediately. But because of $k=0$, he cannot make a move either. So $D$ always wins from this position.

For $n \geq 2$ let us assume the claim holds for $n-1$. So $(\{\{a_{n-1}\},\{b_{n-1}\}\},2^{n-1}-2)$ is a winning position for $D$. We will now play the $\mathcal{M}_{t,r}$-ON game in order to show that $(t(\{\{a_n\},\{b_n\}\}),2^n-2)$ is a winning position for $D$ as well.
During this game, we will use some tactics, that will shorten our proof, by constricting the set of possible moves of $S$ or allowing us to skip certain steps. Firstly, if $S$ during an operator-move picks a node $x$ for $X'$ with $x \in \Box Y$, then $f_D$ ensures $x \in Y'$ for her answer. This node $x$ cannot be removed when tidying up, because there will always be a node of the same colour in the other set, namely $x$ itself. So the sets after this move are not disjoint and, according to Corollary~\ref{loosing condition for S}, $D$ has a winning strategy from this new position. This allows us to ignore all moves of $S$ where he picks such a shared successor of $X$ and $Y$. This is especially helpful for $\mathcal{M}_{t,r}$, because in most cases we can ignore all edges that go down more than one layer. Secondly it is easy to see, that if $D$ can win from $(\{A,B\},k)$, she can also win from $(\{A,B\},k')$ for every $k'\leq k$. So we only have to prove a winning strategy for the highest occurring parameter. This also means that we can ignore all operator-moves where $S$ picks every node as its own successor, since $D$ can answer by doing the same. This is always possible, because $\mathcal{M}_{t,r}$ is reflexive. Then the new position consists of the same pair of sets as the old position, but with the parameter reduced by $1$. Thirdly as long as all nodes within a position are on the same layer, they have the same colour, with the exception of $8_i$ for every $i \in \mathbb{N}$ and $a_0$. So as long as $D$ answers only with nodes of the same layer as the choices of $S$, then no nodes will be removed when tidying up. If that is the case, we will not mention this part of the move. Lastly if both sets $A$ and $B$ of the current position are singletons, then no split-move can be played and we only need to consider operator-moves.

For the remainder of this proof we will identify some more winning positions for player $D$ in segment $n$, finishing with $(\{\{a_n\},\{b_n\}\},2^n-2)$. Since we only examine nodes of the same segment, we can drop the indices for most nodes. The first of these positions is $(\{\{4\},\{5\}\},2^{n-1}-1)$. From this position $S$ has to play an operator-move, where he can either pick $a_{n-1}$ as a successor of $4$ or $b_{n-1}$ as a successor of $5$. Every other successor of $4$ or $5$, like the node $8$, is a shared successor of both of them. If $S$ chooses $a_{n-1}$ then $D$ answers with $\{b_{n-1}\}$ and if he chooses $b_{n-1}$ then she answers with $\{a_{n-1}\}$. So she basically picks the successor $S$ did not pick. Then independent of the choice of $S$, the game continues from $(\{a_{n-1}\},\{b_{n-1}\}\},2^{n-1}-2)$, which according to our induction hypothesis is a winning position for $D$. So $(\{\{4\},\{5\}\},2^{n-1}-1)$ is a winning position for her as well. 

The next position we want to look at is $(\{\{6\},\{7\}\},2^{n-1}-1)$. Here $S$ has to play an operator move picking either $a_{n-1}$ as a successor of $6$ or $b_{n-1}$ as a successor of $7$. Then once again $D$ answers with the node that $S$ did not pick, progressing the game to the position $(\{\{a_{n-1}\},\{b_{n-1}\}\}, 2^{n-1}-2)$ as well. So $(\{\{6\},\{7\}\},2^{n-1}-1)$ is also a winning position for $D$. 

Now we look at $(\{\{2\},\{3\}\},2^{n-1})$. Once again $S$ has to play an operator-move. He can either pick $4$ as a successor of $2$ or $5$ as a successor of $3$. By answering with the respective other node, $D$ can ensure that the game continues from $(\{\{4\},\{5\}\},2^{n-1}-1)$. Then $(\{\{2\},\{3\}\},2^{n-1})$ is a winning position for $D$. 

By an analogous argument $(\{\{1\},\{2\}\},2^{n-1})$ is also a winning position for her. Here $S$ has to pick $6$ as a successor of $1$ or $7$ as a successor of $3$. Independent of his choice, $D$ can guarantee $(\{\{6\},\{7\}\},2^{n-1}-1)$ as the follow-up position.

At position $(\{\{4,7\},\{5,6\}\},2^n-4)$ player $S$ is unable to play an operator-move, since every successor of $4$ or $7$ is also a successor of $5$ or $6$ and vice versa. So he has to play a split-move instead. He can split either $\{4,7\}$ into $\{4\}$ and $\{7\}$ or $\{5,6\}$ into $\{5\}$ and $\{6\}$. Then he has to split the parameter $k=2^n-4$ into $k_1$ and $k_2$. When answering, $D$ can always pick $i$ in such a way that $k_i\leq \lfloor\frac{k}{2}\rfloor = 2^{n-1}-2$. If $S$ picked $X_i$ as $\{4\}$ then $D$ answers with $\{5\}$ and vice versa. On the other hand if $S$ choose $\{6\}$ or $\{7\}$ then $D$ answers with the respective other set. So the game continues from either $(\{\{4\},\{5\}\},k_i)$ or from $(\{\{6\},\{7\}\},k_i)$. We have already shown that $(\{\{4\},\{5\}\},2^{n-1}-1)$ and $(\{\{6\},\{7\}\},2^{n-1}-1)$ are winning positions for $D$. So because of $k_i\leq 2^{n-1}-2$ every possible follow-up position is a winning position for her and thus $(\{\{4,7\},\{5,6\}\},2^n-4)$ is as well.

Now we look into the position $(\{\{2,4\},\{1,5\}\},2^n-4)$. When playing an operator-move, $S$ has to play on $\{1,5\}$. This is because $4$ is a successor of $1$ and every proper successor of $4$ is two or more layers below $1$. So every successor of $4$ is also a successor of $1$ and thus $S$ cannot pick any successor of $4$. He also cannot pick any successor of $5$ besides $5$ itself for similar reasons, since every other successor of $5$ is two or more layers below $2$. So the only option he has left is picking $6$ as a successor of $1$ and using the self-loop of $5$. Then $D$ answers with the set $\{4,7\}$. The new position $(\{\{4,7\},\{5,6\}\},2^n-5)$ is, as already shown, a winning position for $D$. However $S$ could play a split-move instead. Then he has to split either $\{2,4\}$ into $\{2\}$ and $\{4\}$ or $\{1,5\}$ into $\{1\}$ and $\{5\}$. He also splits the parameter $k=2^n-4$ into $k_1$ and $k_2$. $D$ picks $i$ with $k_i\leq \lfloor\frac{k}{2}\rfloor = 2^{n-1}-2$ and then she answers in such a way that the play continues from either $(\{\{1\},\{2\}\},k_i)$ or $(\{\{4\},\{5\}\},k_i)$. So all possible follow-up positions are winning positions for $D$ and thus $(\{\{2,4\},\{1,5\}\},2^n-4)$ is a winning position as well.

The position $(\{\{2,7\},\{3,6\}\},2^n-4)$ is similar to the previously discussed position. If $S$ plays an operator-move, he has to play on $\{3,6\}$, where he has to pick $5$ as a successor of $3$ and $6$ as its own successor. Then $D$ can guarantee that the game continues with her winning position $(\{\{4,7\},\{5,6\}\},2^n-5)$. $S$ may also play a split-move. Then $D$ can ensure $(\{\{2\},\{3\}\},k_i)$ or $(\{\{6\},\{7\}\},k_i)$ as follow-up positions, where $k_i\leq \lfloor\frac{k}{2}\rfloor = 2^{n-1}-2$. Since all of these are winning positions for $D$, so is $(\{\{2,7\},\{3,6\}\},2^n-4)$.

At $(\{\{2\},\{1,3\}\},2^n-3)$, $S$ may play an operator-move. Since every successor of $2$ besides $2$ itself is also a successor of $1$ or $3$, $S$ has to play on $\{1,3\}$ instead. Then he can pick $1$ or $6$ as a successor of $1$ and $3$ or $5$ as a successor of $3$. So excluding the option of playing only on self-loops, he has three possible moves. If $S$ picks $\{1,5\}$, $D$ answers with $\{2,4\}$; if he picks $\{3,6\}$, she answers with $\{2,7\}$ and lastly if he picks $\{5,6\}$ she answer with $\{4,7\}$. Because $(\{\{2,4\},\{1,5\}\},2^n-4)$, $(\{\{2,7\},\{3,6\}\},2^n-4)$ and $(\{\{4,7\},\{5,6\}\},2^n-4)$ are all winning positions for $D$, $S$ cannot win the current position by playing an operator-move. If $S$ plays a split-move instead, he has to split $\{1,3\}$ into $\{1\}$ and $\{3\}$. Then he has to split $k=2^n-3$ into $k_1$ and $k_2$. $D$ answers by picking $i$ in such a way that $k_i\leq \lfloor\frac{k}{2}\rfloor = 2^{n-1}-2$ and picking $Y'=\{2\}$. The game continues with either $(\{\{1\},\{2\}\},k_i)$ or $(\{\{2\},\{3\}\},k_i)$, so importantly the game continues from a winning position for $D$. Hence $(\{\{2\},\{1,3\}\},2^n-3)$ is a winning position for her as well.

Lastly we look at $(\{\{a_n\},\{b_n\}\},2^n-2)$. In this position $S$ has to play an operator-move. All successors of $b_n$ are either shared successors of $a_n$ or $b_n$ itself. So $S$ has to pick a successor of $a_n$ and the only option there is $2$. Then $D$ simply answers with $\{1,3\}$ and the game is moved to the position $(\{\{2\},\{1,3\}\},2^n-3)$, which is a winning position for $D$. This proves that $D$ wins the $\mathcal{M}_{t,r}$-ON game from position $(\{\{a_n\},\{b_n\}\},2^n-2)$ and thus finishes the proof of the lemma.
\end{proof}

\begin{thm}\label{succinctness T,R}
The logic ML$[\leftrightarrow]$ is exponentially more operator-succinct and exponentially more succinct than ML on the class of Kripke structures with a transitive and reflexive accessibility relation.
\end{thm}

\begin{proof} 
Let $n \in \mathbb{N}$ be some natural number. Then $\psi_n$ is the ML$[\leftrightarrow]$-formula and $g_1,g_2$ are the polynomial functions given in Lemma~\ref{biimplication T,R}. Then we know $\vert \psi_n \vert_O =g_1(n)\geq n$ and $\vert \psi_n \vert= g_2(n)\geq n$. We also know that $\psi_n$ separates the two transitive and reflexive Kripke structures $(\mathcal{M}_{t,r},a_n)$ and $(\mathcal{M}_{t,r},b_n)$ from one another. Let $\chi_n$ be some ML-formula equivalent to $\psi_n$ on the class of Kripke structures with a transitive and reflexive accessibility relation. This formula $\chi_n$ has to be a separating formula for $(\mathcal{M}_{t,r},a_n)$ and $(\mathcal{M}_{t,r},b_n)$ as well. So, according to Lemma~\ref{no biimplication T,R}, $\vert \chi_n \vert_O \geq h(n)$ for the exponential function $h$. This implies $\vert \chi_n \vert \geq h(n)$ as well.  We define $f_1 \colon \mathbb{N} \rightarrow \mathbb{N}$ as $f_1(n)=h\left(\lfloor (g_1)^{-1}(n)\rfloor\right)$. Then $\vert \chi_n \vert_O \geq h(n) \geq h\left(\lfloor (g_1)^{-1}(\vert \psi_n \vert_O)\rfloor\right)=f_1\left(\vert\psi_n\vert_O\right)$ holds. So ML$[\leftrightarrow]$ is $f_1$-times more operator-succinct than ML and thus exponentially more operator-succinct. We also define $f_2 \colon \mathbb{N} \rightarrow \mathbb{N}$ as $f_2(n)=h\left(\lfloor (g_2)^{-1}(n)\rfloor\right)$. We know $\vert \chi_n \vert \geq f_2\left(\vert\psi_n\vert\right)$ holds. This implies that ML$[\leftrightarrow]$ is $f_2$-times more succinct than ML, so it is exponentially more succinct as well.
\end{proof}

We can use the existence of this succinctness gap to show that for some formulas the representation through a DAG can be exponentially smaller than the formula itself.

\begin{cor}
The logic DAG(ML) is exponentially more operator-succinct and exponentially more succinct than ML on the class of Kripke structures with a transitive and reflexive accessibility relation.
\end{cor}

\begin{proof}
The main idea here is that when working with DAGs the bi-implication can be removed with only minimal size-increase. We already know that $\varphi \leftrightarrow \psi$ is equivalent to $(\varphi \land \psi) \lor (\lnot \varphi \land \lnot \psi)$. This equivalence can be used to remove all bi-implications from any ML-formula. However this procedure leads to an exponential increase in formula-size. But this is not the case for DAGs. Instead we have the following:
\begin{equation*}
\vert DAG((\varphi \land \psi) \lor (\lnot \varphi \land \lnot \psi))\vert \leq \vert DAG(\varphi) \vert + \vert DAG(\psi) \vert + 5 \leq \vert \varphi \vert + \vert \psi \vert +5 \leq \vert \varphi \leftrightarrow \psi \vert+4
\end{equation*}
So for every $\leftrightarrow$ removed we only get a constant increase in DAG-size. As a very rough estimation for every ML$[\leftrightarrow]$-formula $\varphi$ there exists an equivalent ML-formula $\psi$ with $\vert DAG(\psi)\vert \leq 5 \cdot \vert \varphi \vert$. This linear size increase does not matter when compared to the exponential succinctness gap between ML$[\leftrightarrow]$ and ML. So this gap also exists between DAG(ML) and ML.
\end{proof}

From Theorem~\ref{succinctness T,R} it is easy to infer, that the succinctness gap also exists on all classes of structures, which are supersets of the class of Kripke structures with a transitive and reflexive accessibility relation. Of course the same is true for the succinctness gap between DAG(ML) and ML. We say an accessibility relation $R$ is \emph{serial}, if it satisfies the condition $\forall x \exists y: R(x,y)$. It is called \emph{weakly dense}, if it satisfies $\forall x \forall y:(R(x,y) \rightarrow \exists z:(R(x,z) \land R(z,y))$. Both of these conditions are true for every reflexive relation.

\begin{cor}\label{superset classes t,r}
The logics ML$[\leftrightarrow]$ and DAG(ML) are exponentially more operator-succinct and exponentially more succinct than ML on the following classes of pointed Kripke structures:
\begin{enumerate}
\item The class of all pointed Kripke structures.
\item The class of all pointed Kripke structures with a transitive accessibility relation.
\item The class of all pointed Kripke structures with a reflexive accessibility relation.
\item The class of all pointed Kripke structures with a serial accessibility relation.
\item The class of all pointed Kripke structures with a weakly dense accessibility relation.
\end{enumerate}
\end{cor}

\subsection{Symmetrical and reflexive graphs}\label{section S,R}
The graph $\mathcal{M}_{s,r}$ is also constructed as an extension of $\mathcal{M}=(W,R,V)$. We do not change the nodes or their labels. We only add edges to $\mathcal{M}$. For every edge $(x,y) \in R$ we add its inverse $(y,x)$ to our new graph. For nodes $x$ and $y$, that belong to the same level, we also add the edges $(x,y)$ and $(y,x)$. This explicitly includes all possible self-loops. So formally $R_{s,r}=R \cup \{(y,x)\mid (x,y)\in R\}\cup\{(x,y) \mid \exists i \in \mathbb{N} \colon \text{$x$ and $y$ belong to layer $i$}\}$. It is easy to see that $\mathcal{M}_{s,r}$ is both symmetrical and reflexive.

\begin{lem}\label{biimplication S,R}
There are two polynomial functions $g_1,g_2 \colon \mathbb{N}\rightarrow \mathbb{N}$, so that for every $n \in \mathbb{N}$ there is an ML$[\leftrightarrow]$-formula $\psi_n$, which separates $(\mathcal{M}_{s,r},a_n)$ from $(\mathcal{M}_{s,r},b_n)$, with $\vert \psi_n\vert_O=g_1(n)\geq n$ and $\vert \psi_n \vert =g_2(n)\geq n$.
\end{lem}

\begin{proof} Once again this is proven by adapting $\varphi_n$ from Claim~\ref{phi_i}. All edges in $\mathcal{M}$ go down exactly one layer. But all of the edges added to $R$ during the construction of $R_{s,r}$ either go up one layer or stay on the same layer. Since we used the labels $q_0$, $q_1$ and $q_2$ in an alternating pattern, the current layer, the layer one below and the layer one above are all coloured differently. This means we can use these labels to make sure that only the original edges from $R$ are considered. We define $\psi_n$ via $\psi_0:=p_0$ and $\psi_i:=\Diamond(q_1\land \Box(\lnot q_2\lor((\Diamond(p_1\land q_0)\leftrightarrow (\Diamond(\psi_{i-1}\land q_0)))))$. For all nodes $w$ of layer $3n$ we know that $(\mathcal{M}_{s,r},w)\models\psi_n$ holds if, and only if, $(\mathcal{M},w)\models\varphi_n$. So, by Claim~\ref{phi_i}, $\psi_n$ is a separating formula for $(\mathcal{M}_{s,r},a_n)$ and $(\mathcal{M}_{s,r},b_n)$ with $\vert\psi_n\vert_O=\vert\psi_{n-1}\vert_O+4$ and $\vert \psi_n\vert=\vert\psi_{n-1}\vert+15$. So $g_1(n)=4 \cdot n$ and $g_2(n)=15 \cdot n +1$ fulfil all conditions of the lemma.
\end{proof}

\begin{lem}\label{no biimplication S,R}
There is an exponential function $h: \mathbb{N} \rightarrow \mathbb{Z}$, so that for every $n \in \mathbb{N}$ and for every ML-formula $\chi_n$, which separates $\{(\mathcal{M}_{s,r},a_n)\}$ from $\{(\mathcal{M}_{s,r},b_n)\}$, we have $\vert \chi_n \vert_O > h(n)$.
\end{lem}

\begin{proof} The exponential function we use is $h(n)=2^n-2$, the same function we used during the proof of Lemma~\ref{no biimplication T,R}. Then $\vert \chi_0 \vert_O > -1 =h(0)$ holds. Next we need to show that for every $n \geq 1$ player $D$ has a winning strategy in the $\mathcal{M}_{s,r}$-ON game starting from $(\{\{a_n\},\{b_n\}\},h(n))$. For this purpose we define the following sets of positions of the $\mathcal{M}_{s,r}$-ON game.

\begin{itemize}
\item $W_D^0:=\bigl\{(\{A,B\},k) \mid A,B\subseteq W, A\cap B \neq \emptyset, k\in \mathbb{N}\bigr\}$
\item $W_D^1:=\bigl\{(\{\{a_i\},\{b_i\}\},k) \mid i \geq 1, k \leq 2^i-2\bigr\}$
\item $W_D^2:=\bigl\{(\{\{2_i\},\{1_i,3_i\}\},k) \mid i \geq 2, k \leq 2^i-3\bigr\}$
\item $W_D^3:=\bigl\{(\{\{4_i,7_i\},\{5_i,6_i\}\},k) \mid i \geq 2, k \leq 2^i-4\}$
\item $W_D^4:=\bigl\{(\{\{1_i\},\{2_i\}\},k) \mid i \geq 2, k \leq 2^{i-1}\bigr\}$
\item $W_D^5:=\bigl\{(\{\{1_i\},\{3_i\}\},k) \mid i \geq 2, k \leq 2^{i-1}\bigr\}$
\item $W_D^6:=\bigl\{(\{\{2_i\},\{3_i\}\},k) \mid i \geq 2, k \leq 2^{i-1}\bigr\}$
\item $W_D^7:=\bigl\{(\{\{4_i\},\{5_i\}\},k) \mid i \geq 2, k \leq 2^{i-1}-1\bigr\}$
\item $W_D^8:=\bigl\{(\{\{6_i\},\{7_i\}\},k) \mid i \geq 2, k \leq 2^{i-1}-1\bigr\}$
\end{itemize}

Then $W_D:=\bigcup\limits_{0 \leq j \leq 8} W_D^j$ is defined as the union of these sets. Our aim is to show that all positions in $W_D$ are winning positions for $D$, because then $(\{\{a_n\},\{b_n\}\},2^n-2)\in W_D^1$ has to be one of her winning positions. We prove this fact by showing that player $D$ can stop player $S$ from leaving $W_D$. Since $W_D$ does not contain any position $(\{A,B\},k)$ with $A = B = \emptyset$, this implies that $S$ cannot win the $\mathcal{M}_{s,r}$-ON game from any position in $W_D$. So now we need to examine every position in $W_D$ and every move that $S$ can make and show that player $D$ can always keep the play in $W_D$.

In a position with $k=0$ player $S$ cannot make a move. So in particular he cannot leave $W_D$. So from now on we may assume $k>1$.

We start with $W_D^0$. So let $(\{A,B\},k)\in W_D^0$ be some position and let $w\in A \cap B$ be some node shared between $A$ and $B$. For $k\geq 2$, $S$ can play a split-move in this position. He picks $X\in\{A,B\}$ and then splits this set into $X_1$ and $X_2$. Then the shared node $w$ has to be in one of these two sets, w.l.o.g. we can assume $w\in X_1$. Then $D$ can ensure $w\in Y'$ for her answer and the game continues from the position $(t(\{X_1,Y'\}),k_1)$. The node $w$ is also a shared node for $X_1$ and $Y'$. For this reason $w$ cannot be removed by tidying up, because there will always be a node of the same colour in the respective other set, namely $w$ itself. So in the position after the move, the two sets still have $w$ as a shared node. Then this new position lies in $W_D^0$ as well and thus in $W_D$. On the other hand $S$ might play an operator-move instead. In this case he picks $X \in \{A,B\}$ and then for every $x \in X$ he picks a successor $f(x)$. Because of $w \in Y$, we know $f(w)\in \Box Y$. So $D$ can ensure $f(w)\in Y'$. Then $f(w)$ is a shared node for $X'$ and $Y'$ and thus it is not removed by tidying up. This means that the follow-up position $(t(\{X',Y'\}),k-1)$ is also a member of $W_D^0$ and of $W_D$. So $D$ can ensure that the $\mathcal{M}_{s,r}$-ON game cannot move from a position inside of $W_D^0$ to a position outside of $W_D$.

During an operator-move, if $S$ tries to pick some node $x'$ as $f(x)$ with $x' \in \Box Y$, player $D$ can ensure $x'\in Y'$. So $x'$ is a shared node for the new position and thus this position is a part of $W_D^0$. Then $S$ cannot leave $W_D$ by playing like this. So from now on we may ignore all of his moves where he picks some shared successor. This is especially helpful, because in $\mathcal{M}_{s,r}$ all nodes of the same layer are connected to one another. So we can ignore all moves where $S$ tries to pick some node that is on the same layer as any node in $Y$. This includes $S$ using any self-loops.

The next set we want to look at is $W_D^1$. So let $(\{\{a_i\},\{b_i\}\},k)$ be a position with $i\geq 1$ and $k\leq 2^i-2$. For $i=1$ we know $k=0$. As already mentioned $S$ cannot make another move for $k=0$, meaning he cannot leave $W_D$. For $i\geq 2$ he cannot play a split-move, since both sets have only one element each. So he has to play an operator-move instead. Recall that the nodes $a_i$ and $b_i$ belong to segment $i+1$ as well as to segment $i$ and since $\mathcal{M}_{s,r}$ is a symmetrical graph, these nodes also have successors in the layer above, layer $3i+1$.
Then $S$ can pick $4_{i+1}$, $6_{i+1}$ or $2_i$ as a successor of $a_i$. Alternatively he may pick $5_{i+1}$ or $7_{i+1}$ as a successor of $b_i$. Every other successor of $a_i$ or $b_i$ is a shared successor. If he choose either $4_{i+1}$ or $5_{i+1}$, then $D$ answers with a set that contains only the respective other node. This leads to the follow-up position $(\{\{4_{i+1}\},\{5_{i+1}\}\},k-1)$. Because of $k-1 \leq 2^i-3$ this position belongs to $W_D^7$. Alternatively if $S$ picked $6_{i+1}$ or $7_{i+1}$, then $D$ also answers with the respective other node. Then the follow-up position is $(\{\{6_{i+1}\},\{7_{i+1}\}\},k-1)$, which belongs to $W_D^8$. Lastly if $S$ decided to move from $a_i$ to $2_i$, then $D$ answers with the set $\{1_i,3_i\}$ which progresses the play to the position $(\{\{2_i\},\{1_i,3_i\}\},k-1)$. This position belongs to $W_D^2$. So independent of the choices made by $S$, the play always continues from some position within $W_D$.

For $W_D^2$ we have to look at positions $(\{\{2_i\},\{1_i,3_i\}\},k)$ with $i \geq 2$ and $k \leq 2^i-3$. Since every successor of $2_i$ is also a successor of $1_i$ or $3_i$, player $S$ cannot leave $W_D$ by playing an operator-move on $\{2_i\}$. Similarly when playing on $\{1_i,3_i\}$ we only have to look at $S$ picking $b_i$ or $6_i$ as a successor of $1_i$, as well as $b_i$ or $5_i$ as a successor of $3_i$. This is because every other successor is also a successor of $2_i$. If $S$ picked $b_i$ at all, then $D$ answers with $\{a_i\}$. If he choose $5_i$ or $6_i$ in addition to $b_i$, then these nodes will be removed when tidying up, because they are labelled differently from $a_i$ and $b_i$. So in this case the game continues with $(\{\{a_i\},\{b_i\}\},k-1)$, a position from $W_D^1$. On the other hand, if $S$ picked both $5_i$ and $6_i$ as successors, $D$ answers with $\{4_i,7_i\}$. Then the follow-up position is $(\{\{4_i,7_i\},\{5_i,6_i\}\},k-1)$ from $W_D^3$. So $S$ cannot leave $W_D$ by playing an operator-move at $W_D^2$. Lastly we look at what happens when $D$ plays a split-move. He has to split $\{1_i,3_i\}$ into $\{1_i\}$ and $\{3_i\}$ and $k$ into $k_1$ and $k_2$. Then $D$ can answer in such a way that $k_j \leq \lfloor \frac{k}{2} \rfloor \leq 2^{i-1}$ and the follow-up position is either $(\{\{1_i\},\{2_i\}\},k_j)\in W_D^4$ or $(\{\{2_i\},\{3_i\}\},k_j)\in W_D^6$. With this we have shown that $S$ cannot leave $W_D$ in one move from $W_D^2$.

We continue by analysing $W_D^3$. Let $(\{\{4_i,7_i\},\{5_i,6_i\}\},k)$ be some position with $i\geq 2$ and $k\leq 2^i-4$. If $S$ plays an operator-move from this position, he has to pick $2_i$ as a successor of both $4_i$ and $7_i$, since every other successor is a shared successor. To this, $D$ answers with the set $\{1_i,3_i\}$. Then the follow-up position $(\{\{2_i\},\{1_i,3_i\}\},k-1)$ belongs to $W_D^2$. However for $k\geq 2$, $S$ can also play a split-move. In this case he has to split either $\{4_i,7_i\}$ into $\{4_i\}$ and $\{7_i\}$ or $\{5_i,6_i\}$ into $\{5_i\}$ and $\{6_i\}$. He also splits $k$ into $k_1$ and $k_2$. Let $k_j$ be the smaller of these two. Then we know $k_j \leq \lfloor \frac{k}{2}\rfloor\leq 2^{i-1}-2$. Then $D$ answers by picking the side with $k_j$. The set she picks depends on what choice was made by $S$ for $X_j$. She answers with $\{5_i\}$ to $\{4_i\}$ and vice versa. Similarly she picks $\{6_i\}$ as an answer to $\{7_i\}$ and the other way around. So the follow-up position is either $(\{\{4_i\},\{5_i\}\},k_j)$, which is a member of $W_D^7$, or $(\{\{6_i\},\{7_i\}\},k_j)$, which is a member of $W_D^8$. So $S$ cannot leave $W_D$ from $W_D^3$ either.

The next set is $W_D^4$. So we look at a position $(\{\{1_i\},\{2_i\}\},k)$ with $i \geq 2$ and $k\leq 2^{i-1}$. In this position $S$ cannot play a split-move, since both sets contain only one node each. This will be the case for all sets from this point on, so we will stop mentioning it going forward. Now player $S$ has to play an operator-move, where he picks $b_i$ or $6_i$ as a successor of $1_i$ or $7_i$ as a successor of $2_i$. If he picked $b_i$, $D$ answers with $\{a_i\}$. This moves the game to $(\{\{a_i\},\{b_i\}\},k-1)$. Because of $i \geq 2$, which implies $2^{i-1}\leq 2^i-2$, this position belongs to $W_D^1$. On the other hand, if $S$ choose $6_i$ or $7_i$, then $D$ answers with the set only containing the respective other node. Then the follow-up position is $(\{\{6_i\},\{7_i\}\},k-1)$, which belongs to $W_D^8$. Hence $D$ can force any play from $W_D^4$ to stay in $W_D$.

We continue our proof with $W_D^5$. The positions within this set are of the shape $(\{1_i\},\{3_i\}\},k)$ with $i\geq 2$ and $k\leq 2^{i-1}$. When playing an operator-move, $S$ may pick $4_i$ or $6_i$ as a successor of $1_i$ or he may pick $5_i$ or $7_i$ as a successor of $3_i$. Then $D$ answers with $\{5_i\}$ to $4_i$, with $\{7_i\}$ to $6_i$, with $\{4_i\}$ to $5_i$ and lastly with $\{6_i\}$ to $7_i$. Then the play continues with $(\{\{4_i\},\{5_i\},k-1)$, a position from $W_D^7$, or with $(\{\{6_i\},\{7_i\}\},k-1)$, a position from $W_D^8$. This finishes the case of $W_D^5$.

Now we look at the set $W_D^6$, which contains all positions $(\{\{2_i\},\{3_i\}\},k)$ with $i\geq 2$ and $k\leq 2^{i-1}$. Here $S$ may pick $4_i$ as a successor of $2_i$ or alternatively $b_i$ or $5_i$ as a successor of $3_i$. If he picked $b_i$ then $D$ answers with $\{a_i\}$. Then the follow-up position $(\{\{a_i\},\{b_i\}\},k-1)$ is a member of $W_D^1$. If $S$ choose $4_i$ or $5_i$ instead, then $D$ answers by picking the singleton of the respective other node. In this case the game moves on to the position $(\{\{4_i\},\{5_i\}\},k-1)$, which belongs to $W_D^7$.

The set $W_D^7$ consists of all positions $(\{\{4_i\},\{5_i\}\},k)$ with $i \geq 2$ and $k \leq 2^{i-1}-1$. Then $S$ can play an operator-move by picking $1_i$, $2_i$ or $a_{i-1}$ as a successor of $4_i$. Alternatively he may also pick $3_i$ or $b_{i-1}$. If $S$ choose $2_i$ or $3_i$, then $D$ answers with $\{3_i\}$ or $\{2_i\}$ respectively. This creates the follow-up position $(\{\{2_i\},\{3_i\}\},k-1)$, which is a part of $W_D^6$. If $S$ picked $1_i$, then $D$ answers with $\{3_i\}$, creating $(\{\{1_i\},\{3_i\}\},k-1)$ instead. This position belongs to $W_D^5$. On the other hand if $S$ picked $a_{i-1}$ or $b_{i-1}$, then $D$ answers with the set only containing the respective other node. In this case the follow-up position $(\{\{a_{i-1}\},\{b_{i-1}\}\},k-1)$ belongs to $W_D^1$.

Lastly we look at the set $W_D^8$. This set contains all positions $(\{\{6_i\},\{7_i\}\},k)$ with $i \geq 2$ and $k \leq 2^{i-1}-1$. From such a position, $S$ may play an operator-move where he picks either $1_i$ or $a_{i-1}$ as a successor of $6_i$ or $2_i$, $3_i$ or $b_{i-1}$ as a successor of $7_i$. If he choose $1_i$ or $2_i$, then $D$ answers with the respective other node as a singleton. The follow-up position $(\{\{1_i\},\{2_i\}\},k-1)$ is a member of $W_D^4$. If $S$ choose $3_i$ instead, then $D$ answers with $\{1_i\}$ and the new position is $(\{\{1_i\},\{3_i\}\},k-1)$, a member of $W_D^5$. On the other hand if $S$ picked $a_{i-1}$ or $b_{i-1}$, then $D$ picks $\{b_{i-1}\}$ and $\{a_{i-1}\}$ respectively. Then the game continues with $(\{\{a_{i-1}\},\{b_{i-1}\}\},k-1)$, a position from $W_D^1$. So all follow-up positions belong to $W_D$.

So we have shown that $D$ can stop $S$ from leaving $W_D$. Also there is no immediate winning position for $S$ in $W_D$. Then all position in $W_D$, including $(\{\{a_n\},\{b_n\}\},2^n-2)$, have to be winning positions for player $D$. This is why there cannot be a separating formula for $\{a_n\}$ and $\{b_n\}$ with $2^n-2$ or less modal operators.
\end{proof}

\begin{thm}\label{operator-succinctness S,R}
The logic ML$[\leftrightarrow]$ is exponentially more operator-succinct and exponentially more succinct than ML on the class of Kripke structures with a symmetrical and reflexive accessibility relation.
\end{thm}

\begin{proof} The proof of this theorem can be done analogous to the proof of Theorem~\ref{succinctness T,R}. Simply replace all mentions of transitivity with reflexivity and use Lemma~\ref{biimplication S,R} and Lemma~\ref{no biimplication S,R} instead of Lemma~\ref{biimplication T,R} and Lemma~\ref{no biimplication T,R}, respectively.
\end{proof}

As previously shown the succinctness gap between ML$[\leftrightarrow]$ and ML itself implies a succinctness gap between the representation through DAGs and through formulas.
Also since the succinctness gap exists when looking only at symmetrical and reflexive graphs, it has to exist for any class of graphs that contains all symmetrical and reflexive graphs.

\begin{cor}
The logic DAG(ML) is exponentially more operator-succinct and exponentially more succinct than ML on the class of Kripke structures with a symmetrical and reflexive accessibility relation.
\end{cor}

\begin{cor}
The logics ML$[\leftrightarrow]$ and DAG(ML) are exponentially more operator-succinct and exponentially more succinct than ML on the class of Kripke structures with a symmetrical accessibility relation.
\end{cor}

\subsection{Transitive and symmetrical graphs}
So far in this paper we have shown that adding bi-implications allows us to write more succinct formulas. Even when only looking at special classes of Kripke structures, like those with a transitive accessibility relation and those with a symmetrical one. In this section we are going to show that the same does not hold if we combine these two conditions. More precisely we show that there is no exponential succinctness gap between ML and ML$[\leftrightarrow]$ on the class of Kripke structures with a transitive and symmetrical accessibility relation.

We start with a few remarks about our notation in this section. Let $\mathcal{S}_{S5}$ be the class of all Kripke structures with a transitive, symmetrical and reflexive accessibility relation. We write $\varphi \equiv_{T,S} \psi$ if $\varphi$ and $\psi$ are equivalent on all pointed Kripke structures with a transitive and symmetrical accessibility relation. Similarly we use $\equiv_{\mathcal{S}_{S5}}$ when talking about structures from $\mathcal{S}_{S5}$ and $\equiv_E$ for graphs with an empty accessibility relation.

All connected, transitive and symmetrical graphs, that are not reflexive as well, contain a single node and no edges.
So $\varphi \equiv_{T,S} \psi$ holds if, and only if, both $\varphi \equiv_{\mathcal{S}_{S5}} \psi$ and $\varphi \equiv_E \psi$ hold. This fact offers us a way to find formulas that are equivalent on transitive and symmetrical structures, which we will use to prove the following Theorem.

\begin{thm}
For every ML$[\leftrightarrow]$-formula $\varphi$ there is an ML-formula $\psi$ with $\varphi\equiv_{T,S}\psi$ and the size of $\psi$ is polynomial in the size of $\varphi$.
\end{thm}

\begin{proof} In \cite{Schwarz} it is shown, that ML$[\leftrightarrow]$ has polynomial translations with regards to $\mathcal{S}_{S5}$ in ML. So for every ML$[\leftrightarrow]$-formula $\varphi$ there is an ML-formula $\psi_1$ with $\varphi \equiv_{\mathcal{S}_{S5}} \psi_1$ and the size of $\psi_1$ is polynomial in the size of $\varphi$. We now look at the graph with only one node and no edges. We know $\Diamond \chi \equiv_E \bot$ as well as $\Box \chi \equiv_E \top$ for all formulas $\chi$. We can use this to replace all operators in $\varphi$, obtaining a operator free formula $\psi'$ with $\psi' \equiv_E \varphi$. Since this formula does not contain any operators, we can treat it as a formula of propositional logic. So as described by Pratt in \cite{Pratt}, we can use Spira's construction from \cite{spira} in order to construct an ML-formula $\psi_2$ with $\psi_2 \equiv \psi' \equiv_E \varphi$. The size of $\psi_2$ is polynomial in the size of $\psi'$, which in turn is bounded from above by the size of $\varphi$. Now we just need to combine these formulas to obtain $\psi$. In order to do so we use the formula $\Diamond \top$, which is true on all graphs from $\mathcal{S}_{S5}$ and false for all edgeless graphs, as well as its negation $\Box \bot$. We construct the wanted ML-formula as $\psi := (\psi_1 \land \Diamond \top) \lor (\psi_2 \land \Box \bot)$. This formula fulfils the following:
\begin{equation*}
\psi\equiv_{\mathcal{S}_{S5}}(\psi_1 \land \Diamond \top) \lor (\psi_2 \land \Box \bot) \equiv_{\mathcal{S}_{S5}}(\psi_1 \land \top) \lor (\psi_2 \land \bot) \equiv_{\mathcal{S}_{S5}} \psi_1 \equiv_{\mathcal{S}_{S5}} \varphi
\end{equation*}
\begin{equation*}
\psi\equiv_E(\psi_1 \land \Diamond \top) \lor (\psi_2 \land \Box \bot) \equiv_E(\psi_1 \land \bot) \lor (\psi_2 \land \top) \equiv_E \psi_2 \equiv_E \varphi
\end{equation*}
Which also implies $\psi \equiv_{T,S}\varphi$. The size of this formula is $\vert \psi_1 \vert + \vert \psi_2 \vert + 7$, so the sum of two polynomials over $\vert \varphi \vert$ and a constant, which is also polynomial in the size of $\varphi$.\end{proof}

\section{The operator-number game for computation tree logic}
\begin{defi}
Let $\mathcal{M}=(W,R,V)$ be a serial and finitely labelled graph, $A_0,B_0 \subseteq W$ two sets of nodes and $k_0\in \mathbb{N}$. The $\mathcal{M}$-\emph{Operator}-\emph{Number} game for computation tree logic, or $\mathcal{M}$-ON game for CTL for short, played on $A_0$ and $B_0$ with parameter $k_0$ has two players $S$ and $D$. An \emph{$S$-position} $(\{A,B\},k)$ of the game consists of a tidy pair of sets of nodes $A,B \subseteq W$, as well as a parameter $k \in \mathbb{N}$. The starting position is $(t(\{A_0,B_0\}),k_0)$. A \emph{turn for $S$} starts from an $S$-position and proceeds in one of the following four ways:
\begin{enumerate}
\item \textbf{Split-move:} $S$ may only play this type of move if $k \geq 2$. He picks a set $X \in \{A,B\}$ and splits it into two disjoint and non-empty sets $X_1$ and $X_2$. Then he chooses two parameters $k_1,k_2\geq 1$ with $k_1+k_2=k$.
Afterwards $D$ plays by taking the set $Y$ with $\{X,Y\}=\{A,B\}$ and picking a subset $Y' \subseteq Y$ of it. Then she chooses $i \in \{1,2\}$ and the game continues from the $S$-position $(t(\{X_i,Y'\}),k_i)$.

\item \textbf{Operator-move:} $S$ may only play this type of move if $k\geq 1$. He picks a set $X \in \{A,B\}$ and then picks a function $f:X\rightarrow\Box X$, where $f(x)\in \Box x$ for every $x\in X$. Let $X'=\{f(x)\mid x \in X\}$ be the image of $X$ under $f$. Afterwards $D$ plays by taking the set $Y$ with $\{X,Y\}=\{A,B\}$ and then choosing a set $Y'\subseteq \Box Y$ of successors of the nodes from $Y$. The game continues from the position $(t(\{X',Y'\}),k-1)$.

\item \textbf{EU-move:} $S$ may only play this type of move if $k \geq 1$. He picks a set $X \in \{A,B\}$ and then for every node $x \in X$ he picks a finite path $(x_j^n)_{0 \leq j \leq \ell_n}$ with $x_0^n=x$. Let $X_2=\bigcup\limits_{x_0^n \in X}\{x^n_{\ell_n}\}$ be the set of all final nodes of these paths and $X_1=\bigcup\limits_{x_0^n \in X} \{x_j^n \mid 0 \leq j < \ell_n\}$ the set of all other nodes of these paths. He also splits $k$ into two new parameters $k_1$ and $k_2$ with $k=k_1+k_2+1$. Then player $D$ answers by picking some finite paths $(y_j^m)_{0 \leq j \leq \ell_m}$ with $y_0^m\in Y$ for $\{X,Y\}=\{A,B\}$. For each path chosen by $D$ player $S$ has to pick some $j_m$ with $0 \leq j_m \leq \ell_m$. Then for $j_m < \ell_m$ he puts $y_{j_m}^m$ into $Y_1$ and for $j_m=\ell_m$ he puts $y_{j_m}^m$ into $Y_2$. Lastly $D$ picks $i\in \{1,2\}$ and the game continues with the $S$-position $(t(\{X_i,Y_i\}),k_i)$.

\item\textbf{AU-move:} $S$ may only play this move if $k \geq 1$. He picks a set $X \in \{A,B\}$ and then for every node $x \in X$ he picks either a finite path $(x_j^n)_{0\leq j \leq \ell_n}$ or an infinite path $(x_j^n)_{0 \leq j}$ with $x_0^n=x$. Then he collects all nodes of these paths into $X_1$. In addition all final nodes of the finite paths are also gathered into $X_2$. Then he splits the parameter $k$ into $k_1$ and $k_2$ with $k=k_1+k_2+1$. Player $D$ answers by picking some finite paths $(y_j^m)_{0 \leq j \leq \ell_m}$ and some infinite paths $(y_j^m)_{0\leq j}$, each with $y_0^m\in Y$. For every one of these paths player $S$ picks a node $y_j^m$. For $j=\ell_m$ he can choose to either put $y_j^m$ either into $Y_1$ or $Y_2$. If $y_j^m$ is not the final node of its path or if the path was infinite, the node is always added to $Y_1$. Lastly $D$ picks $i \in \{1,2\}$. Then the game continues with position $t(\{X_i,Y_i\}),k_i)$.
\end{enumerate}
Then our game is a reachability game, where $S$ has to reach a position $(\{A,B\},k)$ with $A=B=\emptyset$.
\end{defi}

The main idea of the $\mathcal{M}$-ON game for CTL is to extend the $\mathcal{M}$-ON game for ML by adding two new moves that correspond to the two new operators $EU$ and $AU$. These two new moves differ from the other moves in the fact that they cannot be divided into a single $S$-move and a single $D$-move. Instead these moves begin with $S$ making a choice, then $D$ has to answer, $S$ has to make a second choice and finally $D$ has to give a second answer. Because of this we do not properly define intermediate $D$-positions for these moves. In order to stay consistent we also remove the mention of $D$-positions from the definitions of the operator- and split-moves as well. However these moves are still functionally identical to those in the game for ML.

In a similar fashion to the $\mathcal{M}$-ON game for ML the parameter $k$ is reduced in every move. So we can once again use Zermelo's Theorem to state that for every position one of the players, either $S$ or $D$, has a positional winning strategy. Since we do not use $D$-positions for this game, we need to slightly alter our definitions for positional strategies. Now for every choice a player $P \in \{S,D\}$ has to make, a positional strategy has to give a valid choice that $P$ can make. For picking this choice the strategy may only take into account the current $S$-position, as well as all choices already made by either player during the current move, if there are any.

In contrast to the game for ML, the $\mathcal{M}$-ON game for CTL requires the graph $\mathcal{M}$ to be serial. We choose to add this condition because of the definition of the $EU$- and $AU$-operators. These operators look for infinite paths, but for each path they only care about some finite prefix of that path. In a serial graph every finite path can be extended into an infinite path. So it suffices to examine these finite prefixes in the knowledge that all of them belong to some infinite path. It is possible to define a version of our game that works on non-serial graphs. We only need to enforce that both players can only pick a finite path, if that path can be extended into an infinite one. However we decided that simply requiring $\mathcal{M}$ to be serial was more convenient.

While the main idea behind the EU-move should be rather obvious, this is most likely not the case for the AU-move. Because of this we want to use this paragraph to convey the idea behind the AU-move. For this let us examine a position $(\{A,B\},k)$ of the $\mathcal{M}$-ON game for modal logic, where $\Box \varphi$ is a separating formula. Now player $S$ cannot show directly that $\Box \varphi$ holds in $A$. Instead he has to exploit that $\Diamond \lnot \varphi$ holds in $B$ and he plays an operator-move on $B$, where he only picks successors $b'$ with $b'\models \lnot \varphi$. The AU-move of the $\mathcal{M}$-ON game for CTL works in a similar fashion. Instead of showing that on every path starting in $A$ $\varphi$ holds until $\psi$ holds, $S$ has to show that for every node in $B$ there is a path where $\varphi$ until $\psi$ does not hold. There are two ways in which a path may contradict $\varphi$ until $\psi$. Firstly there can be paths where $\varphi$ does not hold even though $\psi$ did not hold at any previous node either. In this case $S$ has to mark the beginning of that path up to the node where $\varphi$ does not hold. Secondly there can be paths where $\psi$ never holds, even if $\varphi$ is always true. Then $S$ has to mark the entire infinite path. In this way $S$ can prove that $\lnot (\varphi AU \psi)$ holds in $B$.

Now we want to show that our $\mathcal{M}$-ON game can characterize the number of operators necessary for a separating CTL-formula. The overall structure of our proof is basically identical to the proof for the game for ML. We will even skip the proofs of some lemmas, if we can just reuse the proof of the corresponding lemma from Section~\ref{section game for ML}.

\begin{lem}\label{tidy separating formulas CTL} (Lemma \ref{tidy separating formulas})
Let $\mathcal{M}=(W,R,V)$ be a finitely labelled graph and $A,B,A',B' \subseteq W$ be sets of nodes with $t(\{A,B\})=\{A',B'\}$. If there is a separating CTL-formula $\varphi'$ for $(\mathcal{M},A')$ and $(\mathcal{M},B')$, then there is also a separating CTL-formula $\varphi$ for $(\mathcal{M},A)$ and $(\mathcal{M},B)$ with $\vert \varphi \vert_O=\vert \varphi'\vert_O$.
\end{lem}

\begin{lem}\label{inseparable tidy pairs CTL} (Lemma \ref{inseparable tidy pairs})
Let $\mathcal{M}=(W,R,V)$ be a finitely labelled graph, $A,B \subseteq W$ be sets of nodes that form a tidy pair and let $\varphi$ be a CTL-formula without any temporal operators. Then $(\mathcal{M},A)\models \varphi$ if, and only if, $(\mathcal{M},B)\models \varphi$.
\end{lem}

\begin{lem}\label{from strategy to formula CTL} (Lemma \ref{from strategy to formula})
Let $\mathcal{M}=(W,R,V)$ be a serial and finitely labelled graph, $A,B \subseteq W$ two sets of nodes and $k\in \mathbb{N}$.
If $S$ has a winning strategy in the $\mathcal{M}$-ON game played from $A$ and $B$ with parameter $k$, then there is a CTL-formula $\varphi$ separating $(\mathcal{M},A)$ and $(\mathcal{M},B)$ with $\vert \varphi \vert_O \leq k$.
\end{lem}

\begin{proof}
This proof can be seen as an extension of the proof of Lemma~\ref{from strategy to formula}. So we will skip over some explanations that were already given there. Let $f_S$ be the winning strategy of player $S$ for the position $(\{A,B\},k)$. Then by induction over $k$ we construct $\varphi$ as a separating formula for $(\mathcal{M},A)$ and $(\mathcal{M},B)$. For the base case we look at positions with $k=0$. In this case $\varphi = \top$ is a separating formula. As our induction hypothesis we assume that this lemma holds for all positions $(\{A',B'\},k')$ with $k' < k$. 

For the induction step we look at positions $(\{A,B\},k)$ and we make a case-by-case analysis based on the first move of $f_S$. For split- and operator-moves we can simply reuse our proof from Lemma \ref{from strategy to formula}. So if that first move is a split-move, then there are some CTL-formulas $\varphi_1$ and $\varphi_2$, so that $\varphi=\varphi_1 \lor \varphi_2$ is a separating formula with $\vert \varphi \vert_O\leq k$. On the other hand if $f_S$ prescribes an operator-move on $A$, then there exists some CTL-formula $\varphi'$, so that every node in $A$ has a successor at which $\varphi'$ holds, while no node in $B$ has such a successor. Then $\varphi = EX \varphi'$ is a separating formula with $\vert \varphi' \vert_O \leq k$.

If $f_S$ dictates an EU-move, we may assume that $S$ plays on $A$. Then for every node $a \in A$ he picks some path $(a_j^n)_{0 \leq j \leq \ell_n}$ with $a_0^n=a$ and collects all final nodes of these paths into $A_2$ and all non-final nodes into $A_1$. Then player $S$ chooses the parameters $k_1$ and $k_2$ with $k=k_1+k_2+1$ according to his strategy. Since $f_S$ is a winning strategy it has to win against the strategy of player $D$, where she picks every possible path that begins in $B$. Then for every one of these paths $f_S$ prescribes a node for $S$ to put into either $B_1$ or $B_2$. Next $D$ has to pick $i\in \{1,2\}$. Depending on her choice the game continues from either $(t(\{A_1,B_1\}),k_1)$ or $(t(\{A_2,B_2\}),k_2)$. For both of these positions $f_S$ must still be a winning strategy. So by our induction hypothesis and Lemma~\ref{tidy separating formulas CTL} there are CTL formulas $\varphi_1$ and $\varphi_2$, where $\varphi_i$ is a separating formula for $(\mathcal{M},A_i)$ and $(\mathcal{M},B_i)$ with $\vert \varphi_i\vert_O \leq k_i$. W.l.o.g. we may assume $(\mathcal{M},A_i)\models \varphi_i$. So for every node $a$ from $A$ there is a finite path $(a_j^n)_{0 \leq j \leq \ell_n}$, where all $a_j^n$ with $j < \ell_n$ belong to $A_1$, so $(\mathcal{M},a_j^n)\models \varphi_1$. Also for every one of these paths the last node $a_{\ell_n}^n$ belongs to $A_2$, so $(\mathcal{M},a_{\ell_n}^n)\models \varphi_2$. This implies $(\mathcal{M},A)\models \varphi_1 EU \varphi_2$. 

Now let $(b_j^m)_{0 \leq j \leq \ell_m}$ be some arbitrary finite path with $b_0^m \in B$. This path was chosen by $D$ during her answer. So there has to be a node $b_j^m$ that $S$ picked and added to $B_1$ or $B_2$. For $b_j^m \in B_1$ we know $j < \ell_m$ and $(\mathcal{M},b_j^m)\models \lnot \varphi_1$. For $b_j^m \in B_2$ on the other hand we know $j=\ell_m$ and $(\mathcal{M},b_j^m)\models \lnot \varphi_2$. In either case this path cannot be a witness for $\varphi_1 EU \varphi_2$. Since $(b_j^m)_{0 \leq j \leq \ell_m}$ was arbitrary, this implies $(\mathcal{M},B)\models \lnot (\varphi_1 EU \varphi_2)$. Then $\varphi:= \varphi_1 EU \varphi_2$ is a separating formula for $(\mathcal{M},A)$ and $(\mathcal{M},B)$ with $\vert \varphi \vert_O = \vert \varphi_1 \vert_O + \vert \varphi_2 \vert_O +1 \leq k_1 + k_2 +1 =k$.

Finally we look at $f_S$ prescribing an $AU$-move. We assume that $S$ plays on $A$. Then for every $a \in A$ he picks either a finite path $(a_j^n)_{0\leq j \leq \ell_n}$ or an infinite path $(a_j^n)_{0 \leq j}$. Then he takes all nodes from these paths and puts them into $A_1$. $S$ also gathers the final nodes of all finite paths and collects them into $A_2$. Next he picks two new parameters $k_1$ and $k_2$ with $k=k_1+k_2+1$. Since $f_S$ is a winning strategy it has to win against every possible move of $D$. In particular $S$ has to win if $D$ answers by picking every possible finite or infinite path that starts in $B$. For every one of these paths $f_S$ prescribes some node for $S$ to add to one of his sets. If that node is the last node of its finite path, $f_S$ also specifies the set, $B_1$ or $B_2$, that that node is added to. Otherwise the node is simply added to $B_1$. As the last part of the move $D$ picks $i \in \{1,2\}$ and the next position of the game is either $(t(\{A_1,B_1\}),k_1)$ or $(t(\{A_2,B_2\}),k_2)$. Both of these have to be winning positions, so by our induction hypothesis and Lemma~\ref{tidy separating formulas CTL} there are formulas $\varphi_1$ and $\varphi_2$ with $\vert \varphi_1 \vert_O \leq k_1$ and $\vert \varphi_2 \vert_O \leq k_2$ so that $\varphi_1$ separates $(\mathcal{M},A_1)$ from $(\mathcal{M},B_1)$, while $\varphi_2$ separates $(\mathcal{M},A_2)$ from $(\mathcal{M},B_2)$. We may assume w.l.o.g. $(\mathcal{M},A_i)\models \varphi_i$ for all $i \in \{1,2\}$. 

Let $a\in A$ be some arbitrary node from $A$. If $S$ picked a finite path $(a_j^n)_{0 \leq j \leq \ell_n}$ with $a_0^n=a$, then all nodes of this path belong to $A_1$ and $\varphi_1$ has to hold at every node. Additionally the last node of the path, $a_{\ell_n}^n$, belongs to $A_2$ as well. This implies $(\mathcal{M},a_{\ell_n}^n)\models \varphi_1 \land \varphi_2$. So there is a finite path starting from $a$ where $\varphi_1$ holds at every node and $\varphi_1 \land \varphi_2$ holds at the last node, meaning $(\mathcal{M},a)\models \varphi_1 EU (\varphi_1 \land \varphi_2)$. On the other hand if $S$ picked an infinite path $(a_j^n)_{0 \leq j}$ for $a$, then every node of that path is a member of $A_1$ and thus $\varphi_1$ has to hold. This implies $(\mathcal{M},a)\models EG \varphi_1$. Now we know $(\mathcal{M},A)\models \bigl(\varphi_1 EU (\varphi_1 \land \varphi_2)\bigr) \lor \bigl( EG(\varphi_1)\bigr)$. This formula is equivalent to $\lnot\bigl((\lnot \varphi_2)AU (\lnot \varphi_1)\bigr)$.

Next we examine some arbitrary finite path $(b_j^m)_{0 \leq j \leq \ell_m}$ with $b_0^m \in B$. Player $S$ chose a node $b_j^m$ from this path and for $j<\ell_m$ he added it to $B_1$. In this case we know $(\mathcal{M},b_j^m)\models \lnot \varphi_1$. For $j=\ell_m$ player $S$ put the node into either $B_1$ or $B_2$. Then we have $(\mathcal{M},b_j^m)\models \lnot \varphi_1$ or $(\mathcal{M},b_j^m)\models \lnot \varphi_2$. In either case this path cannot be a witness for $\varphi_1 EU (\varphi_1 \land \varphi_2)$. Since $(b_j^m)_{0 \leq j \leq \ell_m}$ was some arbitrary path , we know $(\mathcal{M},B)\models \lnot \bigl( \varphi_1 EU (\varphi_1 \land \varphi_2) \bigr)$. We can also look at some arbitrary infinite path $(b_j^m)_{0\leq j}$ with $b_0^m \in B$. On this path there has to be a node $b_j^m$ with $b_j^m \in B_1$ and thus $(\mathcal{M},b_j^m)\models \lnot \varphi_1$. So this path cannot be a witness of $EG \varphi_1$, which implies $(\mathcal{M},B)\models \lnot EG \varphi_1$. With this we have shown $(\mathcal{M},B)\models \lnot \bigl(\varphi_1 EU (\varphi_1 \land \varphi_2)\bigr) \lor \bigl( EG(\varphi_1)\bigr)$. So $\varphi=\bigl((\lnot \varphi_2) AU (\lnot \varphi_1)\bigr)$ is a separating formula for $(\mathcal{M},A)$ and $(\mathcal{M},B)$ with $\vert \varphi \vert_O = \vert \varphi_1 \vert_O + \vert \varphi_2 \vert_O +1\leq k_1+k_2+1=k$.

This finishes our proof of the first implication.\end{proof}

\begin{lem}\label{from formula to strategy CTL} (Lemma \ref{from formula to strategy})
Let $\mathcal{M}=(W,R,V)$ be a serial and finitely labelled graph, $\varphi$ a CTL-formula, $A,B \subseteq W$ two sets of nodes and $k\in \mathbb{N}$, so that $\varphi$ is a separating formula for $(\mathcal{M},A)$ and $(\mathcal{M},B)$ with $\vert \varphi \vert_O\leq k$. Then $S$ has a winning strategy in the $\mathcal{M}$-ON game played from $A$ and $B$ with parameter $k$.
\end{lem}

\begin{proof}
This proof expands on the proof of Lemma~\ref{from formula to strategy}. So once again we skip some steps that were already explained during that proof.
To begin with we may assume that the separating formula $\varphi$ only contains the connectives $\lor$ and $\lnot$, as well as the operators $EX$, $EU$ and $AU$, since otherwise we can use the following well known equivalences in order to remove all occurrences of the other connectives and operators:

\begin{tabular}{r c l}
$\varphi \land \psi$ & $\equiv$ & $\lnot(\lnot \varphi \lor \lnot \psi)$\\
$EG \varphi$ & $\equiv$ & $\lnot \bigl(\top AU (\lnot \varphi)\bigr)$\\
$EF \varphi$ & $\equiv$ & $\top EU \varphi$\\
$AX \varphi$ & $\equiv$ & $\lnot EX \lnot \varphi$\\
$AG \varphi$ & $\equiv$ & $\lnot \bigl(\top EU (\lnot \varphi)\bigr)$\\
$AF \varphi$ & $\equiv$ & $\top AU \varphi$
\end{tabular} 

None of these changes increase the number of temporal operators, so $\vert\varphi\vert_O\leq k$ still holds.
We may assume that $\{A,B\}$ is already tidy. So the starting position of the $\mathcal{M}$-ON game is $(\{A,B\},k)$. Now we use the separating formula $\varphi$ in order to construct a winning strategy $f_S$ for $S$ via induction over the formula size of $\varphi$. For the base case we look at all atomic formulas. In this case the current position has to be an immediate winning position for player $S$ already, so $S$ has a winning strategy. As our induction hypothesis we assume that the lemma holds for all formulas $\psi$ with $\vert \psi \vert \leq i$.

For our induction step let $\varphi$ be a separating CTL-formula for $(\mathcal{M},A)$ and $(\mathcal{M},B)$ with $\vert \varphi \vert =i+1$ and $\vert \varphi \vert_O \leq k$. Now we have to differentiate based on the outermost operator of $\varphi$.

If the outermost operator is $\lnot$, $\lor$ or an $EX$-operator, which is the same as an $\Diamond$-operator from modal logic, the we can simply reuse our proof from Lemma \ref{from formula to strategy}.
So for $\varphi=\lnot\psi$, $\psi$ is also a separating formula for our sets, so by our induction hypothesis $S$ already has a winning strategy. For $\varphi=\psi \lor \chi$ if one of these sub-formulas contains no operators, the other one can be used as a shorter separating formula. Otherwise player $S$ can win by playing a split-move. 
Lastly for $\varphi=EX \psi$ player $S$ can win by playing an operator-move.

The next case is $\varphi=\psi EU \chi$ with $\vert \psi \vert + \vert \chi \vert + 1 = i+1$ and $\vert \psi \vert_O + \vert \chi \vert_O + 1 \leq k$. This implies $k\geq 1$. We assume $(\mathcal{M},A)\models \psi EU \chi$ and $(\mathcal{M},B)\models \lnot ( \psi EU \chi)$. Then for every node $a\in A$ there is a finite path $(a_j^n)_{0 \leq j \leq \ell_n}$ with $a_0^n=a$, $(\mathcal{M},a_j^n)\models \psi$ for all $j<\ell_n$ and $(\mathcal{M},a_{\ell_n}^n)\models\chi$. On the other hand no such path starting in $B$ can exist. In this position $f_S$ prescribes an EU-move on $A$. Then for every node $a \in A$ player $S$ picks a finite path that is a witness for $(\mathcal{M},a)\models \psi EU \chi$. For every such path we know that every node except for the last one has to model $\psi$. With this we know $(\mathcal{M},A_1)\models \psi$. Also the last node of each path has to model $\chi$. So we know $(\mathcal{M},A_2)\models \chi$ as well. Next $S$ has to split the parameter $k$. He does so by defining $k_1:=\vert\psi\vert_O$ and $k_2=k-k_1-1\geq \vert \chi \vert_O$. Player $D$ answers by picking some finite paths $(b_j^m)_{0\leq j \leq \ell_m}$. Because of $(\mathcal{M},B)\models \lnot( \psi EU \chi)$ we know that none of these paths can be a witness for $\psi EU \chi$. So for every path there has to be a node $b_j^m$ with $j < \ell_m$ and $(\mathcal{M},b_j^m)\models \lnot\psi$, or with $j = \ell_m$ and $(\mathcal{M},b_j^m)\models \lnot\chi$. For each path player $S$ selects such a node. Then for $j < \ell_m$ he puts them into $B_1$. Otherwise he puts them into $B_2$. This ensures $(\mathcal{M},B_1)\models \lnot \psi$ and $(\mathcal{M},B_2)\models \lnot \chi$. So $\psi$ is a separating formula for $(\mathcal{M},A_1)$ and $(\mathcal{M},B_1)$ and $\chi$ is a separating formula for $(\mathcal{M},A_2)$ and $(\mathcal{M},B_2)$. As the last step of the move player $D$ has to decide on whether the game continues from $(t(\{A_1,B_1\}),k_1)$ or from $(t(\{A_2,B_2\}),k_2)$. For the first of these positions $\psi$ is a separating formula with $\vert \psi \vert_O \leq k_1$ and for the second $\chi$ is a separating formula with $\vert \chi \vert_O \leq k_2$. So by our induction hypothesis $S$ has winning strategies for these two positions. Then the strategy $f_S$ as described here is a winning strategy for player $S$ from $(\{A,B\},k)$.

Finally we need to look at $\varphi = \psi AU \chi$ with $\vert \psi \vert + \vert \chi \vert +1 = i+1$ and $\vert \psi \vert_O + \vert \chi \vert_O + 1 = k$. Once again we may assume $(\mathcal{M},A)\models \psi AU \chi$. With this we know $(\mathcal{M},B) \models \lnot (\psi AU \chi)$, which is equivalent to $(\mathcal{M},B)\models \bigl((\lnot \chi)EU(\lnot \psi \land \lnot \chi)\bigr) \lor EG (\lnot \psi)$. So for every node $b \in B$ we know  $(\mathcal{M},b)\models (\lnot \chi) EU (\lnot \psi \land \lnot \chi)$ or $(\mathcal{M},b)\models EG (\lnot \psi)$. For nodes with $(\mathcal{M},b)\models (\lnot \chi) EU (\lnot \psi \land \lnot \chi)$ there has to be a finite path $(b_j^m)_{0 \leq j \leq \ell_m}$ with $b_0^m=b$, $(\mathcal{M},b_j^m)\models \lnot \chi$ for all $j < \ell_m$ and $(\mathcal{M},b_{\ell_m}^m)\models \lnot \psi \land \lnot \chi$. This implies that $\lnot\chi$ has to hold at every node of such a path. At all other nodes $b$ with $b \in B$ we know that $(\mathcal{M},b)\models EG (\lnot \chi)$ has to hold. For these nodes there has to be an infinite path $(b_j^m)_{0 \leq j}$ with $b_0^m=b$ and $(\mathcal{M},b_j^m)\models \lnot \chi$ for all $j \geq 0$. In this position the strategy $f_S$ prescribes an AU-move on $B$. During this move player $S$ picks for every node $b\in B$ the previously mentioned finite or infinite path. Then all nodes of these paths are collected into $B_1$. Since all nodes picked by $S$ model $\lnot \chi$, we know $(\mathcal{M},B_1)\models \lnot\chi$. Additionally all final nodes of the finite paths are gathered into $B_2$. These final nodes have to model $\lnot \psi \land \lnot \chi$, so in particular they have to model $\lnot \psi$. From this we know $(\mathcal{M},B_2)\models \lnot\psi$. Player $S$ also splits the parameter $k$ by picking $k_1=\vert \chi \vert_O$ and $k_2=k - k_1 -1$, which guarantees $k_2 \geq \vert \psi \vert_O$. Then player $D$ answers by selecting some finite paths $(a_j^n)_{0 \leq j \leq \ell_n}$ as well as some infinite paths $(a_j^n)_{0 \leq j}$ with $a_j^n\in A$. Because of $(\mathcal{M},A)\models \psi AU \chi$ we know that for every infinite path $(a_j^n)_{0 \leq j}$ there has to be a node $(a_j^n)$ with $(\mathcal{M},a_j^n)\models \chi$. So for every infinite path $S$ can pick such a node and put it into $A_1$. For finite paths $(a_j^n)_{0\leq j \leq \ell_n}$ there are two possibilities. The first one is that the path already contains a node at which $\chi$ holds. In this case $S$ adds that node to $A_1$. The second possibility is that $\chi$ holds at no node of that path. In this case we know that at every node of the path $\psi$ has to hold. This includes $a_{\ell_n}$, the last node of the finite path chosen by $D$. Then $S$ picks said last node and puts it into $A_2$. By playing in this way $S$ can guarantee $(\mathcal{M},A_1)\models \psi$ and $(\mathcal{M},A_2)\models \chi$. Lastly player $D$ decides if the game continues form $(t(\{A_1,B_1\}),k_1)$ or $(t(\{A_2,B_2\}),k_2)$. The formula $\psi$ separates $A_1$ and $B_1$ with $\vert \psi \vert_O \leq k_1$, while $\chi$ is a separating formula for $A_2$ and $B_2$ with $\vert \chi \vert_O \leq k_2$. By our induction hypothesis these positions are winning positions for $S$ and so $(\{A,B\},k)$ is as well.
\end{proof}

Lemma~\ref{from strategy to formula CTL} and Lemma~\ref{from formula to strategy CTL} can be combined into the following Correctness- and Completeness-Theorem.

\begin{thm}\label{correctness theorem CTL} (Theorem \ref{correctness theorem})
Let $\mathcal{M}=(W,R,V)$ be a finitely labelled serial graph, $A,B \subseteq W$ two sets of nodes and $k\in \mathbb{N}$.
Then $S$ has a winning strategy in the $\mathcal{M}$-ON game played from $A$ and $B$ with parameter $k$ if, and only if, there is a CTL-formula $\varphi$ with $\vert \varphi \vert_O \leq k$, that is a separating formula for $(\mathcal{M},A)$ and $(\mathcal{M},B)$.
\end{thm}

\begin{cor}\label{subset-positions}
Let $\mathcal{M}=(W,R,V)$ be a finitely labelled graph, $A,A',B,B'\subseteq W$ sets of nodes with $A'\subseteq A$ and $B' \subseteq B$ and let $k,k' \in \mathbb{N}$ be two parameters with $k\leq k'$, so that $S$ has a winning strategy for the $\mathcal{M}$-ON game for CTL from $(t(\{A,B\}),k)$. Then he also has a winning strategy for the $\mathcal{M}$-ON game for CTL from $(t(\{A',B'\}),k')$.
\end{cor}

\begin{proof}
An equivalent for this corollary for modal logic holds as well. However we did not state that version in Section~\ref{section game for ML}, because we did not need it for our proofs in Section~\ref{ML}.

According to Theorem~\ref{correctness theorem CTL} if $S$ has a winning strategy for the $\mathcal{M}$-ON game from $(t(\{A,B\}),k)$ there has to be some CTL-formula $\varphi$ that is a separating formula for $(\mathcal{M},A)$ and $(\mathcal{M},B)$ with $\vert \varphi \vert_O \leq k \leq k'$. W.l.o.g. we may assume $(\mathcal{M},A)\models \varphi$ and $(\mathcal{M},B)\models \lnot \varphi$. This implies that $(\mathcal{M},A')\models \varphi$ and $(\mathcal{M},B')\models \lnot\varphi$ hold as well. So $\varphi$ is also a separating formula for $A'$ and $B'$. Then because of Theorem~\ref{correctness theorem CTL} player $S$ must have a winning strategy for the $\mathcal{M}$-ON game from $(t(\{A',B'\}),k')$.
\end{proof}

\begin{cor}\label{loosing condition for S CTL} (Corollary \ref{loosing condition for S})
Let $\mathcal{M}=(W,R,V)$ be a finitely labelled graph, $A,B\subseteq W$ be two finite sets of nodes with $A \cap B \neq \emptyset$. Then $D$ has a winning strategy for the $\mathcal{M}$-ON game for CTL played on $A$ and $B$ with any parameter $k \in \mathbb{N}$.
\end{cor}

\section{Succinctness of CTL}
Our goal for this section is to show that there is an exponential succinctness gap between CTL and CTL$[\leftrightarrow]$. In order to do so we proceed in a fashion similar to that of our proofs for ML. First we construct a graph $\mathcal{M}_c$ that is based on the graph $\mathcal{M}$ from Section~\ref{ML}, as well as two infinite sequences of nodes $a_0,a_1,...$ and $b_0,b_1,...$. Then we need to show that for every $n \in \mathbb{N}$ there is a separating CTL$[\leftrightarrow]$-formula for $(\mathcal{M}_c,a_n)$ and $(\mathcal{M}_c,b_n)$ of size polynomial in $n$. However we can show something even stronger: There is even a separating ML$[\leftrightarrow]$-formula of polynomial size.

The graph $\mathcal{M}_c=(W_c,R_c,V_c)$ is defined as an extension of $\mathcal{M}=(W,R,V)$. One new node, $c$, is added to $W$, meaning $W_c=W\cup\{c\}$. At this node no propositional formula holds, so $V_c(c)=\emptyset$. For all other nodes the labelling function of $\mathcal{M}_c$ is the same as that of $\mathcal{M}$, so $V_c(w)=V(w)$ for all $w\in W$. Then for every pair of nodes $w,w'\in W$, where $w$ is at least two layers above $w'$, we add an edge from $w$ to $w'$. Lastly for every node $w \in W \cup \{c\}$ we add a new edge from $w$ to $c$. This implicitly includes the self-loop at $c$. Because of these new edges $\mathcal{M}_c$ has a serial accessibility relation. Also because $\mathcal{M}$ is finitely labelled, so is $\mathcal{M}_c$. So we can use the $\mathcal{M}_c$-ON game.

Now we want to show that we can use the bi-implication in order to separate our structures with small formulas.

\begin{lem}\label{bi-implication CTL}
There are two polynomial functions $g_1,g_2\colon \mathbb{N}\rightarrow \mathbb{N}$, so that for every $i\in \mathbb{N}$ there is an ML$[\leftrightarrow]$-formula $\psi_i$, which separates $(\mathcal{M}_c,a_i)$ from $(\mathcal{M}_c,b_i)$, with $\vert \psi_i \vert_O=g_1(i)\geq i$ and $\vert \psi_i \vert =g_2(i)\geq i$.
\end{lem}

\begin{proof}
During the construction of $\mathcal{M}_c$ we added edges between all nodes that are two or more layers apart. All of these edges were also added when constructing $\mathcal{M}_{t,r}$. So we can use the same separating formula $\psi_i$ from the proof of Lemma~\ref{biimplication T,R} defined by $\psi_i:=p_0$ for $i=0$ and $\psi_i:=\Diamond \chi_{3i-1}\land \Box (\lnot \chi_{3i-2} \lor ((\Diamond(p_1\land \chi _{3(i-1)}) \leftrightarrow (\Diamond(\psi_{i-1}\land \chi_{3(i-1)})))))$ for all $i > 0$. The graph $\mathcal{M}_c$ contains the additional node $c$. However this node does not have any labels. So $c$ cannot be part of a path on which the labels $q_0$, $q_1$ and $q_2$ appear in an alternating pattern. Because of this $c$ does not change the evaluation of $\chi_j$ at any node for any $j$. Then $\psi_i$ is still a separating formula for $(\mathcal{M}_c,a_i)$ and $(\mathcal{M}_c,b_i)$ with $\vert \psi_i \vert_O= 24 \cdot i^2 + 10 \cdot i=:g_1(i)$ and $\vert \psi_i\vert=36 \cdot i^2 + 17 \cdot i + 1 =:g_2(i)$.
\end{proof}

Our next goal is to show that the structures $(\mathcal{M}_c,a_i)$ and $(\mathcal{M}_c,b_i)$ cannot be separated by small CTL-formulas without bi-implications. In order to prove this fact, we will use the $\mathcal{M}_c$-ON game and show that player $D$ has a winning strategy. In preparation we need to establish some techniques that we will use during that proof. 
In particular we need a way of dealing with the new EU- and AU-moves. The main problem with these types of moves is the sheer number of options that player $S$ has. For every node in $X$ he can pick any of the paths starting at that node and during the later part of the move, he also gets the choice of which nodes to add to $Y_1$ or $Y_2$. It is not feasible for us to go through all possible choices. So instead we have to reduce the number of moves we need to look at. For this purpose we introduce the following lemma.

\begin{lem}\label{3 properties}
Let $A,B \subseteq W_c$ be some sets of nodes and $k$ some natural number. If $(\{A,B\},k)$ is a winning position for player $S$, then $S$ has a winning strategy with the following three properties:
\begin{enumerate}[P1]
\item During an EU- or AU-move, $S$ only picks paths with two nodes or less.
\item $S$ does not play an EU- or AU-move in a position $(\{A',B'\},k')$ with $\vert A'\vert = \vert B' \vert=1$.
\item During an EU- or AU-move in a position $(\{A',B'\},k')$, $S$ never picks a path $(a,x)$ with $a \in A'$ and $x \in \Box b$ for every $b \in B'$.
\end{enumerate}
\end{lem}

\begin{proof}
This proof is done via induction over the parameter $k$. For our base case we look at positions with $k=0$. If $(\{A,B\},0)$ is a winning position, we know $A=B=\emptyset$. Then the winning strategy $f_S$ has to be the strategy with no moves. This strategy satisfies all three properties for trivial reasons. Now as our induction hypothesis we assume that for all winning positions of $S$ with parameter $k' < k$ there exists a winning strategy for player $S$ with all three properties. For our induction step we look at a position $(\{A,B\},k)$ that is a winning position for $S$. Let $f_S$ be some winning strategy. Now we look at the first move of this strategy. 

If this first move does not conflict with any of our three properties, $S$ can play this first move according to $f_S$. For every possible follow-up position $(\{A',B'\},k')$ after this move we know $k' < k$. Also because $f_S$ is a winning strategy, we know that each of these new positions has to be a winning position for player $S$. So according to our induction hypothesis for every such position, player $S$ has a winning strategy with $P1$, $P2$ and $P3$. Now we can construct a winning strategy for $S$ from the position $(\{A,B\},k)$ that satisfies all three properties. In this strategy $S$ copies the first move of $f_S$ and then for each possible follow-up position he copies the strategy given by our induction hypothesis.

Next we assume that the first move of $f_S$ conflicts with atleast one property. In this case we have to modify $f_S$ into a new strategy $f_S'$ that fits all three properties. For this it suffices to replace the first move of the strategy with a new move that does not conflict with any of these properties and only leads to winning positions for $S$. Then it follows from our induction hypothesis that for every one of these follow-up positions, $S$ has a winning strategy that satisfies $P1$, $P2$ and $P3$. All of these strategies can be combined into the wanted winning strategy for $(\{A,B\},k)$. In this way we can reduce a problem of strategies to a problem for moves. 

Now for every property $P \in \{P1,P2,P3\}$ we give a transformation that takes a move that conflicts with $P$, but agrees with all previous properties, and modifies it into a move that agrees with all properties up to $P$. Then by using these three transformations where necessary and consecutively we can transform any possible move into a move that satisfies all properties.

\textit{Transformation for P1:} If the first move of $f_S$ conflicts with $P1$, then that move has to be either an EU- or an AU-move. First we look at $S$ playing an EU-move from $(\{A,B\},k)$ on a set $X \in \{A,B\}$. Then for every $x \in X$ player $S$ picks a path $(x_j^n)_{0 \leq j \leq \ell_n}$ with $x=x_0^n$. At least one of these paths contains more than two nodes. This creates the sets $X_1=\bigcup\limits_{x_0^n \in X} \{x_j^n \mid 0 \leq j < \ell_n\}$ and $X_2= \bigcup\limits_{x_0^n \in X} \{x_{\ell_n}^n\}$. Then $S$ chooses new parameter $k_1$ and $k_2$ with $k_1+k_2+1=k$. Since $f_S$ is a winning strategy, player $S$ has to win if $D$ picks every possible finite path $(y_j^m)_{0 \leq j \leq \ell_m}$. For each of these paths $f_S$ prescribe a node $y_j^m$ for $S$ to add to $Y_1$ or $Y_2$. Lastly $D$ picks $i\in \{1,2\}$. Independent of her choice the resulting position $(t(\{X_i,Y_i\}),k_i)$ has to be a winning position for $S$.

The graph $\mathcal{M}_c$ is transitive. So for every path $(x_j^n)_{0 \leq j \leq \ell_n}$ with $\ell_n > 0$ there is an edge going from $x_0^n$ to $x_{\ell_n}^n$. Let $f_S'$ be the strategy, where $S$ picks for every $x\in X$ the shortened path $(x_0^n,x_{\ell_n}^n)$ with only two nodes. For $\ell_n=0$ he picks the one-node-path $(x_0^n)$ instead. This strategy leads to the sets $X_1'=\bigcup\limits_{x_0^n \in X}=\{x_0^n \mid \ell_n > 0\}\subseteq X_1$ and $X_2'= \bigcup\limits_{x_0^n \in X} \{x_{\ell_n}^n\} = X_2$. Then $S$ picks the same parameters $k_1$ and $k_2$ he choose when playing according to $f_S$. Player $D$ answers by picking some paths $(y_j^m)_{0 \leq j \leq \ell_m}$. The strategy $f_S'$ once again copies $f_S$ when marking a node from each of these paths and adding them to $Y_1'$ or $Y_2'$. This way of playing ensures $Y_1' \subseteq Y_1$ and $Y_2' \subseteq Y_2$. Then $D$ picks $i \in \{1,2\}$ and the game continues with $(t(\{X_i',Y_i'\}),k_i)$. We know $X_i'\subseteq X_i$ and $Y_i'\subseteq Y_i$. So according to Corollary~\ref{subset-positions}, because $(t(\{X_i,Y_i\}),k_i)$ is a winning position for player $S$, $(t(\{X_i',Y_i'\}),k_i)$ is as well. Also because of our induction hypothesis, player $S$ has a winning strategy for $(t(\{X_i',Y_i'\}),k_i)$ that agrees with $P1$, $P2$ and $P3$. Then $f_S'$ is a winning strategy for player $S$ from $(\{A,B\},k)$, where $P1$ holds for the first move.

Our line of reasoning for the AU-move is very similar, as long as $S$ only picks finite paths. For every such path, he can pick the shortened path with only the first and last node instead. In this case it does not matter that $D$ may also now choose infinite paths, since we only need to modify the strategy of $S$.

However $S$ may also pick infinite paths $(x_j^n)_{0\leq j}$.
In the graph $\mathcal{M}_c$ almost all edges go down one or more layers. The only exception is the self-loop at the node $c$. There is only a finite number of layers below any given node, so every path can only contain a finite number of these edges going down. Then every infinite path in $\mathcal{M}_c$ has to reach $c$ after some finite number of steps and only use the self-loop at that node afterwards. So if $S$ picked any infinite path, we know $c \in X_1$. Since $f_S$ is a winning strategy, $S$ has to win, even if 
player $D$ picks for every node $y \in Y$ the infinite path that starts at $y$, moves to $c$ in one step and then keeps using the self-loop at $c$. This is always possible, because the node $c$ is a successor of every node in $\mathcal{M}_c$. For each of these paths player $S$ in accordance with $f_S$ has to mark a node and add it to $Y_1$. If he marks the node $c$ at all, that node will be a shared node of $X_1$ and $Y_1$. So by deciding on $i=1$, player $D$ can move the game to a position with a shared node, which according to Corollary~\ref{loosing condition for S CTL} is a winning position for her. However this contradicts our assumption that $f_S$ is a winning strategy. So we may assume that for every path $S$ marks the starting node $y$ instead. Then we know $Y \subseteq Y_1$. We also know that $X \subseteq X_1$ has to hold for every AU-move. According to Corollary~\ref{subset-positions}, because $(\{X_1,Y_1\},k_1)$ is a winning position for $S$, $(\{X,Y\},k_1)$, which is the same as $(\{A,B\},k_1)$, is as well. So $S$ has a winning strategy $f_S'$ for this position and with our induction hypothesis we may assume that $f_S'$ satisfies $P1$, $P2$ and $P3$. Now we want to use this strategy $f_S'$ as a winning strategy for the position $(\{A,B\},k)$. In order to do so we only need to tweak how $S$ picks parameters. This is possible without any problems, because $S$ can ensure that for every position he can reach from $(\{A,B\},k)$ the parameter is at least as high as in the corresponding position reachable from $(\{A,B\},k_1)$. Since none of the properties $P1$, $P2$ or $P3$ cares about the parameter, this new strategy still satisfies all three properties.

\textit{Transformation for P2:} Next we look at a strategy $f_S$, where the first move of that strategy agrees with $P1$, but conflicts with $P2$. So $f_S$ prescribes an EU- or AU-move at $(\{A,B\},k)$ with $\vert A \vert = \vert B \vert=1$. First we assume $S$ plays an EU-move. Then $S$ picks a set $X \in \{A,B\}$. Let $x_0, y_0 \in W_c$ be the nodes with $X=\{x_0\}$ and $Y=\{y_0\}$. For the node $x_0$ he picks a path $(x_j)_{0\leq j \leq \ell}$. Because $f_S$ satisfies $P1$, we know $\ell \leq 1$. $S$ also picks new parameters $k_1$ and $k_2$. In the case of $\ell=0$ we know $X_2=\{x_0\}=X$. The strategy $f_S$ has to win if $D$ answers with the path $(y_0)$, which guarantees $Y_2=\{y_0\}=Y$. So $(\{X_2,Y_2\},k_1)$, which is the same as $(\{A,B\},k_2)$, has to be a winning position for $S$. Then by our induction hypothesis $S$ has a winning strategy with $P1$ to $P3$. As previously discussed $S$ can also use that strategy as a winning strategy for $(\{A,B\},k)$ by only slightly changing his choices of parameters.

On the other hand for $\ell=1$ we know $X_1=\{x_0\}$ and $X_2=\{x_1\}$. Since $f_S$ is a winning strategy it has to win if $D$ picks for every node $y'\in \Box y_0$ the path starting at $y_0$, ending at $y'$ and containing no other nodes. For each of these paths player $S$ has to mark a node and add it to $Y_1$ or $Y_2$. If $S$ marks the starting node of any of these paths, we know $Y_1=\{y_0\}$. Then, similar to our previous case, $D$ can ensure $(\{A,B\},k_1)$ as the next position and once again $S$ can use his winning strategy for that position in order to win from $(\{A,B\},k)$. 

Lastly if $S$ never marks the starting node of a path, he has to mark the last node of every path. This leads to $Y_2=\Box y_0$. So $(t(\{\{x_1\},\Box y_0\}),k_2)$ with $k_2 \leq k-1$ must be a winning position for player $S$. In this case instead of playing an EU-move, $S$ can also win by playing an operator-move, where he picks $x_1$ as a successor of $x_0$. Then $D$ has to choose some $Y_1 \subseteq \Box y_0$. According to Corollary~\ref{subset-positions} the resulting position $(t(\{\{x_1\},Y_1\}),k-1)$ has to be a winning position for $S$. Also because the first move of this strategy is an operator-move, we get a winning strategy for player $S$, that has all three properties.

Now we examine $f_S$ prescribing an AU-move at $(\{A,B\},k)$ instead. Once again $S$ begins by deciding on $X \in \{A,B\}$. Let $x_0,y_0 \in W_c$ be the nodes with $X=\{x_0\}$ and $Y=\{y_0\}$. Then player $S$ picks a path with starting node $x_0$. Because of $P1$ we know that he picks a finite path with two nodes at most. If the path contains only one node, $x_0$, then $D$ can answer with the path that contains only $y_0$ and once again $(\{A,B\},k_1)$ with $k_1<k$ has to be a winning position for $S$. So we can use his winning strategy from that position for $(\{A,B\},k)$. 

Next we look at $S$ picking a finite path with exactly two nodes. Let $x_1$ be the second node of that path. Then we know $X_1=\{x_0,x_1\}$ and $X_2=\{x_1\}$. Player $S$ also picks new parameters $k_1,k_2$ with $k=k_1+k_2+1$. Then $f_S$ has to win against player $D$ picking for every node $y' \in \Box y_0$ the finite path going directly from $y_0$ to $y'$. From each of these paths $S$ has to take a node and add it to either $Y_1$ or $Y_2$. If $f_S$ makes him take the starting node $y_0$ for any of these paths, we know $Y \subseteq Y_1$, as well as $X \subseteq X_1$. So $(t(\{X_1,Y_1\},k_1)$ and thus $(\{A,B\},k_1)$ has to be a winning position for $S$. As previously argued, this implies that he has a winning strategy with all three properties for $(\{A,B\},k)$.

On the other hand if $S$ always picks the last node for each path, we know $Y_1 \cup Y_2 = \Box y_0$. Because $f_S$ is a winning strategy, both $(t(\{X_1,Y_1\}),k_1)$ and $(t(\{X_2,Y_2\}),k_2)$ have to be winning positions for $S$. 

If $Y_1$ is empty or contains only nodes $y'$ with $V_c(y')\neq V_c(x_1)$, then $Y_2$ has to contain all nodes from $\Box y_0$ with the same colour as $x_1$. In this case player $S$ has an alternative winning strategy. At the position $(\{A,B\},k)$ he can play an operator-move, where he picks $x_1$ as a successor of $x_0$. Then $D$ has to answer with a set $Y'\subseteq \Box y_0$ and the play continues with $(t(\{\{x_1\},Y'\}),k-1)$.
While it is possible that some nodes are in $Y'$, but not in $Y_2$, every such node has to be of a different colour than $x_1$. So all of them will be removed when tidying up. Because of this the position $(t(\{\{x_1\},Y'\}),k-1)$ has to consist of a subset of $X_2=\{x_1\}$ and a subset of $Y_2$. So as stated in Corollary~\ref{subset-positions} this new position is a winning position for $S$ and thus the strategy described here is a winning strategy for him. Also the first move of this strategy satisfies all of our properties.

If $Y_2$ is empty or contains only nodes with a colour different from that of $x_2$, we can argue in an analogous fashion that $S$ can also win by playing an operator move, where he picks $x_1$ as a successor of $x_0$. In this case the set $Y'$ after the tidy up has to be a subset of $Y_1$ instead and $\{x_1\} \subseteq X_1$ holds as well. So once again $(t(\{\{x_1\},Y'\}),k-1)$ has to be a winning position for $S$.

Lastly it is possible that both $Y_1$ and $Y_2$ contain at least one node with the same colour as $x_1$. Then for both $i \in \{1,2\}$ we know $k_i>0$, since otherwise $(t(\{X_i,Y_i\}),k_i)$ would be a winning position for $D$ instead. In this case there is another winning strategy for $S$: He can first play an operator-move, where he once again picks $x_1$ as a successor of $x_0$. Then $D$ answers by choosing some subset of $\Box y_0$. As long as $D$ picked at least one node with the same colour as $x_1$, the play moves to a position $(\{\{x_1\},Y'\},k-1)$ with $Y' \subseteq \Box y_0$. For $Y' \subseteq Y_1$ or $Y' \subseteq Y_2$ this position is already a winning position for $S$ according to Corollary~\ref{subset-positions}. Otherwise $S$ can continue by playing a split-move and splitting $Y'$ into $Y' \cap Y_1$ and $Y' \cap Y_2$. He also splits $k-1$ into $k_1$ and $k_2$. Next player $D$ has to choose some subset $X'$ of $\{x_1\}$, as well as a $i \in \{1,2\}$. If she picks $X'=\emptyset$, the resulting position is a trivial winning position for $S$. If she picks $X'=\{x_1\}$, the game moves to $(t(\{\{x_1\},Y'\cap Y_i\}),k_i)$. This position is a winning position for player $S$, because so is $(t(\{\{x_1\},Y_i\}),k_i)$.

None of the new first moves introduced during the transformation for $P2$ is an EU- or AU-move. Because of this these new moves trivially agree with property $P1$ as well.

\textit{Transformation for P3:} In this final section of our proof we look at moves that agree with $P1$ and $P2$, but conflict with $P3$.
So first let $f_S$ prescribe an EU-move at $(\{A,B\},k)$, where $S$ picks a path $(a,x)$ with $a \in A$ and $x \in \Box b$ for every $b \in B$. We may assume that he plays on $A$, since otherwise $a$ is  a shared node between $A$ and $B$. So according to Corollary~\ref{loosing condition for S CTL} the current position would be a winning position for $D$. Player $S$ also has to choose a path for every other node in $A$. Let $A_2 \subseteq A$ be the set of nodes for which he choose a one-node-path and let $A_1 = A \setminus A_2$ be the set of the remaining nodes, including $a$. Then we know $A_1 \subseteq X_1$ and $A_2 \subseteq X_2$.
Player $S$ also has to choose $k_1$ and $k_2$ with $k_1+k_2+1=k$. He has to win, even if $D$ picks for every $b \in B$ the one-node-path $(b)$, as well as the two-node-path $(b,x)$. Next from each of these paths player $S$ has to add a node to either $Y_1$ or $Y_2$. If he picks $x$ at all, he has to add it to $Y_2$. However we also know $x \in X_2$, so the node $x$ would be a shared node for the position $(t(\{X_2,Y_2\},k_2)$, making it a winning position for $D$. This contradicts our assumption that $f_S$ is a winning strategy. So instead $S$ has to mark the node $b$ from every path $(b,x)$. Then we know $B \subseteq Y_1$. Also for every one-node-path $(b)$ player $S$ has to add the node $b$ to $Y_2$. With this we know $B \subseteq Y_2$ as well. Lastly $D$ gets to pick either $(t(\{X_1,Y_1\}),k_1)$ or $(t(\{X_2,Y_2\},k_2)$ as the next position. Since $f_S$ is a winning strategy, both of these position have to be winning positions for $S$.

It is possible that $A_1$ is empty. Then we know $A_2=A$ and thus $A\subseteq X_2$. In this case Corollary~\ref{subset-positions} implies that $(\{A,B\},k_2)$ has to be a winning position for $S$. So once again we may assume that $S$ has a winning strategy for $(\{A,B\},k_2)$ that satisfies all three properties and  with only slight changes to the used parameters $S$ can use that strategy to win $(\{A,B\},k)$ as well.
On the other hand if $A_2$ is empty, we know $A \subseteq X_1$. Then we can argue in an analogous fashion that $(\{A,B\},k_1)$ has to be a winning position for $S$ and construct the new winning strategy from there.

Lastly if neither set is empty, $S$ can play a split-move, where he splits $A$ into $A_1$ and $A_2$, as well as $k$ into $k_1+1$ and $k_2$. Then player $D$ answers with some set $Y' \subseteq B \subseteq Y_1 \cap Y_2$ and picks either $(t(\{A_1,Y'\}),k_1+1)$ or $(t(\{A_2,Y'\}),k_2)$ as the new position. According to Corollary~\ref{subset-positions} the first of these two is a winning position for $S$, because $(t(\{X_1,Y_1\}),k_1)$  is one, and the second is a winning position, because so is $(t(\{X_2,Y_2\}),k_2)$. Then this new strategy is a winning strategy for $S$ and because the first move is a split-move we may assume that this strategy fits all three properties.

Next we look at an AU-move at $(\{A,B\},k)$, where $S$ picks a path $(a,x)$ with $a \in A$ and $x \in \Box b$ for every $b \in B$. Once again we can assume that $S$ plays on $A$. He also picks new parameters $k_1$ and $k_2$. The strategy $f_S$ has to win against $D$ picking the path $(b,x)$ for every node $b \in B$. For each of these paths $S$ is forced to add $b$ to $Y_1$ in order to avoid $x$ as a shared node. So we know $B \subseteq Y_1$. For an AU-move $A \subseteq X_1$ holds as well. Since $f_S$ is a winning strategy, $(t(\{X_1,Y_1\}),k_1)$ has to be a winning position for $S$. Then Corollary~\ref{subset-positions} shows that $(\{A,B\},k_1)$ is one of his winning positions as well. So $S$ can use his strategy from that game in order to win from $(\{A,B\},k)$ as well.

In total we have shown that we can replace every winning strategy for player $S$ with a winning strategy that satisfies the properties $P1$, $P2$ and $P3$.
\end{proof}

From the contraposition of this lemma we know that in order to show that $S$ does not have a winning strategy, it suffices to show that he does not have a winning strategy with the properties $P1$, $P2$ and $P3$. So when we try to show that player $D$ has a winning strategy we can assume that $S$ only picks paths of length two or less during an EU- or AU-move, he never plays such a move at a position with only singleton sets and he never picks a path that ends in a node that is a shared successor of all nodes in $Y$. This drastically reduces the number of $S$-moves we need to examine.

Next we have to deal with a troublesome tactic that player $S$ can use. At the position $(\{\{4_i,7_i\},\{5_i,6_i\}\},2^i-4)$ in the $\mathcal{M}_c$-ON game for modal logic, player $S$ has to play a split-move, since every operator-move leads to a winning position for $D$. However in the $\mathcal{M}_c$-ON game for computation tree logic, $S$ can avoid this split entirely. Instead he can play an EU-move on $\{4_i,7_i\}$, where he picks the paths $(4_i)$ and $(7_i,b_{i-1})$. He also picks $k_1=0$ and $k_1=2^i-5$. Player $D$ answers with the paths $(5_i)$ and $(6_i,a_{i-1})$. Then $S$ chooses the last node of each of these paths. This ensures $Y_1=\emptyset$, so $D$ is forced to move the play to $(\{\{4_i,b_{i-1}\},\{5_i,a_{i-1}\}\},2^i-5)$. In this way $S$ can simulate an operator-move on $7_i$, while 'freezing' the other node $4_i$. Afterwards $S$ can use further EU-moves in order to simulate his strategy for the pair $\{\{b_{i-1}\},\{a_{i-1}\}\}$ while continuously 'freezing' $4_i$ and $5_i$. 

While this strategy avoids the split, it doesn't allow $S$ to win this game. Basically playing like this is for $S$ like first having to win from $(\{\{6_i\},\{7_i\}\},k')$ for some $k' \in \mathbb{N}$ and then afterwards still having to win from $(\{\{4_i\},\{5_i\}\},2^i-4-k')$. If $S$ could win like this, he could also win by simply playing a split-move first. However $S$ playing like this is still inconvenient. Mainly because it introduces many new positions that can be reached and that we would need to check. However instead of going through all of these we instead use the following lemma that states that playing a split-move in such a position is optimal for $S$. This makes it way easier for us to argue that these positions are winning positions for $D$.

\begin{lem}\label{special positions}
Let $x_1,y_1\in W_c$ be some nodes that belong to the same layer of $\mathcal{M}_c$ and satisfy $V_c(x_1)=V_c(y_1)$. Also let $X_2,Y_2\subseteq W_c$ some sets of nodes with $X_2 \subseteq \Box y_1$ and $Y_2 \subseteq \Box x_1$ and let $k \in \mathbb{N}$ be a parameter, so that $(\{\{x_1\}\cup X_2,\{y_1\}\cup Y_2\},k)$ is a winning position for $S$ in the $\mathcal{M}_c$-ON game for CTL. Then there exist parameters $k_1$, $k_2$ with $k_1 + k_2 = k$, so that $(\{\{x_1\},\{y_1\}\},k_1)$ and $(t(\{X_2,Y_2\}),k_2)$ are winning positions for $S$ as well. 
\end{lem}

\begin{proof}
Before we get to the proper proof of this lemma, we want to make a short comment about the positions we look at. As stated in the lemma, the nodes $x_1$ and $y_1$ have to belong to the same layer. All edges in $\mathcal{M}_c$ go down at least one layer, with the exception of the single self-loop at the node $c$. So because the nodes in $Y_2$ are successors of $x_1$, all of them have to be on lower layers than $x_1$ and $y_1$. The same holds for all nodes in $X_2$, since they are all successors of $y_1$. So we are interested in positions where there are two special nodes, one on each side, that are above all other nodes, in such a way that each special node is the predecessor of all other nodes on the other side.

We validate this lemma via induction over the parameter $k$. For the base case we look at $k=0$. From the position $(\{x_1\}\cup X_2,\{y_1\}\cup Y_2\}),0)$ player $S$ cannot make a move. Then because of $\{x_1\}\cup X_2 \neq \emptyset$ player $S$ cannot win from this position. So there is no position with parameter $k=0$ that satisfies all conditions of the lemma. In this case the lemma is trivially true.

As our induction hypothesis we assume that the lemma holds for all parameters $k'$ with $k' < k$.
For the induction step we look at a position $(\{x_1\}\cup X_2,\{y_1\}\cup Y_2\}),k)$, which satisfies all conditions of the lemma. Since this position is a winning position for $S$, there has to be a strategy $f_S$, that is a winning strategy for player $S$.
We make a case-by-case analysis for the first move of that strategy. All conditions of the lemma are fully symmetrical. So w.l.o.g. we may assume that $S$ always plays on $\{x_1\}\cup X_2$. Because if he plays on $\{y_1\}\cup Y_2$, we can simply rename all of our sets and nodes so that he plays on $\{x_1\} \cup X_2$ instead.

Before we look at the first type of move, there is a special case we need to explain first. For most of the following explanations it is necessary that the set $S$ plays on contains at least one other element besides $x_1$. However there is the possibility of $X_2=\emptyset$. In this case we know $t(X_2,Y_2)=\{\emptyset,\emptyset\}$. Then $(t(\{X_2,Y_2\}),k_2)$ is a winning position for any $k_2 \in \mathbb{N}$, in particular for $k_2=0$. Then we get $k_1=k$ and according to Corollary~\ref{subset-positions} the position $(\{\{x_1\},\{y_1\}\},k_1)$ has to be a winning position for $S$, just like $(\{\{x_1\}\cup X_2,\{y_1\}\cup Y_2\},k)$. Therefore we may assume $X_2\neq \emptyset$ and $Y_2 \neq \emptyset$ for the remainder of this proof.

First we look at $f_S$ prescribing a split-move. Then he splits $\{x_1\}\cup X_2$ into two subsets. One of these two sets has to contain $x_1$. We refer to these new sets as $\{x_1\}\cup X_3$ and $X_4$. He also has to split the parameter $k$ into $k_3$ and $k_4$ with $0 < k_3,k_4 < k$. Player $S$ has to win, even if player $D$ picks $Y'=\{y_1\} \cup Y_2$. In this case the next position is either $(t(\{\{x_1\}\cup X_3,\{y_1\}\cup Y_2\}),k_3)$ or $(t(\{X_4,\{y_1\}\cup Y_2\}),k_4)$. Both of these positions have to be winning positions for player $S$. Because $\{\{x_1\}\cup X_2,\{y_1\}\cup Y_2\}$ is a tidy pair of sets, for every node in $X_2$ there has to be a node in $\{y_1\} \cup Y_2$ with the same labels. So during tidy-up no node from $X_3$ or $X_4$ is removed. Also because of $V_c(x_1)=V_c(y_1)$ neither $x_1$ nor $y_1$ can be removed from the first position. So there are some set $Y_3 \subseteq Y_2$ and $Y_4 \subseteq \{y_1\}\cup Y_2$ with $t(\{\{x_1\}\cup X_3,\{y_1\}\cup Y_2\})=\{\{x_1\}\cup X_3,\{y_1\}\cup Y_3\}$ and $t(\{X_4,\{y_1\}\cup Y_2\})=\{X_4, Y_4\}$. The position $(\{\{x_1\}\cup X_3,\{y_1\}\cup Y_3\},k_3)$ satisfies all conditions of the lemma and because of $k_3 < k$ we can use our induction hypothesis. So there are parameters $k_5,k_6$ with $k_5 + k_6 = k_3$, so that $(\{\{x_1\},\{y_1\}\},k_5)$ and $(t(\{X_3,Y_3\}),k_6)$ are winning positions for player $S$. So by fixing $k_1=k_5$, we have shown that $S$ can win from $(\{\{x_1\},\{y_1\}\},k_5)$.

Next we have to show that $(t(\{X_2,Y_2\},k_2)$ with $k_2=k-k_1=k_4+k_6$ is a winning position for $S$ as well. Let $X_5\subseteq X_2$ and $Y_5 \subseteq Y_2$ be the sets with $t(\{X_2,Y_2\})=\{X_5,Y_5\}$. It is possible that $X_5\subseteq X_3$. Then all nodes in $Y_5$ are nodes from $Y_2$ that have the same labels as some nodes from $X_3$. This implies $Y_5\subseteq Y_3$. During the tidy-up of $\{X_3,Y_3\}$ we remove all nodes from $X_3$ without a corresponding node in $Y_3$, as well as all nodes from $Y_3$ without a corresponding node in $X_3$. However since $\{X_5,Y_5\}$ is already tidy, none of these removed nodes can be a part of $X_5$ or $Y_5$. So these sets are even subsets of the tidied up versions of $X_3$ and $Y_3$. Then according to Corollary~\ref{subset-positions}, because $(t(\{X_3,Y_3\}),k_6)$ is a winning position for $S$, so is $(\{X_5,Y_5\},k_4+k_6)$. It is also possible that $X_5 \subseteq X_4$. In this case we can argue in a similar fashion that because $(\{X_4,Y_4\},k_4)$ is a winning position for $S$, $(\{X_5,Y_5\},k_4+k_6)$ is as well.

The last possibility is that both $X_5$ contains nodes from both $X_3$ and $X_4$.
In this case $S$ can play a split-move where he splits $X_5$ into $X_3 \cap X_5$ and $X_4 \cap X_5$ and he splits $k_1$ into $k_6$ and $k_4$. Then $D$ answers with some set $Y'\subseteq Y_2$ and decides on whether the game continues with $(t(\{X_3\cap X_5,Y'\}),k_6)$ or with $(t(\{X_4 \cap X_5,Y'\}),k_4)$. Let $X_6 \subseteq X_3 \cap X_5$, $X_7 \subseteq X_4 \cap X_5$ and $Y_6, Y_7 \subseteq Y'$ be the sets with $t(\{X_3 \cap X_5,Y'\})=\{X_6,Y_6\}$ and $t(\{X_4 \cap X_5,Y'\})=\{X_7,Y_7\}$. We know $X_6 \subseteq X_3$. The set $Y_3$ contains all nodes from $Y_2$ for which there is a node with the same labels in $X_3$. On the other hand $Y_6$ contains only nodes from $Y_2$ for which there is a node with the same labels in $X_3 \cap X_5$. So $Y_6 \subseteq Y_3$ has to hold. According to Corollary~\ref{subset-positions} because $(\{X_3,Y_3\},k_6)$ is a winning position for $S$, the position $(\{X_6,Y_6\},k_6)$ is a winning position for him as well. We also know $X_7\subseteq X_4$. Since $Y_4$ contains all nodes from $Y_2$ with a corresponding node in $X_4$ and $Y_7$ contains only nodes from $Y_2$ with a corresponding node in $X_4 \cap X_5$, we know $Y_7 \subseteq Y_4$ as well. So since $(\{X_4,Y_4\},k_4)$ is a winning position for $S$, $(\{X_7,Y_7\},k_4)$ is as well. With this we have shown that $S$ has a winning strategy for $(\{\{x_1\},\{y_1\}\},k_1)$ and $(t(\{X_2,Y_2\},k_2)$ with $k_1+k_2=k$.

Now we look at $f_S$ prescribing an operator-move. We may once again assume that $S$ plays on $\{x_1\} \cup X_2$. Let $x_2 \in X_2$ be some node. During his move player $S$ has to pick a node $x_2' \in \Box x_2$ and add it to $X'$. Now we know $(y_1,x_2)\in R_c$ and $(x_2,x_2')\in R_c$. Because the accessibility relation $R_c$ is transitive we also know $(y_1,x_2')\in R_c$. So player $D$ can pick $x_2'$ as a part of her answer, ensuring $x_2'\in X' \cap Y'$. This node $x_2'$ cannot be removed during tidy up. So the sets of the position after this move cannot be disjoint. Then by Corollary~\ref{loosing condition for S} that position is a winning position for player $D$. So player $S$ cannot win the game from the current position by playing an operator-move.

Next we examine what happens when $S$ plays an EU-move on $\{x_1\}\cup X_2$. Because of Lemma~\ref{3 properties} we can assume that this move satisfies all three properties stated in that lemma.
In particular we know that for every node in $\{x_1\}\cup X_2$ $S$ picks a path with one or two nodes. Then $S$ collects all final nodes of these paths into $X_4$ and all remaining nodes into $X_3$. Then $S$ also decides on new parameters $k_3$ and $k_4$ with $k=k_3+k_4+1$.

At this point we have to make an additional differentiation based on the path $S$ picked for $x_1$. First we assume that $S$ picked the one-node-path $(x_1)$. In this case we know $x_1 \in X_4$. Player $S$ has to win independent of the strategy played by $D$. So in particular he has to win if $D$ picks the one-node-path $(y_1)$ as well as every possible finite path that begins in $Y_2$. For every one of these paths $f_S$ has to prescribe a node for $S$ to add to either $Y_3$ or $Y_4$. For the path $(y_1)$ player $S$ has to pick the node $y_1$. Since that node is the last node of his path we know $y_1 \in Y_4$. Let $X_5 \subseteq X_4$ and $Y_5 \subseteq Y_4$ be the sets with $t(\{X_4,Y_4\})=\{\{x_1\}\cup X_5,\{y_1\}\cup Y_5\}$. Then player $D$ decides if the play continues with $(t(\{X_3,Y_3\}),k_3)$ or with $(\{\{x_1\}\cup X_5,\{y_1\}\cup Y_5\},k_4)$. So both of these positions are winning positions for player $S$. Every node in $X_5$ must be the last node of some one-node- or two-node-path starting in $X_2$. So these nodes belong to $X_2 \cup \Box X_2$ and since $R_c$ is transitive, they are all successors of $y_1$. By an analogous argument all nodes of $Y_5$ are successors of $x_1$ as well. Together with $k_4<k$ this implies that we can use our induction hypothesis. There have to be parameters $k_5,k_6$ with $k_5+k_6=k_4$ so that $(\{\{x_1\},\{y_1\}\},k_5)$ and $(t(\{X_5,Y_5\}),k_6)$ are winning positions for player $S$. Then for $k_1=k_5$ we know that $S$ wins from $(\{\{x_1\},\{y_1\}\},k_1)$.

Now we want to show that $(t(\{X_2,Y_2\}),k_2)$ with $k_2=k_3+k_6$ is a winning position for $S$ as well. For this let $X_6\subseteq X_2$ and $Y_6 \subseteq Y_2$ be the sets with $t(\{X_2,Y_2\})=\{X_6,Y_6\}$. In this position $S$ can play an EU-move, where for every node in $X_6$ he selects the same path he picked when playing on $(\{\{x_1\}\cup X_2,\{y_1\}\cup Y_2\},k)$. Let $X_7$ be the set of all non-final nodes of these paths and $X_8$ the set of the final nodes. Then this strategy ensures $X_7 \subseteq X_3$ and $X_8 \subseteq X_4$. The node $x_1$ cannot be a successor of a node from $X_2$, since it is on a higher layer than all of $X_2$. So this node cannot be a part of the paths chosen by $S$ and we know $x_1\notin X_8$. By a similar argument we know $y_1 \notin Y_8$. Player $S$ also takes $k_3$ as the first new parameter and $k_6$ as the second one. Player $D$ answers by picking some finite paths that start in $Y_6$. Every one of these paths was also chosen by her during the EU-move from $(\{\{x_1\}\cup X_2, \{y_1\} \cup Y_2\},k)$. So $S$ can once again copy his strategy $f_S$ when picking a node from every path. In this way he can ensure $Y_7 \subseteq Y_3$ and $Y_8\subseteq Y_4$. As the last part of the move player $D$ decides on whether the play continues with $(t(\{X_7,Y_7\}),k_3)$ or with $(t(\{X_8,Y_8\},k_6)$. Let $X_9\subseteq X_8$ and $Y_9 \subseteq Y_8$ be the sets with $t(\{X_8,Y_8\})=\{X_9,Y_9\}$. If a node is removed when tidying up some pair of sets, then that node also has to be removed when tidying up any pair of subsets of these sets. So the subset-relation between $X_7$ and $X_3$ as well as between $Y_7$ and $Y_3$ persists even through tidy up. Then according to Corollary~\ref{subset-positions}, because $(t(\{X_3,Y_3\}),k_1)$ is a winning position for $S$, so is $(t(\{X_7,Y_7\}),k_1)$. In a similar fashion we can argue $X_9\subseteq \{x_1\}\cup X_5$ and thus $X_9 \subseteq X_5$, as well as $Y_9\subseteq Y_5$. Recall that $\{\{x_1\}\cup X_5,\{y_1\}\cup Y_5\}$ is already tidy. So during the tidy-up process on $\{X_5,Y_5\}$ the only nodes that are removed are either nodes from $X_5$ that have the same labels as $y_1$ while no node in $Y_5$ has the same labels, or nodes from $Y_5$ with the same labels as $x_1$. However since $\{X_9,Y_9\}$ is tidy and does not contain $x_1$ nor $y_1$, none of these nodes can be a part of $X_9$ or $Y_9$. Then $X_9$ and $Y_9$ are even subsets of the tidied up versions of $X_5$ and $Y_5$. So because $(t(\{X_5,Y_5\}),k_4)$ is a winning position for $S$, $(\{X_9,Y_9\},k_4)$ is as well. That implies that by following the strategy we described, player $S$ wins any play from the position $(t(\{X_2,Y_2\},k_2)$.

We still have to examine what happens when $S$ plays an EU-move on $\{x_1\}\cup X_2$, but the path he picks for $x_1$ contains two nodes. In this case $x_1$ is a non-final node of the path. So we know $x_1 \in X_3$. Then player $D$ can answer in the following way: She looks at one path that starts in $X_2$ and was chosen by player $S$. Let $x_2 \in X_2 \cup \Box X_2$ be the final node of that path.
Then $D$ chooses the path that goes from the node $y_1$ directly to $x_2$. This is always possible, because of $X_2 \subseteq \Box y_1$ and the transitivity of $\mathcal{M}_c$. Player $D$ additionally picks every possible path that starts in $Y_2$. As the next part of the move, player $S$ has to pick a node from every path chosen by $D$ and add it to $Y_3$ or $Y_4$. As the last part of the move player $D$ gets to decide whether $(t(\{X_3,Y_3\}),k_3)$ or $(t(\{X_4,Y_4\}),k_4)$ is the next position. Both of these positions have to be winning positions for player $S$.

Now we want to take a closer look at the node that player $S$ picks from the path $(y_1,x_2)$. If he picks $x_2$, that node has to be added to $Y_4$. However since it is also the last node of a path chosen by $S$, $x_2$ is also in $X_4$. So according to Corollary~\ref{loosing condition for S CTL} the position $(t(\{X_4,Y_4\}),k_4)$ is a winning position for $D$. This contradicts our assumption that $(\{\{x_1\}\cup X_2,\{y_1\}\cup Y_2\},k)$ is a winning position for player $S$. Instead player $S$ has to pick the node $y_1$ and we know $y_1\in Y_3$. Additionally we know $x_1 \in X_3$. Because of $V_c(x_1)=V_c(y_1)$ neither of these nodes can be removed during tidy up. So there are sets $X_5\subseteq X_3$ and $Y_5\subseteq Y_3$ with $t(\{X_3,Y_3\})=\{\{x_1\}\cup X_5,\{y_1\}\cup Y_5\}$. Then every node $x_5\in X_5$ has to belong to the set $X_2 \cup \Box X_2$, which because of the transitivity of $\mathcal{M}_c$, is a subset of $\Box y_1$. Similarly $Y_5$ has to be a subset of $x_1$. So the position $(\{\{x_1\}\cup X_5,\{y_1\}\cup Y_5\},k_3)$ satisfies all conditions of the lemma and according to our induction hypothesis there are some parameters $k_5$ and $k_6$ with $k_5+k_6=k_3$, so that $(\{\{x_1\},\{y_1\}\},k_5)$ and $(t(\{X_5,Y_5\}),k_6)$ are winning positions for player $S$.

So we set $k_1=k_5$ and $k_2=k - k_5= k_4 +k_6 +1$. Next we need to show that $(t(\{X_2,Y_2\}),k_1)$ is also a winning position for $S$. Let $X_6\subseteq X_2$ and $Y_6\subseteq Y_2$ be the sets with $t(\{X_2,Y_2\})=\{X_6,Y_6\}$. Then player $S$ can play an EU-move on $X_6$. For every node in $X_6$ he picks the same path he picked for that node when playing on $(\{\{x_1\}\cup X_2,\{y_1\}\cup Y_2\},k)$. Let $X_7$ and $X_8$ be the sets of non-final and final nodes respectively. This strategy ensures $X_7 \subseteq X_3$ and $X_8 \subseteq X_4$. Player $S$ also decides on $k_6$ as the first new parameter and $k_4$ as the second one. As the next step of the move player $D$ picks some finite paths that begin in $Y_6$. During the EU-move on $(\{\{x_1\}\cup X_2,\{y_1\}\cup Y_2\},k)$ Player $D$ picked every single finite path starting in $Y_2$. So when picking a node from every path picked by $D$, player $S$ can simply copy the strategy he used in that position. By playing in this way he can ensure $Y_7 \subseteq Y_3$ and $Y_8 \subseteq Y_4$. Afterwards player $D$ gets to decide if $(t(\{X_7,Y_7\}),k_4)$ or $(t(\{X_8,Y_8\}),k_2)$ is the follow-up position. Let $X_9\subseteq X_7$ and $X_9 \subseteq Y_7$ be the sets with $t(\{X_7,Y_7\})=\{X_9,Y_9\}$. For this position we know $x_1 \notin X_9$ and $y_1 \notin Y_9$, because the nodes $x_1$ and $y_1$ cannot be successors of $X_2$ or $Y_2$ respectively, because they are on a higher layer than these sets. Also as previously stated all nodes that are removed during the tidy up of a pair of sets will also be removed during the tidy-up of any pair of subsets. This implies that $X_9 \subseteq X_5$ and $Y_9 \subseteq Y_5$ must hold. This subset-relation even holds after some nodes might be removed from $X_5$ or $Y_5$ during the tidy-up of $\{X_5,Y_5\}$. So according to Corollary~\ref{subset-positions}, because $(t(\{X_5,Y_5\}),k_6)$ is a winning position for $S$, so is $(\{X_9,Y_9\},k_6)$. Similarly the subset-relations between $X_8$ and $X_4$ as well as between $Y_8$ and $Y_4$ persists through the tidy-up of $\{X_4,Y_4\}$. So because $(t(\{X_4,Y_4\}),k_4)$ is a winning position for $S$, $(t(\{X_8,Y_8\}),k_4)$ is a winning position for him as well. With this we have shown that player $S$ has a winning strategy for the position $(t(\{X_2,Y_2\}),k_2)$.

As the last segment of this proof we have to examine what happens when the first move of $f_S$ is an AU-move. In general player $S$ has to pick a finite or infinite path for every node in $\{x_1\}\cup X_2$. But because of Lemma~\ref{3 properties} we can assume that he only picks finite paths with one or two nodes. Then $S$ puts all nodes his chosen paths into a set $X_3$, as well as all final nodes into $X_4$. From this we know $x_1\in X_3$. Player $S$ also picks new parameters $k_3$ and $k_4$ with $k=k_3+k_4+1$. Since $f_S$ is a winning strategy, Player $S$ has to win against every possible strategy of $D$. That includes the following: Player $D$ looks at some path that starts in $X_2$ and was chosen by $S$. Let $x_2 \in X_2 \cup \Box X_2$ be the final node of that path. Then $D$ picks the two-node-path $(y_1,x_2)$. The edge from $y_1$ to $x_2$ has to exist, because of $X_2 \subseteq \Box y_1$ and the transitivity of our graph. Additionally player $D$ also marks every path, finite or infinite, that starts in $Y_2$. Next for every path chosen by $D$ player $S$ has to pick a node and add it to $Y_3$, or if it is the last node she can add it to $Y_4$ instead. As the last step of this move player $D$ gets to decide if the play moves to the position $(t(\{X_3,Y_3\}),k_3)$ or to $(t(\{X_4,Y_4\}),k_4)$. Since $f_S$ is a winning strategy, both of these positions have to be winning positions for $S$.

Now we look at the path $(y_1,x_2)$ and at the node $S$ chose from it. Similar to the  EU-case we know that $S$ has to pick $y_1$ itself, since we know $x_2 \in X_3 \cap X_4$. Because $y_1$ is not the last node on this path we get $y_1 \in Y_3$. Since $x_1$ and $y_1$ share the same label, neither of them will be removed when tidying up. So there exist some sets $X_5\subseteq X_3$ and $Y_5 \subseteq Y_3$ with $t(\{X_3,Y_3\})=\{\{x_1\}\cup X_5,\{y_1\}\cup Y_5\}$. Every node in $X_5$ must belong to $X_2$ or be a successor of a node from $X_2$. Together with the transitivity of $\mathcal{M}_c$ this implies $X_5 \subseteq \Box y_1$. By an analogous argument we get $Y_5 \subseteq \Box x_1$. Then $(\{\{x_1\}\cup X_5,\{y_1\}\cup Y_5\},k_3)$ satisfies all conditions of the lemma. So according to our induction hypothesis there are parameters $k_5$ and $k_6$ with $k_3=k_5+k_6$, so that $(\{\{x_1\},\{y_1\}\},k_5)$ and $(t(\{X_5,Y_5\}),k_6)$ are winning positions for $S$ as well. Now we set $k_1=k_5$ and $(\{\{x_1\},\{y_1\}\},k_1)$ is won by $S$.

Lastly we need to show that $(t(\{X_2,Y_2\}),k_2)$ with $k_2=k_4+k_6+1$ is also a winning position for $S$. Let $X_6\subseteq X_2$ and $Y_6 \subseteq Y_2$ be the sets with $t(\{X_2,Y_2\})=\{X_6,Y_6\}$. Then at $(\{X_6,Y_6\},k_2)$ player $S$ can play an AU-move. For every node in $X_6$ he picks the same path he picked during his move on $(\{\{x_1\}\cup X_2,\{y_1\}\cup Y_2\},k)$. Let $X_7$ be the set of all nodes he picked and $X_8$ the set of all final nodes of these paths. Then this strategy guarantees $X_7 \subseteq X_3$ and $X_8\subseteq X_4$. Player $S$ decides on $k_6$ as the first parameter and $k_4$ as the second one. Next player $D$ has to answer with some finite or infinite paths that start in $Y_6$. All of these paths were also chosen by $D$ during the AU-move on $(\{\{x_1\}\cup X_2,\{y_1\}\cup Y_2\},k)$. So player $S$ can once again copy his strategy from that position when picking a node from every path. From this we know $Y_7\subseteq Y_3$ and $Y_8 \subseteq Y_4$. Then player $D$ picks $(t(\{X_7,Y_7\}),k_6)$ or $(t(\{X_8,Y_8\}),k_4)$ as the follow-up position. Let $X_9 \subseteq X_7$ and $Y_9 \subseteq Y_7$ be the sets with $t(\{X_7,Y_7\})=\{X_9,Y_9\}$. Every node that is removed during the tidy-up of $\{X_3,Y_3\}$, has to be removed during the tidy-up of $\{X_7,Y_7\}$ as well. Since we also know $x_1 \notin X_7$ and $y_1 \notin Y_7$ this implies $X_9 \subseteq X_5$ and $Y_9 \subseteq Y_5$. So because $(t(\{X_5,Y_5\}),k_6)$ is a winning position for $S$, so is $(t(\{X_9,Y_9\}),k_6)$, which is the same as $(\{X_9,Y_9\},k_4)$. Similarly because $S$ has a winning strategy for $(t(\{X_4,Y_4\}),k_4)$, he also has one for $(t(\{X_8,Y_8\}),k_4)$. So by playing as described here player $S$ can guarantee a winning position of his as the next position. Thus $(t(\{X_2,Y_2\}),k_2)$ is a winning position for him.

With this we have shown that for every possible first move of his winning strategy for the position $(\{\{x_1\}\cup X_2,\{y_1\}\cup Y_2\},k)$ player $S$ can also win from the positions $(\{\{x_1\},\{y_1\}\},k_1)$ and $(t(\{X_2,Y_2\}),k_2)$ for some $k_1,k_2$ with $k=k_1+k_2$.
\end{proof}

In the following proof we are going to use Lemma~\ref{special positions} in order to show that certain positions of the $\mathcal{M}_c$-ON game are winning positions for player $D$. If there are $k_1$ and $k_2$ with $k_1+k_2=k$ so that $(\{\{x_1\},\{y_1\}\},k_1)$ and $(t(\{X_2,Y_2\}),k_2)$ are winning positions for player $D$, then all positions with the same sets but lower parameters are also winning positions for her. Then for all $k_1'$ and $k_2'$ with $k_1'+k_2'=k$ we know that $k_1' \leq k_1$ or $k_2' \leq k_2$ has to hold. So at least one position of $(\{\{x_1\},\{y_1\}\},k_1')$ or $(t(\{X_2,Y_2\}),k_2')$ cannot be a winning position for $S$. Then because of Lemma~\ref{special positions} the position $(\{\{x_1\}\cup X_2,\{y_1\}\cup Y_2\},k)$ cannot be a winning position for $S$. So it has to be a winning position for $D$ instead.

\begin{lem}\label{no bi-implication ctl}
There is an exponential function $h\colon \mathbb{N}\rightarrow \mathbb{N}$, so that for every $i \in \mathbb{N}$ and for every CTL-formula $\chi_i$ which separates $(\mathcal{M}_c, a_i)$ from $(\mathcal{M}_c,b_i)$, we have $\vert \chi_i \vert_O > h(i)$.
\end{lem}

\begin{proof}
Once again we define $h(i)=2^i-2$. Obviously $\vert \chi_0 \vert_O\ > -1=h(0)$ holds. So we may assume $i\geq 1$ from here on. Now we only need to show that player $D$ has a winning strategy $f_D$ in the $\mathcal{M}_c$-ON game from the position $(t(\{\{a_i\},\{b_i\}\}),h(i))$, which is the same as $(\{\{a_i\},\{b_i\}\},2^i-2)$. We do this via induction over $i$. As a base case we use $i=1$. Then $((\{a_1\},\{b_1\}\},0)$ is a winning position for player $D$, because player $S$ cannot make a move.
Now let $i\geq 2$. As our induction hypothesis we assume that our claim holds for $i-1$. So $(\{\{a_{i-1}\},\{b_{i-1}\}\},2^{i-1}-2)$ is a winning position for $D$.

We begin by examining the position $(\{\{4_i\},\{5_i\}\},2^{i-1}-1)$. In this position both sets contain only one node each. Because of Lemma~\ref{3 properties} we can assume that $S$ plays according to property $P2$ and he does not play an EU- or AU-move at the current position. So we only need to check what operator-moves $S$ can play. He can choose $a_{i-1}$ as a successor of $4_i$ or $b_{i-1}$ as a successor of $5_i$. Every other successor is a shared successor. Player $S$ could also play these two moves in this position during the $\mathcal{M}_{t,r}$-ON game for modal logic we used during the proof of Lemma~\ref{no biimplication T,R}. We can also copy the strategy that $D$ played in that game. So if $S$ choose $b_{i-1}$, $D$ answers with $\{a_{i-1}\}$ and if $S$ choose $a_{i-1}$, $D$ answers $\{b_{i-1}\}$. In either case the follow-up position is $(\{\{a_{i-1}\},\{b_{i-1}\}\},2^{i-1}-2)$, which according to our induction hypothesis is a winning position for $D$. This implies that $(\{\{4_i\},\{5_i\}\},2^{i-1}-1)$ is a winning position for her as well.

The next position we look at is $(\{\{6_i\},\{7_i\}\},2^{i-1}-1)$. The only moves we need to look at from this position are $S$ playing an operator-move and picking either $a_{i-1}$ as a successor of $6_i$ or $b_{i-1}$ as a successor of $7_i$. Then $D$ answers with $\{b_{i-1}\}$ or $\{a_{i-1}\}$ respectively, which ensures $(\{\{a_{i-1}\},\{b_{i-1}\}\},2^{i-1}-2)$ as the follow-up position. So the current position is a winning position for $D$.

For $(\{\{1_i\},\{2_i\}\},2^{i-1})$ we only need to examine the operator-moves, where $S$ picks $6_i$ as a successor of $1_i$ or $7_i$ as a successor of $2_i$. In either case $D$ can simply pick the respective other node in order to ensure that the play continues with $(\{\{6_i\},\{7_i\}\},2^{i-1}-1)$, a winning position for $D$.

In the position $(\{\{2_i\},\{3_i\}\},2^{i-1})$ player $S$ can play an operator-move where he picks either $4_i$ as a successor of $2_i$ or $5_i$ as a successor of $3_i$. Analogous to the previous positions, $D$ simply answers with $\{5_i\}$ or $\{4_i\}$ respectively in order to move the play to the position $(\{\{4_i\},\{5_i\}\},2^{i-1}-1)$. So the current position is also a winning position for $D$.

There are some positions for which we do not need to check every possible move in order to show that they are winning positions for $D$. Instead we can use Lemma~\ref{special positions}. The first of these positions is $(\{\{4_i,b_{i-1}\},\{5_i,a_{i-1}\}\},2^i-5)$. The nodes $4_i$ and $5_i$ belong to the same layer of $\mathcal{M}_c$ and we know $V_c(4_i)=V_c(5_i)=\{q_0\}$. Also there is an edge from $4_i$ to $a_{i-1}$ and one from $5_i$ to $b_{i-1}$. So for $x_1=4_i$, $X_2=\{b_{i-1}\}$, $y_1=5_i$ and $Y_2=\{a_{i-1}\}$ this position satisfies all conditions of Lemma~\ref{special positions}. Assuming this position was a winning position for $S$, there would have to be parameters $k_1$ and $k_2$ with $k_1+k_2=2^i-5$, so that both $(\{\{4_i\},\{5_i\}\},k_1)$ and $(\{\{a_{i-1}\},\{b_{i-1}\}\},k_2)$ are winning positions for $S$ as well. However this would require both $k_1 > 2^{i-1}-1$ and $k_2 > 2^{i-1}-2$, contradicting $k_1+k_2=2^i-5$. So our assumption has to be wrong and $(\{\{4_i,b_{i-1}\},\{5,a_{i-1}\}\},2^i-5)$ has to be a winning position for $D$ instead. In a similar fashion we can also argue that $(\{\{6_i,b_{i-1}\},\{7_i,a_{i-1}\}\},2^i-5)$ is a winning position for $D$, because so are $(\{\{6_i\},\{7_i\}\},2^{i-1}-1)$ and $(\{\{a_{i-1}\},\{b_{i-1}\}\},2^{i-1}-2)$. The next position we look at is $(\{1_i,b_{i-1}\},\{2_i,a_{i-1}\}\},2^i-5)$. We know that both $(\{\{1_i\},\{2_i\}\},2^{i-1})$ and $(\{\{a_{i-1}\},\{b_{i-1}\}\},2^{i-1}-2)$ are winning positions for $D$, so this position is as well. The last position of this type we need to examine here is $(\{\{2_i,b_{i-1}\},\{3_i,a_{i-1}\}\},2^i-5)$. This position is a winning position for $D$, because of $(\{\{2_i\},\{3_i\}\},2^{i-1})$ and $(\{\{a_{i-1}\},\{b_{i-1}\}\},2^{i-1}-2)$.

Now we examine $(\{\{4_i,7_i\},\{5_i,6_i\}\},2^i-4)$. In this position player $S$ cannot win by playing an operator-move, because every successor of $4_i$ or $7_i$ is also a successor of $5_i$ or $6_i$ and vice versa. However in this position both sets contain more than one node each, so we have to check the other types of moves as well. $S$ can also play a split-move, where he either splits $\{4_i,7_i\}$ into $\{4_i\}$ and $\{7_i\}$ or $\{5_i,6_i\}$ into $\{5_i\}$ and $\{6_i\}$. He also has to split $2^i-4$ into $k_1$ and $k_2$. Let $k'$ be the smaller of these parameters. Then we know $k' \leq  \frac{2^i-4}{2} = 2^{i-1}-2$. Then player $D$ chooses the side corresponding to $k'$ and she answers in such a way that the resulting position is either $(\{\{4_i\},\{5_i\}\},k')$ or $(\{\{6_i\},7_i\}\},k')$. For all possible values of $k'$ these are winning positions for $D$. So $S$ cannot win from $(\{\{4_i,7_i\},\{5_i,6_i\}\},2^i-4)$ by playing a split-move.

This position is the first position in which we have to look at $S$ playing an EU- or AU-move. However we can still assume that such a move conforms to the properties $P1$ and $P3$ from Lemma~\ref{3 properties}. So we only have to look at those choices of $S$, where he only picks paths with two nodes or less and he does not pick a path that ends in some shared successor of $Y$.

First we look at an EU-move on $\{4_i,7_i\}$. In this case $S$ has to pick a finite path with starting node $4_i$ and one with starting node $7_i$. He also has to decide on two new parameters $k_1$ and $k_2$ with $k_1+k_2+1=2^i-4$. There are only two successors of $4_i$ that are not also successors of $5_i$ and $6_i$. Those being the nodes $8_i$ and $a_{i-1}$. So including the one-node-path $(4_i)$ we only have to look at three paths with starting node $4_i$. Similarly for $7_i$ we only need to look at the paths $(7_i)$, $(7_i,b_{i-1})$ and $(7_i,9_i)$.
These three paths each can be combined for a total of nine different options for $S$. In order to show that none of these options allows him to win the play, we examine Figure~\ref{EU-move on 4,7}. In this table we omit the index $i$ from the nodes $4_i$, $5_i$, $6_i$, $7_i$, $8_i$ and $9_i$ as well as the index $i-1$ from the nodes $a_{i-1}$ and $b_{i-1}$, because it is clear from context what node we mean. The first column contains the two paths chosen by player $S$, while the second one states the paths that $D$ picks as an answer. Next for every path player $S$ has to add a node to either $Y_1$ or $Y_2$. We give his possible choices in the third column. While doing so we omit some choices that are obviously bad for $S$. In particular we ignore all choices that lead to positions with a shared node, since according to Corollary~\ref{loosing condition for S CTL} every such position is a winning position for $D$. As the last part of the move player $D$ gets to choose either $(t(\{X_1,Y_1\}),k_1)$ or $(t(\{X_2,Y_2\}),k_2)$ as the follow-up position. However this choice might depend on the values player $S$ selected for $k_1$ and $k_2$. This is why we state some condition for these values in the fourth column. While doing so we ensure that all possible values for $k_1$ and $k_2$ with $k_1+k_2+1=2^i-4$ are covered. Then in the fifth column we present the follow-up position that results from this move. Lastly in the final column we give what we call a simplified position. This is a position that according to Corollary~\ref{subset-positions} is easier to win for player $S$ than the follow-up position. More precisely if $S$ has a winning strategy for the follow-up position, he also has a winning strategy for the simplified position. We can use this simplified position in order to show that the corresponding follow-up position cannot be won by $S$. In most cases this is quite simple, because we have already shown that the simplified position is a winning position for $D$. Then the contraposition of Corollary~\ref{subset-positions} states that the follow-up position is as well. The one simplified position were we need to argue differently is $(\{\{4_i,7_i\},\{5_i,6_i\}\},2^i-5)$. This position is basically the same as the current position only with a reduced parameter. So by playing the corresponding move, player $S$ does not make any progress. More precisely if $S$ could win by playing this move, he could also win by skipping this move and then using his winning strategy for the simplified position at the current position instead.

\begin{figure}[h]
\begin{tabular}{| l | l | l | l | l | l |}
\hline
$S$-paths & $D$-paths & chosen nodes & parameter & follow-up position & simplified position\\
\hline\hline
$(4);(7)$&$(5);(6)$ & $5\in Y_2; 6\in Y_2$&$k_2<2^i-4$&$(\{\{4,7\},\{5,6\}\},k_2)$&$(\{\{4,7\},\{5,6\}\},2^i-5)$\\
\hline
$(4,8);(7)$&$(5,8);(6)$&$5 \in Y_1; 6 \in Y_2$&$k_1 < 2^{i-1}-2$&$(\{\{4\},\{5\}\},k_1)$&$(\{\{4\},\{5\},2^{i-1}-1)$\\
\cline{4-6}
& & & $k_2< 2^{i-1}-2$&$(\{\{6\},\{7\}\},k_2)$&$(\{\{6\},\{7\}\},2^{i-1}-1)$\\
\hline
$(4,a);(7)$&$(5,b);(6)$&$5\in Y_1; 6 \in Y_2$&$k_1 < 2^{i-1}-2$&$(\{\{4\},\{5\}\},k_1)$&$(\{\{4\},\{5\}\},2^{i-1}-1)$\\
\cline{4-6}
& & & $k_2 < 2^{i-1}-2$&$(\{\{6\},\{7\}\},k_2)$&$(\{\{6\},\{7\}\},2^{i-1}-1)$\\
\cline{3-6}
& &$b\in Y_2; 6 \in Y_2$&$k_2 < 2^i-4$&$(\{\{6,b\},\{7,a\}\},k_2)$&$(\{6,b\},\{7,a\}\},2^i-5)$\\
\hline
$(4);(7,b)$&$(5);(6,a)$&$5 \in Y_2; 6 \in Y_1$&$k_1 < 2^{i-1}-2$&$(\{\{6\},\{7\}\},k_1)$&$(\{\{6\},\{7\}\},2^{i-1}-1)$\\
\cline{4-6}
& & &$k_2 < 2^{i-1}-2$&$(\{\{4\},\{5\}\},k_2)$&$(\{\{4\},\{5\}\},2^{i-1}-1)$\\
\cline{3-6}
& &$5 \in Y_2; a \in Y_2$&$k_2 < 2^i-4$&$(\{\{4,b\},\{5,a\}\},k_2)$&$(\{\{4,b\},\{5,a\}\},2^i-5)$\\
\hline
$(4,8);(7,b)$&$(5,8);(6,a)$&$5 \in Y_1; 6 \in Y_1$&$k_1< 2^i-4$&$(\{\{4,7\},\{5,6\}\},k_1)$&$(\{\{4,7\},\{5,6\}\},2^i-5)$\\
\cline{3-6}
& &$5\in Y_1;a\in Y_2$&$k_1 < 2^{i-1}-2$&$(\{\{4,7\},\{5\}\},k_1)$&$(\{\{4\},\{5\}\},2^{i-1}-1)$\\
\cline{4-6}
& & &$k_2 < 2^{i-1}-2$&$(\{\{a\},\{b\}\},k_2)$&$\{\{a\},\{b\}\},2^{i-1}-2)$\\
\hline
$(4,a);(7,b)$&$(5,b);(6,a)$&$5\in Y_1;6\in Y_1$&$k_1<2^i-4$&$(\{\{4,7\},\{5,6\}\},k_1)$&$(\{\{4,7\},\{5,6\}\},2^i-5)$\\
\hline
$(4);(7,9)$&$(5);(6,9)$&$5\in Y_2;6\in Y_1$&$k_1 < 2^{i-1}-2$&$(\{\{6\},\{7\}\},k_1)$&$(\{\{6\},\{7\}\},2^{i-1}-1)$\\
\cline{4-6}
& & &$k_2 < 2^{i-1}-2$&$(\{\{4\},\{5\}\},k_2)$&$(\{\{4\},\{5\}\},2^{i-1}-1)$\\
\hline
$(4,8);(7,9)$&$(5,8);(6,9)$&$5\in Y_1;6\in Y_1$&$k_1<2^i-4$&$(\{\{4,7\},\{5,6\}\},k_1)$&$(\{\{4,7\},\{5,6\}\},2^i-5)$\\
\hline
$(4,a);(7,9)$&$(5,b);(6,9)$&$5\in Y_1;6\in Y_1$&$k_1<2^i-4$&$(\{\{4,7\},\{5,6\}\},k_1)$&$(\{\{4,7\},\{5,6\}\},2^i-5)$\\
\cline{3-6}
& &$b\in Y_2;6 \in Y_1$&$k_1 < 2^{i-1}-2$&$(\{\{4,7\},\{6\}\},k_1)$&$(\{\{6\},\{7\}\},2^{i-1}-1)$\\
\cline{4-6}
& & &$k_2 < 2^{i-1}-2$&$(\{\{a,9\},\{b\}\},k_2)$&$(\{\{a\},\{b\}\},2^{i-1}-2)$\\
\hline
\end{tabular}
\caption{EU-move on $\{4_i,7_i\}$}\label{EU-move on 4,7}
\end{figure}

Next we look into what happens when $S$ tries to play an EU-move on $\{5_i,6_i\}$. Now he has to pick a path starting at $5_i$ and one starting at $6_i$. For the first of these nodes he may pick $(5_i)$, $(5_i,8_i)$ or $(5_i,b_{i-1})$. For the second node his choices are $(6_i)$, $(6_i,a_{i-1})$ and $(6_i,9_i)$. Once again three options per node can be combined in a total of nine different ways. In Figure~\ref{EU-move on 5,6} we examine all of these possible moves and show that none of them leads to a winning position for $S$.

\begin{figure}[h]
\begin{tabular}{| l | l | l | l | l | l |}
\hline
$S$-paths & $D$-paths & chosen nodes & parameter & follow-up position & simplified position\\
\hline\hline
$(5);(6)$ & $(4);(7)$& $4\in Y_2; 7\in Y_2$&$k_2<2^i-4$&$(\{\{4,7\},\{5,6\}\},k_2)$&$(\{\{4,7\},\{5,6\}\},2^i-5)$\\
\hline
$(5,8);(6)$&$(4,8);(7)$&$4 \in Y_1; 7 \in Y_2$&$k_1 < 2^{i-1}-2$&$(\{\{4\},\{5\}\},k_1)$&$(\{\{4\},\{5\},2^{i-1}-1)$\\
\cline{4-6}
& & & $k_2< 2^{i-1}-2$&$(\{\{6\},\{7\}\},k_2)$&$(\{\{6\},\{7\}\},2^{i-1}-1)$\\
\hline
$(5,b);(6)$&$(4,a);(7)$&$4\in Y_1; 7 \in Y_2$&$k_1 < 2^{i-1}-2$&$(\{\{4\},\{5\}\},k_1)$&$(\{\{4\},\{5\}\},2^{i-1}-1)$\\
\cline{4-6}
& & & $k_2 < 2^{i-1}-2$&$(\{\{6\},\{7\}\},k_2)$&$(\{\{6\},\{7\}\},2^{i-1}-1)$\\
\cline{3-6}
& &$a\in Y_2; 7 \in Y_2$&$k_2 < 2^i-4$&$(\{\{6,b\},\{7,a\}\},k_2)$&$(\{6,b\},\{7,a\}\},2^i-5)$\\
\hline
$(5);(6,a)$&$(4);(7,b)$&$4 \in Y_2; 7 \in Y_1$&$k_1 < 2^{i-1}-2$&$(\{\{6\},\{7\}\},k_1)$&$(\{\{6\},\{7\}\},2^{i-1}-1)$\\
\cline{4-6}
& & &$k_2 < 2^{i-1}-2$&$(\{\{4\},\{5\}\},k_2)$&$(\{\{4\},\{5\}\},2^{i-1}-1)$\\
\cline{3-6}
& &$4 \in Y_2; b \in Y_2$&$k_2 < 2^i-4$&$(\{\{4,b\},\{5,a\}\},k_2)$&$(\{\{4,b\},\{5,a\}\},2^i-5)$\\
\hline
$(5,8);(6,a)$&$(4,8);(7,b)$&$4 \in Y_1; 7 \in Y_1$&$k_1< 2^i-4$&$(\{\{4,7\},\{5,6\}\},k_1)$&$(\{\{4,7\},\{5,6\}\},2^i-5)$\\
\cline{3-6}
& &$4\in Y_1;b\in Y_2$&$k_1 < 2^{i-1}-2$&$(\{\{4\},\{5,6\}\},k_1)$&$(\{\{4\},\{5\}\},2^{i-1}-1)$\\
\cline{4-6}
& & &$k_2 < 2^{i-1}-2$&$(\{\{a\},\{b\}\},k_2)$&$\{\{a\},\{b\}\},2^{i-1}-2)$\\
\hline
$(5,b);(6,a)$&$(4,a);(7,b)$&$4\in Y_1;7\in Y_1$&$k_1<2^i-4$&$(\{\{4,7\},\{5,6\}\},k_1)$&$(\{\{4,7\},\{5,6\}\},2^i-5)$\\
\hline
$(5);(6,9)$&$(4);(7,9)$&$4\in Y_2;7\in Y_1$&$k_1 < 2^{i-1}-2$&$(\{\{6\},\{7\}\},k_1)$&$(\{\{6\},\{7\}\},2^{i-1}-1)$\\
\cline{4-6}
& & &$k_2 < 2^{i-1}-2$&$(\{\{4\},\{5\}\},k_2)$&$(\{\{4\},\{5\}\},2^{i-1}-1)$\\
\hline
$(5,8);(6,9)$&$(4,8);(7,9)$&$4\in Y_1;7\in Y_1$&$k_1<2^i-4$&$(\{\{4,7\},\{5,6\}\},k_1)$&$(\{\{4,7\},\{5,6\}\},2^i-5)$\\
\hline
$(5,b);(6,9)$&$(4,a);(7,9)$&$4\in Y_1;7\in Y_1$&$k_1<2^i-4$&$(\{\{4,7\},\{5,6\}\},k_1)$&$(\{\{4,7\},\{5,6\}\},2^i-5)$\\
\cline{3-6}
& &$a\in Y_2;7 \in Y_1$&$k_1 < 2^{i-1}-2$&$(\{\{5,6\},\{7\}\},k_1)$&$(\{\{6\},\{7\}\},2^{i-1}-1)$\\
\cline{4-6}
& & &$k_2 < 2^{i-1}-2$&$(\{\{a\},\{b,9\}\},k_2)$&$(\{\{a\},\{b\}\},2^{i-1}-2)$\\
\hline
\end{tabular}
\caption{EU-move on $\{5_i,6_i\}$}\label{EU-move on 5,6}
\end{figure}

When $S$ plays an AU-move on $\{4_i,7_i\}$ he has to pick a path with starting node $4_i$ and one with starting node $7_i$. Because of Lemma~\ref{3 properties} we only need to look at paths with two nodes or less that do not end in a successor of $5_i$ and $6_i$. So he has the same options he had when playing an EU-move. Then player $D$ simply copies her paths as well. When all chosen paths are the same, there are only two potential differences between an EU- and an AU-move. During an EU-move the set $X_1$ contains all non-final nodes. During an AU-move it contains all of these nodes and in addition it also contains all final nodes as well. So the set $X_1$ can only get bigger when compared to the EU-move, while $X_2$ has to stay the same. Then if player $S$ also marks the same nodes and adds them to the same set, we also get the same $Y_1$ and $Y_2$. We already know that $S$ cannot win by playing an EU-move. So he also cannot win by playing an AU-move, where he plays in the same way he could play during an EU-move. However the second possible difference is what happens when $S$ marks the final node of a $D$-path. During an EU-move he had to add that node to $Y_2$, but during an AU-move he may instead add it to $Y_1$. In this way $S$ can play an AU-move without a corresponding EU-move. We examine these new possibilities in Figure~\ref{AU-move on 4,7}. For all possible choices of $D$-paths player $S$ can just add the starting node of each path to $Y_1$. Here this would mean $5_i \in Y_1$ and $6_i \in Y_1$. In this case we get $Y \subseteq Y_1$. Also $X \subseteq X_1$ is true for every AU-move.
So if $S$ plays like this, we get $(\{X,Y\},k_1)$ as a simplified position. In this way $S$ cannot make any progress. Because of this we skip these moves in our tables.

For some paths chosen by $D$, $S$ has two very similar choices of marking nodes. We explain this with the example of $(5_i)$ and $(6_i)$. Player $D$ picks these paths when $S$ chooses $(4_i)$ and $(7_i)$. Then player $S$ is forced to mark the nodes $5_i$ and $6_i$. But for each node he still has the choice of adding it to $X_1$ or $X_2$. We are mainly interested in what happens if $S$ puts $5_i$ and $6_i$ into different sets. He can do so in two different ways. However both choices lead to essentially the same outcomes. More precisely for $5 \in Y_1$ and $6 \in Y_2$ we get either $(\{\{4_i,7_i\},\{5_i\}\},k_1)$ or $(\{\{4_i,7_i\},\{6_i\}\},k_2)$, depending on what parameter is lower.  For $5 \in Y_2$ and $6 \in Y_1$ on the other hand we get $(\{\{4_i,7_i\},\{6_i\}\},k_1)$ or $(\{\{4_i,7_i\},\{5_i\}\},k_2)$. So the same positions only with parameters swapped. This is because in this case we have $X_1 = X_2$ and our two choices basically just swap $Y_1$ and $Y_2$. So the position $(t(\{X_1,Y_1\}),k_1)$ from the first choice is the same as the position $(t(\{X_2,Y_2\}),k_2)$ from the second choice. On a more general level during an AU-move we always have $X_2 \subseteq X_1$. So it is theoretically possible that we get new positions, because $X_1$ might contain more nodes than $X_2$. So we also need to show that these additional nodes are not necessary for $D$ to win. The easiest way of doing this is to simply look at the simplified position and to verify that it does not contain any of these additional nodes. In total if we can show that player $D$ wins if $S$ puts the last node of the first path into $Y_1$ and the last node of the second path into $Y_2$ without needing the nodes from $X_1 \setminus X_2$, then we also know that $D$ wins if $S$ adds the last node of the first path to $Y_2$ and the last node of the second path to $Y_1$. In this way we can reduce the number of moves we need to check. Then Figure~\ref{AU-move on 4,7} proves that $S$ cannot win by playing an AU-move on $\{4_i,7_i\}$.

\begin{figure}[h]
\begin{tabular}{| l | l | l | l | l | l |}
\hline
$S$-paths & $D$-paths & chosen nodes & parameter & follow-up position & simplified position\\
\hline\hline
$(4);(7)$&$(5);(6)$ & $5\in Y_1; 6\in Y_2$&$k_1<2^{i-1}-2$&$(\{\{4,7\},\{5\}\},k_1)$&$(\{\{4\},\{5\}\},2^{i-1}-1)$\\
\cline{4-6}
& & &$k_2<2^{i-1}-2$&$(\{\{4,7\},\{6\}\},k_2)$&$(\{\{6\},\{7\}\},2^{i-1}-1)$\\
\hline
$(4,a);(7)$&$(5,b);(6)$&$b\in Y_1; 6 \in Y_1$&$k_1 < 2^i-4$&$(\{\{4,7,a\},\{6,b\}\},k_1)$&$(\{\{6,b\},\{7,a\},2^i-5)$\\
\cline{3-6}
& &$b\in Y_1;6 \in Y_2$&$k_1<2^{i-1}-2$&$(\{\{a\},\{b\}\},k_1)$&$(\{\{a\},\{b\}\},2^{i-1}-2)$\\
\cline{4-6}
& & &$k_2<2^{i-1}-2$&$(\{\{6\},\{7\}\},k_2)$&$(\{\{6\},\{7\}\},2^{i-1}-1)$\\
\hline
$(4);(7,b)$&$(5);(6,a)$&$5 \in Y_1; a \in Y_1$&$k_1 < 2^i-4$&$(\{\{4,7,b\},\{5,a\}\},k_1)$&$(\{\{4,b\},\{5,a\}\},2^i-5)$\\
\cline{3-6}
& &$5 \in Y_1; a \in Y_2$&$k_1 < 2^{i-1}-2$&$(\{\{4,7\},\{5\}\},k_1)$&$(\{\{4\},\{5\}\},2^{i-1}-1)$\\
\cline{4-6}
& & &$k_2<2^{i-1}-2$&$(\{\{a\},\{b\}\},k_2)$&$(\{\{a\},\{b\}\},2^{i-1}-2)$\\
\hline
$(4,8);(7,b)$&$(5,8);(6,a)$&$5\in Y_1; a \in Y_1$&$k_1< 2^i-4$&$(\{\{4,7,b\},\{5,a\}\},k_1)$&$(\{\{4,b\},\{5,a\}\},2^i-5)$\\
\hline
$(4,a);(7,9)$&$(5,b);(6,9)$&$b\in Y_1;6\in Y_1$&$k_1<2^i-4$&$(\{\{4,7,a,9\},\{6,b\}\},k_1)$&$(\{\{6,b\},\{7,a\}\},2^i-5)$\\
\hline
\end{tabular}
\caption{AU-move on $\{4_i,7_i\}$}\label{AU-move on 4,7}
\end{figure}

Player $S$ can also play an AU-move, where he picks a path each for $5_i$ and $6_i$. Because of Lemma~\ref{3 properties} we only need to check the same paths as during the EU-move on $\{5_i,6_i\}$. All shortcuts we described for the AU-move on $\{4_i,7_i\}$ still apply here. Then Figure~\ref{AU-move on 5,6} shows that $S$ cannot win with this AU-move either. In total we have shown that the position $(\{\{4_i,7_i\},\{5_i,6_i\}\},2^i-4)$ is a winning position for $D$.

\begin{figure}[h]
\begin{tabular}{| l | l | l | l | l | l |}
\hline
$S$-paths & $D$-paths & chosen nodes & parameter & follow-up position & simplified position\\
\hline\hline
$(5);(6)$ &$(4);(7)$& $4\in Y_1; 7\in Y_2$&$k_1<2^{i-1}-2$&$(\{\{4\},\{5,6\}\},k_1)$&$(\{\{4\},\{5\}\},2^{i-1}-1)$\\
\cline{4-6}
& & &$k_2<2^{i-1}-2$&$(\{\{5,6\},\{7\}\},k_2)$&$(\{\{6\},\{7\}\},2^{i-1}-1)$\\
\hline
$(5,b);(6)$&$(4,a);(7)$&$a\in Y_1; 7 \in Y_1$&$k_1 < 2^i-4$&$(\{\{5,6,b\},\{7,a\}\},k_1)$&$(\{\{6,b\},\{7,a\},2^i-5)$\\
\cline{3-6}
& &$a\in Y_1;7 \in Y_2$&$k_1<2^{i-1}-2$&$(\{\{a\},\{b\}\},k_1)$&$(\{\{a\},\{b\}\},2^{i-1}-2)$\\
\cline{4-6}
& & &$k_2<2^{i-1}-2$&$(\{\{6\},\{7\}\},k_2)$&$(\{\{6\},\{7\}\},2^{i-1}-1)$\\
\hline
$(5);(6,a)$&$(4);(7,b)$&$4 \in Y_1; b \in Y_1$&$k_1 < 2^i-4$&$(\{\{4,b\},\{5,6,a\}\},k_1)$&$(\{\{4,b\},\{5,a\}\},2^i-5)$\\
\cline{3-6}
& &$4 \in Y_1; b \in Y_2$&$k_1 < 2^{i-1}-2$&$(\{\{4\},\{5,6\}\},k_1)$&$(\{\{4\},\{5\}\},2^{i-1}-1)$\\
\cline{4-6}
& & &$k_2<2^{i-1}-2$&$(\{\{a\},\{b\}\},k_2)$&$(\{\{a\},\{b\}\},2^{i-1}-2)$\\
\hline
$(5,8);(6,a)$&$(4,8);(7,b)$&$4\in Y_1; b \in Y_1$&$k_1< 2^i-4$&$(\{\{4,b\},\{5,6,a\}\},k_1)$&$(\{\{4,b\},\{5,a\}\},2^i-5)$\\
\hline
$(5,b);(6,9)$&$(4,a);(7,9)$&$a\in Y_1;7\in Y_1$&$k_1<2^i-4$&$(\{\{5,6,b,9\},\{7,a\}\},k_1)$&$(\{\{6,b\},\{7,a\}\},2^i-5)$\\
\hline
\end{tabular}
\caption{AU-move on $\{5_i,6_i\}$}\label{AU-move on 5,6}
\end{figure}

The next position we have to examine is $(\{\{2_i,4_i\},\{1_i,5_i\}\},2^i-4)$. In this position player $S$ cannot win with an operator-move, since every successor of $4_i$ is also a successor of $1_i$ and every successor of $5_i$ is also a successor of $2_i$. So for every possible choice of $S$, player $D$ can ensure a shared node in the next position.
We continue with what happens when $S$ plays a split-move. He can either split $\{2_i,4_i\}$ into $\{2_i\}$ and $\{4_i\}$ or he can split $\{1_i,5_i\}$ into $\{1_i\}$ and $\{5_i\}$. Next he has to split the parameter $2^i-4$ into $k_1$ and $k_2$. For $k'$, the smaller of these two parameters, we know $k' \leq 2^{i-1}-2$. Player $D$ defends by picking the side corresponding to $k'$. Then with her choice of $Y$ she can ensure that the follow-up position is of the form $(\{\{1_i\},\{2_i\}\},k')$ or $(\{\{4_i\},\{5_i\}\},k')$. So this move can be guaranteed to result in a winning position for $D$.

When playing an EU-move on $\{2_i,4_i\}$, $S$ has to choose a path with starting node $2_i$ and one with starting node $4_i$. Because of Lemma~\ref{3 properties} we can assume that both paths consist of two nodes or less and do not end in a shared successor of $1_i$ and $5_i$. This only leaves the paths $(2_i)$, $(2_i,4_i)$, $(2_i,7_i)$, $(2_i,a_{i-1})$ and $(2_i,9_i)$ for the node $2_i$, as well as $(4_i)$ and $(4_i,a_{i-1})$ for $4_i$. However the path chosen for $(2_i)$ is mostly irrelevant. This is because the node $4_i$ is a successor of $1_i$. So $D$ can choose her path for $1_i$ mostly independently from the path chosen by $S$ for $2_i$. In the following shortened table from Figure~\ref{EU-move on 2,4} the node $x$ is some successor of  $2_i$. It should be noted here that for some choices of $x$ the node $x$ can appear in the follow-up position in addition to the nodes stated in the table. However it is easy to see that adding another node to the position cannot turn it into a winning position for $S$.
 Then this table shows that $S$ cannot win by playing an EU-move on $\{2_i,4_i\}$.

\begin{figure}[h]
\begin{tabular}{| l | l | l | l | l | l |}
\hline
$S$-paths & $D$-paths & chosen nodes & parameter & follow-up position & simplified position\\
\hline\hline
$(2);(4)$&$(1),(5)$&$1 \in Y_2;5\in Y_2$&$k_2<2^i-4$&$(\{\{2,4\},\{1,5\}\},k_2)$&$(\{\{2,4\},\{1,5\}\},2^i-5)$\\
\hline
$(2,x);(4)$&$(1,4);(5)$&$1 \in Y_1;5 \in Y_2$&$k_1<2^{i-1}-2$&$(\{\{1\},\{2\}\},k_1)$&$(\{\{1\},\{2\}\},2^{i-1})$\\
\cline{4-6}
& & &$k_2<2^{i-1}-2$&$(\{\{4\},\{5\}\},k_2)$&$(\{\{4\},\{5\}\},2^{i-1}-1)$\\
\hline
$(2);(4,a)$&$(1);(5,b)$&$1 \in Y_2; 5\in Y_1$&$k_1<2^{i-1}-2$&$(\{\{4\},\{5\}\},k_1)$&$(\{\{4\},\{5\}\},2^{i-1}-1)$\\
\cline{4-6}
& & &$k_2<2^{i-1}-2$&$(\{\{1\},\{2\}\},k_2)$&$(\{\{1\},\{2\}\},2^{i-1})$\\
\cline{3-6}
& &$1 \in Y_2; b \in Y_2$&$k_2<2^i-4$&$(\{\{1,b\},\{2,a\}\},k_2)$&$(\{\{1,b\},\{2,a\}\},2^i-5)$\\
\hline
$(2,x);(4,a)$&$(1,4);(5,b)$&$1\in Y_1;5\in Y_1$&$k_1<2^i-4$&$(\{\{2,4\},\{1,5\}\},k_2)$&$(\{\{2,4\},\{1,5\}\},2^i-5)$\\
\cline{3-6}
& &$1 \in Y_1;b \in Y_2$&$k_1<2^{i-1}-2$&$(\{\{1\},\{2\}\},k_1)$&$(\{\{1\},\{2\}\},2^{i-1})$\\
\cline{4-6}
& & &$k_2<2^{i-1}-2$&$(\{\{a\},\{b\}\},k_2)$&$(\{\{a\},\{b\}\},2^{i-1}-2)$\\
\hline
\end{tabular}
\caption{EU-move on $\{2_i,4_i\}$}\label{EU-move on 2,4}
\end{figure}

Player $S$ can also play an EU-move on $\{1_i,5_i\}$. For the node $1_i$ $S$ can pick the paths $(1_i)$, $(1_i,4_i)$, $(1_i,6_i)$, $(1_i,b_i)$ and $(1_i,9_i)$. However one of these paths, namely $(1_i,4_i)$ is a bad choice for $S$. This is because $D$ can answer with the path $(4_i)$, which leads to a position with $4_i$ as a shared node. So a winning position for $D$. Player $S$ can also chose between $(5_i)$ and $(5_i,b_i)$ as the path for $5_i$. Similar to the EU-move on $\{2_i,4_i\}$, $D$ can sometimes ignore the path chosen for $1_i$. For the move on $\{1_i,5_i\}$, this is only possible if $S$ also picked the path $(5_i,b_i)$, since $b_i$ is also a successor $2_i$. Let $x$ be some successor of $1_i$. Then Figure~\ref{EU-move on 1,5} shows that $S$ cannot win by playing an EU-move on $\{1_i,5_i\}$ either.

\begin{figure}[h]
\begin{tabular}{| l | l | l | l | l | l |}
\hline
$S$-paths & $D$-paths & chosen nodes & parameter & follow-up position & simplified position\\
\hline\hline
$(1);(5)$&$(2);(4)$&$2\in Y_2;4 \in Y_2$&$k_2<2^i-4$&$(\{\{2,4\},\{1,5\}\},k_2)$&$(\{\{2,4\},\{1,5\}\},2^i-5)$\\
\hline
$(1,6);(5)$&$(2,7);(4)$&$2 \in Y_1;4\in Y_2$&$k_1<2^{i-1}-2$&$(\{\{1\},\{2\}\},k_1)$&$(\{\{1\},\{2\}\},2^{i-1})$\\
\cline{4-6}
& & &$k_2 < 2^{i-1}-2$&$(\{\{4\},\{5,6\}\},k_2)$&$(\{\{4\},\{5\}\},2^{i-1}-1)$\\
\cline{3-6}
& &$7\in Y_2;4 \in Y_2$&$k_2<2^i-4$&$(\{\{4,7\},\{5,6\}\},k_2)$&$(\{\{4,7\},\{5,6\}\},2^i-4)$\\
\hline
$(1,b);(5)$&$(2,b);(4)$&$2\in Y_1;4 \in Y_2$&$k_1<2^{i-1}-2$&$(\{\{1\},\{2\}\},k_1)$&$(\{\{1\},\{2\}\},2^{i-1})$\\
\cline{4-6}
& & &$k_2<2^{i-1}-2$&$(\{\{4\},\{5\}\},k_2)$&$(\{\{4\},\{5\}\},2^{i-1}-1)$\\
\hline
$(1,9);(5)$&$(2,9);(4)$&$2 \in Y_1;4 \in Y_2$&$k_1 < 2^{i-1}-2$&$(\{\{1\},\{2\}\},k_1)$&$(\{\{1\},\{2\}\},2^{i-1})$\\
\cline{4-6}
& & &$k_2<2^{i-1}-2$&$(\{\{4\},\{5\}\},k_2)$&$(\{\{4\},\{5\}\},2^{i-1}-1)$\\
\hline
$(1);(5,b)$&$(2);(4,a)$&$2 \in Y_2; 4 \in Y_1$&$k_1 < 2^{i-1}-2$&$(\{\{4\},\{5\}\},k_2)$&$(\{\{4\},\{5\}\},2^{i-1}-1)$\\
\cline{4-6}
& & &$k_2<2^{i-1}-2$&$(\{\{1\},\{2\}\},k_2)$&$(\{\{1\},\{2\}\},2^{i-1})$\\
\cline{3-6}
& &$2 \in Y_2;a \in Y_2$&$k_2< 2^i-4$&$(\{\{1,b\},\{2,a\}\},k_2)$&$(\{\{1,b\},\{2,a\}\},2^i-5)$\\
\hline
$(1,x);(5,b)$&$(2,b);(4,a)$&$2 \in Y_1;4 \in Y_1$&$k_1<2^i-4$&$(\{\{2,4\},\{1,5\}\},k_1)$&$(\{\{2,4\},\{1,5\}\},2^i-5)$\\
\cline{3-6}
& &$2 \in Y_1;a\in Y_2$&$k_1<2^{i-1}-2$&$(\{\{1\},\{2\}\},k_1)$&$(\{\{1\},\{2\}\},2^{i-1})$\\
\cline{4-6}
& & &$k_2<2^{i-1}-2$&$(\{\{a\},\{b\}\},k_2)$&$(\{\{a\},\{b\}\},2^{i-1}-2)$\\
\hline
\end{tabular}
\caption{EU-move on $\{1_i,5_i\}$}\label{EU-move on 1,5}
\end{figure}

The next move we want to look at is the AU-move on $\{2_i,4_i\}$. As stated previously, since we have already shown that $S$ cannot win with an EU-move on $\{2_i,4_i\}$, we only need to look at the AU-moves without a corresponding EU-move. More precisely we only need to look at those moves, where the final node of a path is added to $Y_1$. We can also skip $S$ marking both $1_i$ and $5_i$ and putting them both into $Y_1$, since he cannot make progress this way. Lastly we can also skip certain moves where $S$ marks the final nodes of both paths and puts them into different sets. For all remaining options, Figure~\ref{AU-move on 2,4} shows that none of them allow $S$ to win.

\begin{figure}[h]
\begin{tabular}{| l | l | l | l | l | l |}
\hline
$S$-paths & $D$-paths & chosen nodes & parameter & follow-up position & simplified position\\
\hline\hline
$(2);(4)$&$(1);(5)$&$1\in Y_1;5 \in Y_2$&$k_1<2^{i-1}-2$&$(\{\{1\},\{2\}\},k_1)$&$(\{\{1\},\{2\}\},2^{i-1})$\\
\cline{4-6}
& & &$k_2< 2^{i-1}-2$&$(\{\{4\},\{5\}\},k_2)$&$(\{\{4\},\{5\}\},2^{i-1}-1)$\\
\hline
$(2);(4,a)$&$(1);(5,b)$&$1 \in Y_1; b \in Y_1$&$k_1<2^i-4$&$(\{\{1,b\},\{2,a\}\},k_1)$&$(\{\{1,b\},\{2,a\}\},2^i-5)$\\
\cline{3-6}
& &$1 \in Y_1;b \in Y_2$&$k_1<2^{i-1}-2$&$(\{\{1\},\{2\}\},k_1)$&$(\{\{1\},\{2\}\},2^{i-1})$\\
\cline{4-6}
& & &$k_2<2^{i-1}-2$&$(\{\{a\},\{b\}\},k_2)$&$(\{\{a\},\{b\}\},2^{i-1}-2)$\\
\hline
$(2,4);(4,a)$&$(1,4);(5,b)$&$1\in Y_1;b \in Y_1$&$k_1 < 2^i-4$&$(\{\{1,b\},\{2,a\}\},k_1)$&$(\{\{1,b\},\{2,a\}\},2^i-5)$\\
\hline
$(2,7);(4,a)$&$(1,a);(5,b)$&$1 \in Y_1; b \in Y_1$&$k_1<2^i-4$&$(\{\{1,b\},\{2,a\}\},k_1)$&$(\{\{1,b\},\{2,a\}\},2^i-5)$\\
\hline
$(2,a);(4,a)$&$(1,a);(5,b)$&$1 \in Y_1;b \in Y_1$&$k_1 < 2^i-4$&$(\{\{1,b\},\{2,a\}\},k_1)$&$(\{\{1,b\},\{2,a\}\},2^i-5)$\\
\hline
$(2,9);(4,a)$&$(1,a);(5,b)$&$1 \in Y_1;b \in Y_1$&$k_1< 2^i-4$&$(\{\{1,b\},\{2,a\}\},k_1)$&$(\{\{1,5\},\{2,a\}\},2^i-5)$\\
\hline
\end{tabular}
\caption{AU-move on $\{2_i,4_i\}$}\label{AU-move on 2,4}
\end{figure}

As the last choice for the position $(\{\{2_i,4_i\},\{5_i,6_i\}\},2^i-4)$, player $S$ can play an AU-move on $\{1_i,5_i\}$. Once again we use our known shortcuts for AU-moves. Then Figure~\ref{AU-move on 1,5} shows that all of his moves result in winning positions for $D$. So the current position has to be a winning position for her as well.

\begin{figure}[h]
\begin{tabular}{| l | l | l | l | l | l |}
\hline
$S$-paths & $D$-paths & chosen nodes & parameter & follow-up position & simplified position\\
\hline\hline
$(1);(5)$&$(2);(4)$&$2 \in Y_1;4 \in Y_2$&$k_1< 2^{i-1}-2$&$(\{\{1\},\{2\}\},k_1)$&$(\{\{1\},\{2\}\},2^{i-1})$\\
\cline{4-6}
& & &$k_2<2^{i-1}-2$&$(\{\{4\},\{5\}\},k_2)$&$(\{\{4\},\{5\}\},2^{i-1}-1)$\\
\hline
$(1,6);(5)$&$(2,7);(4)$&$7\in Y_1;4 \in Y_1$&$k_1<2^i-4$&$(\{\{4,7\},\{5,6\}\},k_1)$&$(\{\{4,7\},\{5,6\}\},2^i-4)$\\
\cline{3-6}
& &$7 \in Y_1;4 \in Y_2$&$k_1<2^{i-1}-2$&$(\{\{5,6\},\{7\}\},k_1)$&$(\{\{6\},\{7\}\},2^{i-1}-1)$\\
\cline{4-6}
& & &$k_2<2^{i-1}-2$&$(\{\{4\},\{5,6\}\},k_2)$&$(\{\{4\},\{5\}\},2^{i-1}-1)$\\
\hline
$(1);(5,b)$&$(2);(4,a)$&$2\in Y_1;a \in Y_1$&$k_1<2^i-4$&$(\{\{1,b\},\{2,a\}\},k_1)$&$(\{\{1,b\},\{2,a\}\},2^i-5)$\\
\cline{3-6}
& &$2 \in Y_1;a \in Y_2$&$k_1<2^{i-1}-2$&$(\{\{1\},\{2\}\},k_1)$&$(\{\{1\},\{2\}\},2^{i-1})$\\
\cline{4-6}
& & &$k_2<2^{i-1}-2$&$(\{\{a\},\{b\}\},k_2)$&$(\{\{a\},\{b\}\},2^{i-1}-2)$\\
\hline
$(1,6);(5,b)$&$(2,b);(4,a)$&$2 \in Y_1;a \in Y_1$&$k_1<2^i-4$&$(\{\{1,b\},\{2,a\}\},k_1)$&$(\{\{1,b\},\{2,a\}\},2^i-5)$\\
\hline
$(1,b);(5,b)$&$(2,b);(4,a)$&$2 \in Y_1;a \in Y_1$&$k_1<2^i-4$&$(\{\{1,b\},\{2,a\}\},k_1)$&$(\{\{1,b\},\{2,a\}\},2^i -5)$\\
\hline
$(1,9);(5,b)$&$(2,b);(4,a)$&$2 \in Y_1; a \in Y_1$&$k_1<2^i-4$&$(\{\{1,b,9\},\{2,a\}\},k_1)$&$(\{\{1,b\},\{2,a\}\},2^i-5)$\\
\hline
\end{tabular}
\caption{AU-move on $\{1_i,5_i\}$}\label{AU-move on 1,5}
\end{figure}

Another position we need to look at is $(\{\{2_i,7_i\},\{3_i,6_i\}\},2^i-4)$. Once again player $S$ cannot win by playing an operator-move. This is because every successor of $7_i$ is also a successor $3_i$ and every successor of $6_i$ is also a successor of $2_i$. When playing a split-move, $S$ can split either $\{2_i,7_i\}$ into $\{2_i\}$ and $\{7_i\}$ or $\{3_i,6_i\}$ into $\{3_i\}$ and $\{6_i\}$. Let $k_1$ and $k_2$ be the new parameters chosen by $S$ and let $k'\in \{k_1,k_2\}$ be the smaller of them. Then $D$ can ensure either $(\{\{2_i\},\{3_i\}\},k')$ or $(\{\{6_i\},\{7_i\}\},k')$ as the next position. Because of $k' \leq 2^{i-1}-2$ both of these are winning positions for player $D$.

In this position $S$ can also play an EU-move on $\{2_i,7_i\}$. The only relevant paths he can pick are $(2_i)$, $(2_i,4_i)$, $(2_i,7_i)$, $(2_i,8_i)$ and $(2_i,b_{i-1})$ for $2_i$, as well as $(7_i)$ and $(7_i,b_{i-1})$ for $7_i$. However like in some previous moves we do not need to go through all possible combinations separately. Instead let $x$ be some successor of $2_i$. Then Figure~\ref{EU-move on 2,7} shows that every possible choice leads to a winning position for $D$.

\begin{figure}[h]
\begin{tabular}{| l | l | l | l | l | l |}
\hline
$S$-paths & $D$-paths & chosen nodes & parameter & follow-up position & simplified position\\
\hline\hline
$(2);(7)$&$(3);(6)$&$3 \in Y_2;6 \in Y_2$&$k_2<2^i-4$&$(\{\{2,7\},\{3,6\}\},k_2)$&$(\{\{2,7\},\{3,6\}\},2^i-5)$\\
\hline
$(2,x);(7)$&$(3,7);(6)$&$3 \in Y_1;6 \in Y_2$&$k_1<2^{i-1}-2$&$(\{\{2\},\{3\}\},k_1)$&$(\{\{2\},\{3\}\},2^{i-1}$\\
\cline{4-6}
& & &$k_2<2^{i-1}-2$&$(\{\{6\},\{7\}\},k_2)$&$(\{\{6\},\{7\}\},2^{i-1}-1)$\\
\hline
$(2);(7,b)$&$(3);(6,a)$&$3 \in Y_2; 6 \in Y_1$&$k_1<2^{i-1}-2$&$(\{\{6\},\{7\}\},k_1)$&$(\{\{6\},\{7\}\},2^{i-1}-1)$\\
\cline{4-6}
& & &$k_2<2^{i-1}-2$&$(\{\{2\},\{3\}\},k_2)$&$(\{\{2\},\{3\}\},2^{i-1})$\\
\cline{3-6}
& &$3 \in Y_2;a \in Y_2$&$k_2<2^i-4$&$(\{\{2,b\},\{3,a\}\},k_2)$&$(\{\{2,b\},\{3,a\}\},2^i-5)$\\
\hline
$(2,x);(7,b)$&$(3,b);(6,a)$&$3 \in Y_1;6 \in Y_1$&$k_1<2^i-4$&$(\{\{2,7\},\{3,6\}\},k_1)$&$(\{\{2,7\},\{3,6\}\},2^i-5)$\\
\cline{3-6}
& &$3 \in Y_1;a \in Y_2$&$k_1<2^{i-1}-2$&$(\{\{2\},\{3\}\},k_1)$&$(\{\{2\},\{3\}\},2^{i-1})$\\
\cline{4-6}
& & &$k_2<2^{i-1}-2$&$(\{\{a\},\{b\}\},k_2)$&$(\{\{a\},\{b\}\},2^{i-1}-2)$\\
\hline
\end{tabular}
\caption{EU-move on $\{2_i,7_i\}$}\label{EU-move on 2,7}
\end{figure}

Player $S$ can also play an EU-move on $\{3_i,6_i\}$. Then he has to pick $(3_i)$, $(3_i,5_i)$, $(3_i,7_i)$, $(3_i,8_i)$ or $(3_i,a_{i-1})$ as a path for starting node $3_i$. But if he picks $(3_i,7_i)$, $D$ can guarantee $7_i$ as a shared node. So we can ignore this choice. Player $S$ also has to choose either $(6_i)$ or $(6_i,a_{i-1})$ as a path for starting node $6_i$. In Figure~\ref{EU-move on 3,6} the node $x$ replaces any successor of $3_i$. Then this table shows, that $S$ cannot win with such an EU-move either.

\begin{figure}[h]
\begin{tabular}{| l | l | l | l | l | l |}
\hline
$S$-paths & $D$-paths & chosen nodes & parameter & follow-up position & simplified position\\
\hline\hline
$(3);(6)$&$(2);(7)$&$2 \in Y_2;7 \in Y_2$&$k_2< 2^i-4$&$(\{\{2,7\},\{3,6\}\},k_2)$&$(\{\{2,7\},\{3,6\}\},2^i-5)$\\
\hline
$(3,5);(6)$&$(2,4);(7)$&$2 \in Y_1; 7 \in Y_2$&$k_1< 2^{i-1}-2$&$(\{\{2\},\{3\}\},k_1)$&$(\{\{2\},\{3\}\},2^{i-1})$\\
\cline{4-6}
& & &$k_2<2^{i-1}-2$&$(\{\{5,6\},\{7\}\},k_2)$&$(\{\{6\},\{7\}\},2^{i-1}-1)$\\
\cline{3-6}
& &$4 \in Y_2;7 \in Y_2$&$k_2<2^i-4$&$(\{\{4,7\},\{5,6\}\},k_2)$&$(\{\{4,7\},\{5,6\}\},2^i-4)$\\
\hline
$(3,8);(6)$&$(2,8);(7)$&$2 \in Y_1; 7 \in Y_2$&$k_1<2^{i-1}-2$&$(\{\{2\},\{3\}\},k_1)$&$(\{\{2\},\{3\}\},2^{i-1})$\\
\cline{4-6}
& & &$k_2<2^{i-1}-2$&$(\{\{6\},\{7\}\},k_2)$&$(\{\{6\},\{7\}\},2^{i-1}-1)$\\
\hline
$(3,a);(6)$&$(2,a);(7)$&$2 \in Y_1;7 \in Y_2$&$k_1<2^{i-1}-2$&$(\{\{2\},\{3\}\},k_1)$&$(\{\{2\},\{3\}\},2^{i-1})$\\
\cline{4-6}
& & &$k_2<2^{i-1}-2$&$(\{\{6\},\{7\}\},k_2)$&$(\{\{6\},\{7\}\},2^{i-1}-1)$\\
\hline
$(3);(6,a)$&$(2);(7,b)$&$2 \in Y_2;7 \in Y_1$&$k_1< 2^{i-1}-2$&$(\{\{6\},\{7\}\},k_1)$&$(\{\{6\},\{7\}\},2^{i-1}-1)$\\
\cline{4-6}
& & &$k_2<2^{i-1}-2$&$(\{\{2\},\{3\}\},k_2)$&$(\{\{2\},\{3\}\}2^{i-1})$\\
\cline{3-6}
& &$2 \in Y_2;b \in Y_2$&$k_2<2^i-4$&$(\{\{2,b\},\{3,a\}\},k_2)$&$(\{\{2,b\},\{3,a\}\},2^i-5)$\\
\hline
$(3,x);(6,a)$&$(2,a);(7,b)$&$2 \in Y_1;7 \in Y_1$&$k_1<2^i-4$&$(\{\{2,7\},\{3,6\}\},k_1)$&$(\{\{2,7\},\{3,6\}\},2^i-5)$\\
\cline{3-6}
& &$2 \in Y_1;b \in Y_2$&$k_1<2^{i-1}-2$&$(\{\{2\},\{3\}\},k_1)$&$(\{\{2\},\{3\}\},2^{i-1})$\\
\cline{4-6}
& & &$k_2<2^{i-1}-2$&$(\{\{a\},\{b\}\},k_2)$&$(\{\{a\},\{b\}\},2^{i-1}-2)$\\
\hline
\end{tabular}
\caption{EU-move on $\{3_i,6_i\}$}\label{EU-move on 3,6}
\end{figure}

Next we look at what happens if $S$ plays an AU-move in the current position. Figure~\ref{AU-move on 2,7} shows that $S$ cannot win by playing an AU-move on $\{2_i,7_i\}$, while Figure~\ref{AU-move on 3,6} shows that he cannot win by playing on $\{3_i,6_i\}$ either. With this we have shown that the position $(\{\{2_i,7_i\},\{3_i,6_i\}\},2^i-4)$ is also a winning position for player $D$.

\begin{figure}[h]
\begin{tabular}{| l | l | l | l | l | l |}
\hline
$S$-paths & $D$-paths & chosen nodes & parameter & follow-up position & simplified position\\
\hline\hline
$(2);(7)$&$(3);(6)$&$3 \in Y_1; 6 \in Y_2$&$k_1<2^{i-1}-2$&$(\{\{2\},\{3\}\},k_1)$&$(\{\{2\},\{3\}\},2^{i-1})$\\
\cline{4-6}
& & &$k_2<2^{i-1}-2$&$(\{\{6\},\{7\}\},k_2)$&$(\{\{6\},\{7\}\},2^{i-1}-1)$\\
\hline
$(2);(7,b)$&$(3);(6,a)$&$3 \in Y_1;a \in Y_1$&$k_1<2^i-4$&$(\{\{2,b\},\{3,a\}\},k_1)$&$(\{\{2,b\},\{3,a\}\},2^i-5)$\\
\cline{3-6}
& &$3 \in Y_1;a \in Y_2$&$k_1<2^{i-1}-2$&$(\{\{2\},\{3\}\},k_1)$&$(\{\{2\},\{3\}\},2^{i-1})$\\
\cline{4-6}
& & &$k_2<2^{i-1}-2$&$(\{\{a\},\{b\}\},k_2)$&$(\{\{a\},\{b\}\},2^{i-1}-2)$\\
\hline
$(2,4);(7,b)$&$(3,b);(6,a)$&$3\in Y_1;a \in Y_1$&$k_1<2^i-4$&$(\{\{2,b\},\{3,a\}\},k_1)$&$(\{\{2,b\},\{3,a\}\},2^i-5)$\\
\hline
$(2,7);(7,b)$&$(3,b);(6,a)$&$3\in Y_1;a \in Y_1$&$k_1<2^i-4$&$(\{\{2,b\},\{3,a\}\},k_1)$&$(\{\{2,b\},\{3,a\}\},2^i-5)$\\
\hline
$(2,8);(7,b)$&$(3,b);(6,a)$&$3\in Y_1;a \in Y_1$&$k_1<2^i-4$&$(\{\{2,b\},\{3,a\}\},k_1)$&$(\{\{2,b\},\{3,a\}\},2^i-5)$\\
\hline
$(2,b);(7,b)$&$(3,b);(6,a)$&$3\in Y_1;a \in Y_1$&$k_1<2^i-4$&$(\{\{2,b\},\{3,a\}\},k_1)$&$(\{\{2,b\},\{3,a\}\},2^i-5)$\\
\hline
\end{tabular}
\caption{AU-move on $\{2_i,7_i\}$}\label{AU-move on 2,7}
\end{figure}

\begin{figure}[h]
\begin{tabular}{| l | l | l | l | l | l |}
\hline
$S$-paths & $D$-paths & chosen nodes & parameter & follow-up position & simplified position\\
\hline\hline
$(3);(6)$&$(2);(7)$&$2 \in Y_1; 7\in Y_2$&$k_1<2^{i-1}-2)$&$(\{\{2\},\{3\}\},k_1)$&$(\{\{2\},\{3\}\},2^{i-1})$\\
\cline{4-6}
& & &$k_2<2^{i-1}-2$&$(\{\{6\},\{7\}\},k_2)$&$(\{\{6\},\{7\}\},2^{i-1}-1)$\\
\cline{3-6}
& &$2 \in Y_2;7 \in Y_1$&$k_1<2^{i-1}-2$&$(\{\{6\},\{7\}\},k_1)$&$(\{\{6\},\{7\}\},2^{i-1}-1)$\\
\cline{4-6}
& & &$k_2<2^{i-1}-2$&$(\{\{2\},\{3\}\},k_2)$&$(\{\{2\},\{3\}\},2^{i-1})$\\
\hline
$(3,5);(6)$&$(2,4);(7)$&$4 \in Y_1;7 \in Y_1$&$k_1<2^i-4$&$(\{\{4,7\},\{3,6\}\},k_1)$&$(\{\{4,7\},\{3,6\}\},2^i-4)$\\
\cline{3-6}
& &$4 \in Y_1;7 \in Y_2$&$k_1<2^{i-1}-2$&$(\{\{4\}\{5,6\}\},k_1)$&$(\{\{4\},\{5\}\},2^{i-1}-1)$\\
\cline{4-6}
& & &$k_2<2^{i-1}-2$&$(\{\{5,6\},\{7\}\},k_2)$&$(\{\{6\},\{7\}\},2^{i-1}-1)$\\
\cline{3-6}
& &$4 \in Y_2; 7 \in Y_1$&$k_1<2^{i-1}-2$&$(\{\{5,6\},\{7\}\},k_1)$&$(\{\{6\},\{7\}\},2^{i-1}-1)$\\
\cline{4-6}
& & &$k_2<2^{i-1}-2$&$(\{\{4\},\{5,6\}\},k_2)$&$(\{\{4\},\{5\}\},2^{i-1}-1)$\\
\hline
$(3);(6,a)$&$(2);(7,b)$&$2 \in Y_1;b \in Y_1$&$k_1<2^i-4$&$(\{\{2,b\},\{3,a\}\},k_1)$&$(\{\{2,b\},\{3,a\}\},2^i-5)$\\
\cline{3-6}
& &$2 \in Y_1;b \in Y_2$&$k_1<2^{i-1}-2$&$(\{\{2\},\{3\}\},k_1)$&$(\{\{2\},\{3\}\},2^{i-1})$\\
\cline{4-6}
& & &$k_2<2^{i-1}-2$&$(\{\{a\},\{b\}\},k_2)$&$(\{\{a\},\{b\}\},2^{i-1}-2)$\\
\cline{3-6}
& &$2 \in Y_2;b \in Y_1$&$k_1<2^{i-1}-2)$&$(\{\{a\},\{b\}\},k_1)$&$(\{\{a\},\{b\}\},2^{i-1}-2)$\\
\cline{4-6}
& & &$k_2<2^{i-1}-2$&$(\{\{2\},\{3\}\},k_2)$&$(\{\{2\},\{3\}\},2^{i-1})$\\
\hline
$(3,5);(6,a)$&$(2,a);(7,b)$&$2 \in Y_1;b \in Y_1$&$k_1<2^i-4$&$(\{\{2,b\},\{3,a\}\},k_1)$&$(\{\{2,b\},\{3,a\}\},2^i-5)$\\
\hline
$(3,8);(6,a)$&$(2,a);(7,b)$&$2 \in Y_1;b \in Y_1$&$k_1<2^i-4$&$(\{\{2,b\},\{3,a\}\},k_1)$&$(\{\{2,b\},\{3,a\}\},2^i-5)$\\
\hline
$(3,a);(6,a)$&$(2,a);(7,b)$&$2 \in Y_1;b \in Y_1$&$k_1<2^i-4$&$(\{\{2,b\},\{3,a\}\},k_1)$&$(\{\{2,b\},\{3,a\}\},2^i-5)$\\
\hline
\end{tabular}
\caption{AU-move on $\{3_i,6_i\}$}\label{AU-move on 3,6}
\end{figure}

The next position we examine is $(\{\{2_i\},\{1_i,3_i\}\},2^i-3)$. Here player $S$ can play an operator-move. However because every successor of $2_i$ is also a successor of $1_i$ or $3_i$, he cannot win by playing this move on $\{2_i\}$. When playing an operator-move on $\{1_i,3_i\}$ he has to pick a successor each for $1_i$ and for $3_i$. His only choice that does not lead to an immediate victory for $D$ is picking $6_i$ as a successor of $1_i$ and $5_i$ as a successor of $3_i$. In this case $D$ answers with the set $\{4_i,7_i\}$. Playing like this leads to the position $(\{\{4_i,7_i\},\{5_i,6_i\}\},2^i-4)$. A position we have already shown to be a winning position for player $D$.

When $S$ play a split-move, he has to split the set $\{1_i,3_i\}$ into $\{1_i\}$ and $\{3_i\}$. He also has to split the parameter $2^i-3$ into $k_1$ and $k_2$. Player $D$ always answers with the set $\{2_i\}$. Then she chooses the side with the smaller parameter $k'$. This creates either the position $(\{\{1_i\},\{2_i\}\},k')$ or $(\{\{2_i\},\{3_i\}\},k')$. Since we know $k'< 2^{i-1}-2$, we have already shown that both of these position are winning positions for $D$.

Player $S$ can also play an EU-move. If he plays such a move on $\{2_i\}$, he has to pick a path with starting node $2_i$. Because of Lemma~\ref{3 properties} we are only interested in paths with two nodes or less that end in a node that is not a shared successor of $1_i$ and $3_i$. This leaves only the paths $(2_i)$, $(2_i,4_i)$ and $(2_i,7_i)$. In Figure~\ref{EU-move on 2} we show that every choice leads to a winning position for player $D$.

\begin{figure}[h]
\begin{tabular}{| l | l | l | l | l | l |}
\hline
$S$-paths & $D$-paths & chosen nodes & parameter & follow-up position & simplified position\\
\hline\hline
$(2)$&$(1);(3)$&$1\in Y_2;3 \in Y_2$&$k_2<2^i-3$&$(\{\{2\},\{1,3\}\},k_2)$&$(\{\{2\},\{1,3\}\},2^i-4)$\\
\hline
$(2,4)$&$(1,4);(3,5)$&$1 \in Y_1;3 \in Y_1$&$k_1< 2^i-3$&$(\{\{2\},\{1,3\}\},k_1)$&$(\{\{2\},\{1,3\}\},2^i-4)$\\
\cline{3-6}
& &$1 \in Y_1;5 \in Y_2$&$k_1 < 2^{i-1}-1$&$(\{\{1\},\{2\}\},k_1)$&$(\{\{1\},\{2\}\},2^{i-1})$\\
\cline{4-6}
& & &$k_2< 2^{i-1}-2$&$(\{\{4\},\{5\}\},k_2)$&$(\{\{4\},\{5\}\},2^{i-1}-1)$\\
\hline
$(2,7)$&$(1,6);(3,7)$&$1 \in Y_1;3 \in Y_1$&$k_1<2^i-3$&$(\{\{2\},\{1,3\}\},k_1)$&$(\{\{2\},\{1,3\}\},2^i-4)$\\
\cline{3-6}
& &$6 \in Y_2;3 \in Y_1$&$k_1 < 2^{i-1}-1$&$(\{\{2\},\{3\}\},k_1)$&$(\{\{2\},\{3\}\},2^{i-1})$\\
\cline{4-6}
& & &$k_2< 2^{i-1}-2$&$(\{\{6\},\{7\}\},k_2)$&$(\{\{6\},\{7\}\},2^{i-1}-1)$\\
\hline
\end{tabular}
\caption{EU-move on $\{2_i\}$}\label{EU-move on 2}
\end{figure}

For an EU-move on $\{1_i,3_i\}$ player $D$ has to choose between $(1_i)$ and $(1_i,6_i)$ as a path for $1_i$, as well as between $(3_i)$ and $(3_i,5_i)$ for $3_i$. These two paths each can be combined in a total of four different ways and as Figure~\ref{EU-move on 1,3} shows, none of these combination allow $S$ to win.

\begin{figure}[h]
\begin{tabular}{| l | l | l | l | l | l |}
\hline
$S$-paths & $D$-paths & chosen nodes & parameter & follow-up position & simplified position\\
\hline\hline
$(1),(3)$&$(2)$&$2 \in Y_2$&$k_2<2^i-3$&$(\{\{2\},\{1,3\}\},k_2)$&$(\{\{2\},\{1,3\}\},2^i-4)$\\
\hline
$(1,6);(3)$&$(2);(2,7)$&$2 \in Y_2;2 \in Y_1$&$k_1<2^{i-1}-1$&$(\{\{1\},\{2\}\},k_1)$&$(\{\{1\},\{2\}\},2^{i-1})$\\
\cline{4-6}
& & &$k_2<2^{i-1}-2$&$(\{\{2\},\{3\}\},k_2)$&$(\{\{2\},\{3\}\},2^{i-1})$\\
\cline{3-6}
& &$2\in Y_2; 7 \in Y_2$&$k_2<2^i-3$&$(\{\{2,7\},\{3,6\}\},k_2)$&$(\{\{2,7\},\{3,6\}\},2^i-4)$\\
\hline
$(1);(3,5)$&$(2);(2,4)$&$2 \in Y_2;2 \in Y_1$&$k_1<2^{i-1}-1$&$(\{\{2\},\{3\}\},k_1)$&$(\{\{2\},\{3\}\},2^{i-1})$\\
\cline{4-6}
& & &$k_2<2^{i-1}-2$&$(\{\{1\},\{2\}\},k_2)$&$(\{\{1\},\{2\}\},2^{i-1})$\\
\cline{3-6}
& &$2 \in Y_2;4 \in Y_2$&$k_2< 2^i-3$&$(\{\{2,4\},\{1,5\}\},k_2)$&$(\{\{2,4\},\{1,5\}\},2^i-4)$\\
\hline
$(1,6);(3,5)$&$(2,4);(2,7)$&$2 \in Y_1;2 \in Y_1$&$k_1<2^i-3$&$(\{\{2\},\{1,3\}\},k_1)$&$(\{\{2\},\{1,3\}\},2^i-4)$\\
\cline{3-6}
& &$2 \in Y_1;7 \in Y_2$&$k_1<2^i-3$&$(\{\{2\},\{1,3\}\},k_1)$&$(\{\{2\},\{1,3\}\},2^i-4)$\\
\cline{3-6}
& &$4 \in Y_2;2 \in Y_1$&$k_1<2^i-3$&$(\{\{2\},\{1,3\}\},k_1)$&$(\{\{2\},\{1,3\}\},2^i-4)$\\
\cline{3-6}
& &$4 \in Y_2;7 \in Y_2$&$k_2< 2^i-3$&$(\{\{4,7\},\{5,6\}\},k_2)$&$(\{\{4,7\},\{5,6\}\},2^i-4)$\\
\hline
\end{tabular}
\caption{EU-move on $\{1_i,3_i\}$}\label{EU-move on 1,3}
\end{figure}

For an AU-move on $\{2_i\}$, player $D$ may pick the same paths as during an EU-move on this set. For all options we need to examine, we do so in Figure~\ref{AU-move on 2}. There we can see that none of them lead to a winning position for $S$.

\begin{figure}[h]
\begin{tabular}{| l | l | l | l | l | l |}
\hline
$S$-paths & $D$-paths & chosen nodes & parameter & follow-up position & simplified position\\
\hline\hline
$(2)$&$(1);(3)$&$1 \in Y_1;3 \in Y_2$&$k_1<2^{i-1}-1$&$(\{\{1\},\{2\}\},k_1)$&$(\{\{1\},\{2\}\},2^{i-1})$\\
\cline{4-6}
& & &$k_2<2^{i-1}-2$&$(\{\{2\},\{3\}\},k_2)$&$(\{\{2\},\{3\}\},2^{i-1})$\\
\hline
$(2,4)$&$(1,4);(3,5)$&$1 \in Y_1;5 \in Y_1$&$k_1<2^i-3$&$(\{\{2,4\},\{1,5\}\},k_1)$&$(\{\{2,4\},\{1,5\}\},2^i-4)$\\
\hline
$(2,7)$&$(1,6);(3,7)$&$6 \in Y_1;3 \in Y_1$&$k_1< 2^i-3$&$(\{\{2,7\},\{3,6\}\},k_1)$&$(\{\{2,7\},\{3,6\}\},2^i-4)$\\
\hline
\end{tabular}
\caption{AU-move on $\{2_i\}$}\label{AU-move on 2}
\end{figure}

Now we examine all possible AU-moves on $\{1_i,3_i\}$. For every such move we know $1_i \in X_1$ and $3_i \in X_1$. So if $S$ marks the node $2_i$ and adds it to $Y_1$, the position $(t(\{X_1,Y_1\}),k_1)$ is atleast as difficult for $S$ as the current position according to Corollary~\ref{subset-positions}. So $S$ cannot win by playing this way and we can skip these moves. For the remaining options Figure~\ref{AU-move on 1,3} shows that they all lead to winning positions for $D$. So $(\{\{2_i\},\{1_i,3\}\},2^i-3)$ has to be a winning position for her as well.

\begin{figure}[h]
\begin{tabular}{| l | l | l | l | l | l |}
\hline
$S$-paths & $D$-paths & chosen nodes & parameter & follow-up position & simplified position\\
\hline\hline
$(1,6);(3)$&$(2);(2,7)$&$2 \in Y_2;7 \in Y_1$&$k_1<2^{i-1}-1$&$(\{\{6\},\{7\}\},k_1)$&$(\{\{6\},\{7\}\},2^{i-1}-1)$\\
\cline{4-6}
& & &$k_2<2^{i-1}-2$&$(\{\{2\},\{3\}\},k_2)$&$(\{\{2\},\{3\}\},2^{i-1})$\\
\hline
$(1);(3,5)$&$(2);(2,4)$&$2 \in Y_2;4\in Y_1$&$k_1<2^{i-1}-1$&$(\{\{4\},\{5\}\},k_1)$&$(\{\{4\},\{5\}\},2^{i-1}-1)$\\
\cline{4-6}
& & &$k_2<2^{i-1}-2$&$(\{\{1\},\{2\}\},k_2)$&$(\{\{1\},\{2\}\},2^{i-1})$\\
\hline
$(1,6);(3,5)$&$(2,4);(2,7)$&$4 \in Y_1;7 \in Y_1$&$k_1<2^i-3$&$(\{\{4,7\},\{5,6\}\},k_1)$&$(\{\{4,7\},\{5,6\}\},2^i-4)$\\
\cline{3-6}
& &$4 \in Y_1;7 \in Y_2$&$k_1<2^{i-1}-1$&$(\{\{4\},\{5,6\}\},k_1)$&$(\{\{4\},\{5\}\},2^{i-1}-1)$\\
\cline{4-6}
& & &$k_2<2^{i-1}-2$&$(\{\{5,6\},\{7\}\},k_2)$&$(\{\{6\},\{7\}\},2^{i-1}-1)$\\
\hline
\end{tabular}
\caption{AU-move on $\{1_i,3_i\}$}\label{AU-move on 1,3}
\end{figure}

Finally we get to the position $(\{\{a_i\},\{b_i\}\},2^i-2)$. In this position both sets contain only one node each. So $S$ cannot play a split-move. Also because of property $P2$ from Lemma~\ref{3 properties} we do not need to look into $S$ playing an EU- or AU-move in this position either. If $S$ plays an operator-move, he has to pick a successor of $a_i$ or of $b_i$. The only node that is a successor of one of these nodes but not of both is $2_i$. If $S$ picks $2_i$ as a successor of $a_i$, then $D$ answers with $\{1_i,3_i\}$. This moves the play to the position $(\{\{2_i\},\{1_i,3_i\}\},2^i-3)$, which we have already shown to be a winning position for $D$. Then $(\{\{a_i\},\{b_i\}\},2^i-2)$ has to be a winning position for $D$ as well. Because of this there cannot be a CTL-formula $\chi_i$ that separates $(\mathcal{M}_c,a_i)$ from $(\mathcal{M}_c,b_i)$ with $h(i)=2^i-2$ or less operators.
\end{proof}

\begin{thm}
The logic ML$[\leftrightarrow]$ is exponentially more operator-succinct and exponentially more succinct than CTL on the class of Kripke structures with a serial and transitive accessibility relation.
\end{thm}

\begin{proof}
This follows directly from Lemma~\ref{bi-implication CTL}, Lemma~\ref{no bi-implication ctl} and the fact that $\mathcal{M}_c$ is transitive and serial.
\end{proof}

This succinctness gap differs from all other succinctness gaps presented in this paper. This is because even though ML$[\leftrightarrow]$ is exponentially more succinct than CTL, CTL is also more expressive than ML$[\leftrightarrow]$. Meaning there are CTL-formulas without an equivalent ML$[\leftrightarrow]$. A gap in expressiveness can be seen as something like an infinite succinctness gap, since there are no equivalent formulas of any size. So here we have a case, where some properties can be expressed more succinctly in one logic, while other properties can be expressed more succinctly, or even at all, in the second logic.

This Theorem also implies the existence of multiple other succinctness gaps. First of all every ML$[\leftrightarrow]$-formula can be translated into an equivalent CTL$[\leftrightarrow]$-formula of equal size, because $\Diamond$ has the same semantics as $EX$ and $\Box$ has the same semantics as $AX$. So this exponential succinctness gap also exists between CTL$[\leftrightarrow]$ and CTL itself. We have previously shown that a succinctness gap between the extension of a logic with $\leftrightarrow$ and the logic itself also implies a succinctness gap between the representation as DAGs and the representation as formulas. Lastly all of these succinctness gaps also exist for all supersets of the class of all structures with a serial and transitive accessibility relation.

\begin{cor}
The logics CTL$[\leftrightarrow]$ and DAG(CTL) are exponentially more operator-succinct and exponentially more succinct than  CTL on the following classes of pointed Kripke-structures:
\begin{enumerate}
\item The class of all pointed Kripke structures.
\item The class of all pointed Kripke structures with a serial accessibility relation.
\item The class of all pointed Kripke structures with a transitive accessibility relation.
\item The class of all pointed Kripke structures with a serial and transitive accessibility relation.
\end{enumerate}
\end{cor}

\section{Conclusion}
The main result of this paper is the development of two new model comparison games, the $\mathcal{M}$-Operator-Number game for modal logic and the $\mathcal{M}$-operator-number game for computation tree logic. These games can be used to prove lower bounds for the number of modal or temporal operators necessary to express certain properties in modal logic or computation tree logic respectively. An interesting follow-up question to this might be: For which logics and which parameters can a similar game be designed? If there is a version of the Hella-Vilander game for some logic, it should be possible to adjust that game to only count all occurrences of one specific type of operator. This should work for every type of operator of that logic. The more interesting part of that question is whether or not some process similar to our idea of 'tidying up' is possible. A positive answer is given for FO$^l$ in \cite{Vinall-Smeeth2024}. However additional research is necessary to achieve a better understanding of this topic.

We also give a more complete overview of the succinctness of modal logic with bi-implication on different classes of Kripke structures. For every possible combination of the properties reflexive, transitive and symmetrical, we have either proven or disproven the existence of an exponential succinctness gap between ML and ML$[\leftrightarrow]$ on the corresponding class of structures.
A possible future task could be to look into even more classes of Kripke structures, for example those with a function or a partial function as accessibility relation. Although these examples might require a quite different proof from the ones presented in this paper.

For computation tree logic we have shown, that the bi-implication allows us to write exponentially shorter formulas as well. There are even modal logic formulas using the bi-implication, so that every equivalent computation tree logic formula has to be exponentially larger. This even holds if only serial and transitive graphs are considered. Of course it would be desirable to find out for what other classes this succinctness gap exists. In particular we are interested in the same classes we examined for modal logic. Our proof heavily relies on the transitivity of the graph, which allows us circumvent the complexities of the temporal operators. So for non transitive graphs we might require a different approach.

In future we could also examine how the bi-implications influences formula size in other logics. For modal logic and computation tree logic we have proven a succinctness gap. For propositional logic we know from \cite{Pratt} that there is no such gap. A promising future candidate might be first order logic FO. Alternatively we could look into other temporal logics, like LTL or CTL$^*$.

\newpage

\bibliographystyle{alphaurl}
\bibliography{Literatur}

\end{document}